\documentclass[aip,jcp,preprint,amsmath,amssymb]{revtex4-1}

\pdfoutput=1

\usepackage[utf8]{inputenc}
\usepackage{times}
\usepackage{graphicx}
\usepackage{xspace}
\usepackage{dcolumn}

\usepackage{color}

\newcommand{\FIG}[3]{
\begin{figure}[htbp]
\begin{center}
\includegraphics[width=\columnwidth]{#2}
\end{center}
\caption{#3\label{fig:#1}}
\end{figure}}

\def\kt{k_{\textrm{B}}T\xspace}
\def\const{{\textrm{const}}\xspace}

\newcommand{\Exp}[1]{\exp\left(#1\right)}
\newcommand{\Avg}[1]{\langle #1\rangle}

\newcommand{\dd}[2]{\frac{d}{d#2}#1}

\def\Myo{Myosin {II}\xspace}
\def\myo{myosin {II}\xspace}
\def\atp{ATP\xspace}
\def\adp{ADP\xspace}
\def\pho{P$_{\textrm{i}}$\xspace}

\def\nt{N_{\textrm{t}}\xspace}
\def\nb{N_{\textrm{b}}\xspace}

\def\fdel{F_0\xspace}
\def\fext{F_{\textrm{ext}}\xspace}
\def\fs{F_{\textrm{s}}\xspace}
\def\favg{\Avg{F_{\textrm{ext}}}\xspace}

\def\epp{E_{\textrm{pp}}\xspace}
\def\eel{E^{\textrm{el}}\xspace}
\def\ext{E^{\textrm{ext}}\xspace}

\def\fn{F_n\xspace}
\def\xin{\xi_n\xspace}

\def\xij{x_{ij}\xspace}
\def\fij{F_{ij}\xspace}
\def\eij{E_{ij}\xspace}

\def\xn{x_n\xspace}
\def\zn{z_n\xspace}

\def\zfil{z_{\textrm{fil}}\xspace}
\def\dzfil{{\dot z}_{\textrm{fil}}\xspace}
\def\zavg{{\bar z}\xspace}
\def\xavg{{\bar x}\xspace}

\def\dzavgon{\Delta{\bar z}^{\textrm{on}}\xspace}
\def\dzavgoff{\Delta{\bar z}^{\textrm{off}}\xspace}

\def\dzon{\Delta z^{\textrm{on}}\xspace}
\def\dzoff{\Delta z^{\textrm{off}}\xspace}

\def\vu{v_{\textrm{u}}\xspace}

\def\km{k_{\textrm{m}}\xspace}
\def\kf{k_{\textrm{f}}\xspace}
\def\kfc{k^{\textrm{c}}_{\textrm{f}}\xspace}

\def\pbnd{{\hat p}\xspace}

\def\vb{v_{\textrm{b}}\xspace}
\def\vu{v_{0}\xspace}
\def\veff{v_{\textrm{eff}}\xspace}

\def\dw{d_{\textrm{w}}\xspace}
\def\rd{\rho_{\textrm{d}}\xspace}
\def\rds{\rho^{\rm single}_{\textrm{d}}\xspace}

\def\tdet{T_{10}\xspace}
\def\tatt{T_{01}\xspace}

\def\s{\operatorname{s}}
\def\Hz{\s^{-1}}
\def\pN{\operatorname{pN}}

\def\nm{\operatorname{nm}}
\def\mob{\nm\pN^{-1}\s^{-1}}

\def\eq#1{Eq.~\eqref{eq:#1}\xspace}
\def\eqs#1#2{Eqs.~\eqref{eq:#1} and \eqref{eq:#2}\xspace}
\def\eqr#1#2{Eqs.~\eqref{eq:#1} to \eqref{eq:#2}\xspace}
\def\fig#1{Fig.~\ref{fig:#1}\xspace}
\def\tab#1{Tab.~\ref{tab:#1}\xspace}
\def\scn#1{Sec.~\ref{scn:#1}\xspace}
\def\app#1{Appendix \ref{app:#1}\xspace}

\begin{document}

\date{\today}

\title{Stochastic dynamics of small ensembles of non-processive molecular motors: \\ the parallel cluster model}

\author{Thorsten Erdmann} \affiliation{BioQuant, Heidelberg University, Im Neuenheimer Feld 267, 69120 Heidelberg, Germany and Institute for Theoretical Physics, Heidelberg University, Philosophenweg 19, 69120 Heidelberg, Germany} 
\author{Philipp J.~Albert}\affiliation{BioQuant, Heidelberg University, Im Neuenheimer Feld 267, 69120 Heidelberg, Germany and Institute for Theoretical Physics, Heidelberg University, Philosophenweg 19, 69120 Heidelberg, Germany} 
\author{Ulrich S.~Schwarz}\affiliation{BioQuant, Heidelberg University, Im Neuenheimer Feld 267, 69120 Heidelberg, Germany and Institute for Theoretical Physics, Heidelberg University, Philosophenweg 19, 69120 Heidelberg, Germany} 

\begin{abstract}
Non-processive molecular motors have to work together in ensembles in
order to generate appreciable levels of force or movement. In skeletal
muscle, for example, hundreds of \myo molecules cooperate in thick
filaments. In non-muscle cells, by contrast, small groups with few tens
of non-muscle \myo motors contribute to essential cellular processes
such as transport, shape changes or mechanosensing. Here we introduce a
detailed and analytically tractable model for this important situation.
Using a three-state crossbridge model for the \myo motor cycle and
exploiting the assumptions of fast power stroke kinetics and equal load
sharing between motors in equivalent states, we reduce the stochastic
reaction network to a one-step master equation for the binding and
unbinding dynamics (\textit{parallel cluster model}) and derive the
rules for ensemble movement. We find that for constant external load,
ensemble dynamics is strongly shaped by the catch bond character of
\myo, which leads to an increase of the fraction of bound motors under
load and thus to firm attachment even for small ensembles. This
adaptation to load results in a concave force-velocity relation
described by a Hill relation. For external load provided by a linear
spring, \myo ensembles dynamically adjust themselves towards an
isometric state with constant average position and load. The dynamics of
the ensembles is now determined mainly by the distribution of motors
over the different kinds of bound states. For increasing stiffness of
the external spring, there is a sharp transition beyond which \myo can
no longer perform the power stroke. Slow unbinding from the
pre-power-stroke state protects the ensembles against detachment.
\end{abstract}

\maketitle

\section{Introduction}

Numerous processes in single cells and tissues require the generation of
mechanical force and directed motion. Most of these processes are based
on the activity of molecular motors interacting with the filaments of
the cytoskeleton, that is, motors from the dynein-, kinesin- and
myosin-families interacting with microtubule or actin filaments
\cite{a:LipowskyKlumpp2005}. Examples include separation of chromosomes
and closure of the constriction ring during cell division, intracellular
transport of cargo vesicles and organelles, contraction of muscle cells,
large-scale rearrangements in a developing tissue, and wound closure
after tissue injury. The mechanical energy used in these processes is
gained by hydrolysis of \atp and drives a cycle of conformational
changes in the allosteric motor molecules. Over the last decades, the
way single motor molecules work has been dissected in great quantitative
detail \cite{rv:Howard1997, rv:ValeMilligan2000}. However, it remains a
formidable challenge to understand how molecular motors work in the
physiological context of cells and tissues, where they usually
collaborate in groups \cite{a:GuerinEtAl2010}. Here we theoretically
address one crucial aspect of this situation, namely force generation in
small ensembles of non-processive motors.

A large research effort has been focused on processive motors which stay
attached to the substrate sufficiently long as to not loose contact for
many motor cycles \cite{rv:Vale2003}. For example, the two motor heads
of conventional kinesin typically take more than $100$ steps of $8\nm$
length before the motor unbinds from its microtubule track
\cite{a:SvobodaEtAl1993, a:ValeEtAl1996}. Although this property would
enable processive motors to work alone, experimental evidence suggests
that also processive motors in a physiological context often collaborate
in small groups \cite{a:AshkinEtAl1990}. The main benefit here is that
attaching several motors to the same cargo increases the walk length
dramatically \cite{a:KlumppLipowsky2005a} and allows the cargo to pass
over defects on the track and change reliably between tracks of finite
length \cite{a:ChaiEtAl2009}. Furthermore, because the velocity of a
processive motor typically decreases with the applied load, groups of
motors sharing an external load are able to transport cargo at larger
velocities or to exert larger forces on a cargo or an elastic element.
As an example for the latter, it has been shown that the force necessary
to pull membrane tubes from a lipid vesicle can only be produced by
groups of processive motors \cite{a:KosterEtAl2003, a:LeducEtAl2004}.

In contrast to processive motors, non-processive motors cannot do useful
work single-handedly and therefore necessarily have to operate in groups
in order to generate persistent motion or appreciable levels of force.
The paradigm for a non-processive motor acting in ensembles is \myo in
cross-striated skeletal muscle \cite{b:McMahon1984}. In the muscle
sarcomere, hundreds of \myo motors are assembled in so-called thick
filaments. The arrangement of \myo in a thick filament is of bipolar
order, that is, the \myo motors in the two halves of a thick filament
are oriented in opposing direction. The \myo motors in either half of a
thick filament walk as an ensemble on so-called thin filaments, actin
filaments which are anchored in the two Z-discs bounding the sarcomere.
The arrangement of thick and thin filaments is symmetric with respect to
the mid-plane of the sarcomere so that motor activity leads to muscle
contraction. Investigation of the structure of muscle has been
facilitated by the remarkable precision of the spatial arrangement of
\myo motors in the sarcomere. For example, in frog skeletal muscle each
half of a thick filament contains $294$ \myo motors which are arranged
at a very regular distance of $14.5\nm$ \cite{a:PiazzesiEtAl2007}.

Even before \myo had been biochemically characterized, the first
theoretical description of muscle contraction was already based on the
non-equilibrium binding and unbinding kinetics of myosin to actin which
effectively rectified thermal fluctuations of an elastic element
\cite{a:Huxley1957}. In this early model, myosin was described as an
actin binding site fluctuating in a harmonic potential and binding
preferentially ahead of and unbinding preferentially behind its
equilibrium position, thus inducing a net force displacing the actin
filament. Precise measurements of contraction speed as function of force
in combination with X-ray diffraction and detailed modeling allowed to
identify the chemical and mechanical details of the \myo hydrolysis
cycle and led to the development of the crossbridge model
\cite{a:HuxleySimmons1971, a:GeevesHolmes2005}, which provided a
molecular mechanism for the principle of preferential binding and
unbinding proposed by \textcite{a:Huxley1957}. Those advances
inspired theoretical models analyzing the statistical physics of large
ensembles of \myo motors simultaneously pulling on a single filament
\cite{a:LeiblerHuse1993, a:Duke1999}. With these models it has been
shown that in order to describe the response of skeletal muscle to
varying loading conditions it is essential that the unbinding rate of
\myo from actin is a decreasing function of the applied load. In
contrast to, e.g., the processive motor kinesin, this make \myo a catch
bond rather than a slip bond \cite{a:VeigelEtAl2003, a:GuoGuilford2006}
and leads to the recruitment of additional crossbridges under load.
Experimentally, this has been confirmed in a combination of mechanical
and X-ray techniques \cite{a:PiazzesiEtAl2007}.

Apart from muscle tissue, groups of \myo molecular motors are also
active in the organization of the actin cytoskeleton of non-muscle
tissue cells like fibroblasts. Increasing evidence shows that non-muscle
\myo motors acting in small groups play a crucial role in cell adhesion
and migration, with dramatic implications for development, health and
disease \cite{a:VicenteEtAl2009}. For example, \myo motors contribute to
the retrograde flow of the actin cytoskeleton away from the leading edge
during cell spreading and migration \cite{a:GuptonWatermanStorer2006} as
well as to the maintenance of cortical tension underlying cell shape,
movement and division \cite{a:WakatsukiEtAl2003, a:PaluchEtAl2005}.
Adherent tissue cells in culture tend to form contractile actin bundles
called \textit{stress fibers} which play an important role in force
generation and mechanosensing \cite{a:PellegrinMellor2007}. There exist
several kinds of stress fibers and only some of them (most notably
ventral stress fibers) bear some similarity with skeletal muscle in
being characterized by a sarcomeric organization with alternating
regions of the crosslinker $\alpha$-actinin and \myo
\cite{a:PetersonEtAl2004}.

While skeletal and cardiac muscle show a large degree of order, smooth
muscle is characterized by far more disordered actin-myosin assemblies.
Nevertheless, in all types of muscle, \myo motors work in large groups.
The organization of cytoskeletal actin and myosin structures is, in
general, far more disordered and far more dynamic than the actin-myosin
assemblies in all types of muscle tissue. Most importantly, the number
of \myo motors in the ensembles is much smaller: non-muscle \myo is
usually organized in minifilaments with a bipolar structure similar to
thick filaments but comprising only $10$-$30$ \myo motors, as has been
estimated from the size of minifilaments in electron micrographs
\cite{a:VerkhovskyBorisy1993}. The exact numbers of \myo motors in
different actin modules will vary depending on the cellular conditions
under which minifilaments are formed. Recently, contractile actin
bundles have been reconstituted \textit{in vitro}
\cite{a:ThoresenEtAl2011, a:ThoresenEtAl2013, a:SoaresEtAl2011}. It was
demonstrated that for sufficiently large concentration, \myo
minifilaments are able to contract parallel bundles as well as networks
of actin filaments with random polarity. In these experiments, the
number of \myo molecules in the minifilaments depends on the type of
\myo and the preparation of the minifilaments and was estimated to range
from $56$ for non-muscle \myo \cite{a:ThoresenEtAl2013} to several $100$
for muscle \myo. Due to the small duty ratio of (smooth) muscle \myo of
$0.04$ \cite{a:ThoresenEtAl2013} compared to a duty ratio of $0.23$ for
non-muscle \myo \cite{a:SoaresEtAl2011}, however, the number of \myo
attached to the substrate should be comparable for muscle and non-muscle
minifilaments. For a minifilament with $200$ muscle \myo molecules, the
number of attached motors may well be as low as $8$ \cite{a:ThoresenEtAl2011}.

For the small number of \myo motors in cytoskeletal minifilaments,
stochastic effects are expected to become important and have indeed been
observed in measurements of the tension generated by \myo motors in
reconstituted assays. In three bead assays, an actin filament is held in
two optical traps and \myo attached to the surface of a third bead is
allowed to bind to the actin filament. At elevated \myo concentrations
\cite{a:FinerEtAl1994, a:VeigelEtAl2003, a:DeboldEtAl2005}, several \myo
are able to bind simultaneously and the displacement of the actin
filament against the trap force is observed. In active gels, which are
\emph{in vitro} mixtures of actin filaments, actin crosslinkers and \myo
minifilaments, the fluctuating tension in the actin network induced by
the activity of \myo minifilaments is measured \cite{a:MizunoEtAl2007,
a:SoaresEtAl2011}. In motility assays, \myo motors are distributed over
a surface and the movement of an actin filament against an external
force is followed \cite{a:DukeEtAl1995, a:PlacaisEtAl2009,
a:HexnerKafri2009}. Such reconstituted assays characteristically reveal
noisy trajectories with a gradual build-up of tension followed by an
abrupt release, which is likely due to the detachment of the whole
ensemble of \myo motors, allowing the actin filament to slip. However, a
detailed and analytically tractable model for this important situation
is still missing.

In order to interpret experimental data from cellular and reconstituted
assays in terms of molecular properties, theoretical models are required
which allow to calculate experimentally accessible quantities like duty
ratio and force-velocity relation for small and variable number of
motors in an ensemble. At the same time, known molecular
characteristics, in particular the catch bond character of \myo and
conformational changes as the power stroke, should be included in the
model. Such a model should not only permit to estimate the number of
motors present in the experiment, but also provide a deeper
understanding of the generic principles governing the statistics of
motor ensembles. Here, we present such a model for the non-processive
molecular motor \myo. A short account of some of our results has been
given previously \cite{a:ErdmannSchwarz2012}.

Our work is motivated by the long tradition in modeling force generation
in skeletal muscle. Generic models for molecular motors investigate the
fundamental conditions for the generation of directed motion in a
thermal environment but do not take specific properties of molecular
motors into account. In ratchet models, a particle switches between
diffusive movement in a flat potential and in a periodic potential
\cite{a:JuelicherProst1995}. By breaking detailed balance for the
transitions between the two potential landscapes, directed motion of the
particle can ensue. Ratchet models allow to study generic effects of
cooperativity of a large number of motors such as the emergence of
directed motion in symmetric systems or spontaneous oscillations
\cite{a:JuelicherProst1995, a:JuelicherProst1997}. Diffusion and
switching of the particles is usually described in the framework of a
Fokker-Planck equation. Within this framework, the effect of a finite
ensemble size has been included by assuming a fluctuating drift velocity
with a noise intensity that increases with decreasing ensemble size
\cite{a:PlacaisEtAl2009}. This approach allows to observe effects
specific for finite sized ensembles, such as the reversal of the
direction of motion \cite{a:BadoualEtAl2002}. Beginning with the work of
Huxley \cite{a:Huxley1957}, some molecular characteristics were
introduced by assuming different conformational states of the motor
molecules. With the focus on the large assemblies of motors in muscle,
analytical progress was usually made using mean-field approximations. A
mean-field model for molecular motors with three conformational states,
in which the bound motors moved with given velocity was used to
calculate the force-velocity relation of an ensemble of motors
\cite{a:LeiblerHuse1993}. Adapting the transition rates between the
conformational states allowed to study processive as well as
non-processive motors. Onset of oscillatory behavior of the ensembles
was investigated in a generic two-state model, in which conformational
changes of the motors upon binding and unbinding allowed bound motors to
exert force on their environment \cite{a:VilfanEtAl1999}. Here, the
binding an unbinding rates could be adapted to describe different types
of motors and different force-velocity relations. In a generalization,
an ensemble of molecular motors working against a visco-elastic element
was investigated \cite{a:VilfanFrey2005}. Using a similar mean-field
approach as \textcite{a:LeiblerHuse1993} for a two-state model, the
dynamic behavior of ensembles was investigated, revealing limit cycle
oscillations induced by the coupling to the visco-elastic element. To
describe specific properties of muscle fibers, crossbridge models with
varying degree of detail have been used. Using computer simulations on
large ensemble of \myo with a crossbridge model including explicitly the
power stroke and load dependent unbinding from the post-power-stroke
state, details of the force-velocity relation for muscle fibers could be
fitted very accurately to experimental results and collective phenomena
such as the synchronization of the power stroke under load or the
transient response of muscle to a step change of the external load were
investigated \cite{a:Duke1999, a:Duke2000}. Coupling to an external
elastic element allowed to observe oscillations for the crossbridge
model \cite{a:VilfanDuke2003b}. Recently, a detailed crossbridge model
for \myo was used to describe also the activity of small \myo ensembles
in the cytoskeleton \cite{a:WalcottEtAl2012}. The activity of small
groups of \myo in motility assays was studied using computer
simulations. Large ensembles were described in a mean-field approach. By
comparison of the results to experiments, parameters of the model could
be determined. The focus of the model was on the description of the \atp
dependence of the transition rates. Therefore, the crossbridge model
included two separate post-power-stroke states of the bound motors but
did not include a bound pre-power-stroke state so that it did not
explicitly described the power stroke.

In order to study effects of molecular details for ensembles of \myo
motors, we use a crossbridge model with three states as a starting
point, which was originally used for skeletal muscle \cite{a:Duke2000,
a:VilfanDuke2003b}. Unlike \textcite{a:WalcottEtAl2012}, we use a pre-
and a post-power-stroke state as the two bound states, so that the power
stroke of \myo is included explicitly. To reduce the complexity of the
analytical description, we make two approximations: $(i)$ we assume that
molecular motors in equivalent conformational states have equal strain
and $(ii)$ we exploit a separation of time scales in the \myo cycle and
assume that there is thermal equilibrium of the bound states. The
partial mean-field approximation of the first assumption still
distinguishes between the two different bound states. It can be
justified by the small duty ratio of \myo motors which leads to a narrow
distribution of the strains of bound motors. The assumption of local
thermal equilibrium between bound states reduces the system to a
two-state model. The effective properties of these states, however,
still depend on the distribution over the two bound states. The two
approximations allow us to derive a one-step master equation for the
binding dynamics of the motors, which explicitly includes the effects of
strain-dependent rates and small system size. A one-step master equation
has been introduced before for transport by finite-sized ensembles of
processive motors with slip bond behavior \cite{a:KlumppLipowsky2005a},
but not for non-processive motors with catch bond behavior. Together
with rules for the displacement of en ensemble upon binding and
unbinding, the one-step master equation fully characterizes the dynamics
of ensembles. We investigate two paradigmatic loading conditions for the
ensemble: constant loading and linear loading, in which the external
load depends linearly on the position of the ensemble. For constant
external load, we can solve the one-step master equation for the
stationary states and derive binding properties and force-velocity
relation from these. For linear external load, the movement of the
ensemble feeds back to the load dependent binding rates, so that we have
to use computer simulations to analyze this case. In both loading
scenarios, we find that the motor ensemble adapts its dynamical state to
the external conditions in a way which is reminiscent of its
physiological function.

\section{Model}

\subsection{Crossbridge model for single non-processive motors}

\FIG{Fig01}{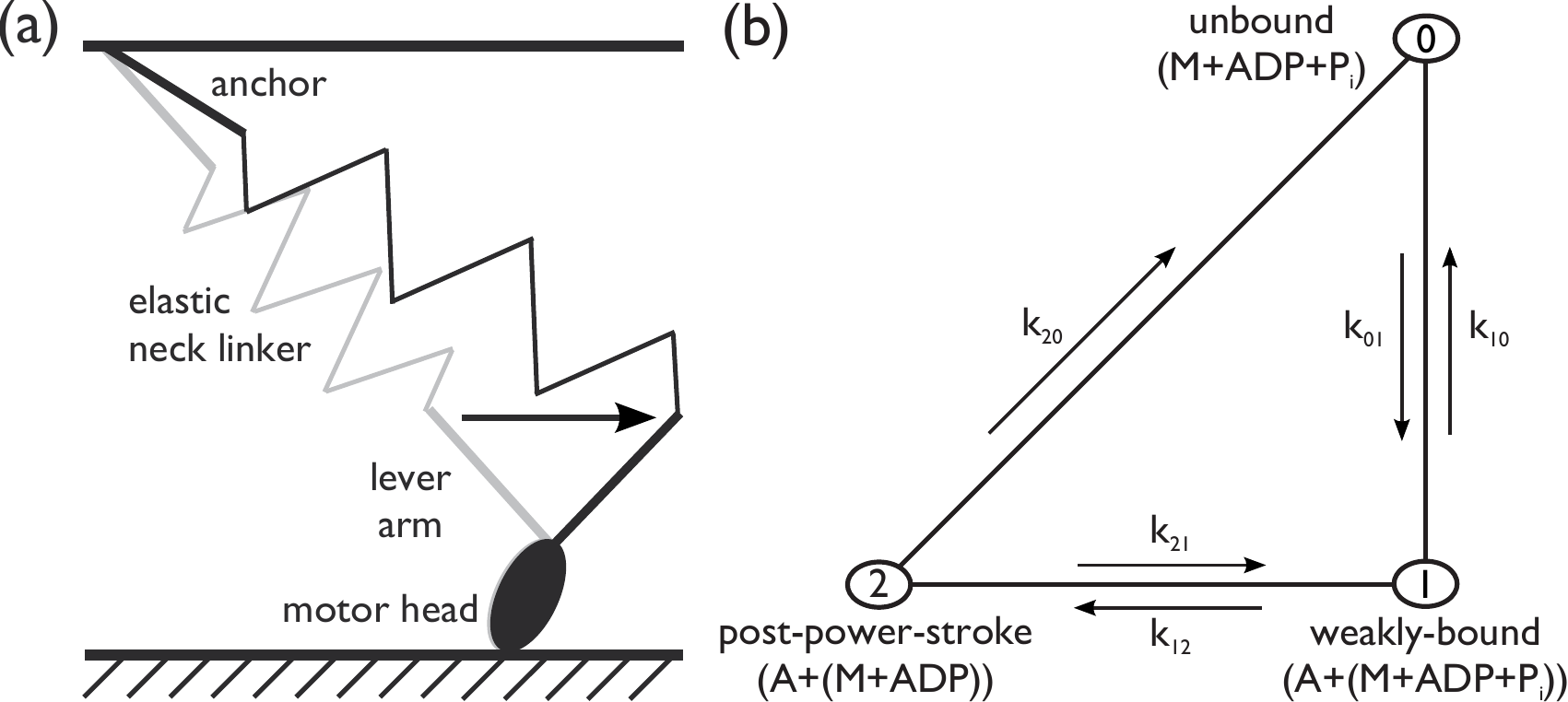}{Crossbridge model for non-processive \myo motors
molecules. (a) Mechanical elements of \myo. (b) \Myo motor cycle with
three discrete mechano-chemical states. In the unbound state $(0)$ the
motor head binds \adp and \pho and the lever arm is in the primed
conformation (gray in (a)). Binding to the substrate brings the motor to
the weakly-bound (pre-power-stroke) state $(1)$ with unchanged
mechanical conformation. After release of the \pho group, the lever arm
swings forward into the stretched conformation (black in (a)). This
power stroke brings the motor to the post-power-stroke state $(2)$.
Replacing \adp by \atp, unbinding from the substrate and \atp hydrolysis
brings the motor back to the unbound state $(0)$. Because of the
consumption of \atp, the last transition is irreversible. All other
transitions are reversible.}

To describe the mechanism of force generation by non-processive
molecular motors, we use a crossbridge model for \myo
\cite{a:HuxleySimmons1971, a:Duke1999}. Variants of cross-bridge models
differ by number and type of conformational states they include
\cite{a:Duke1999, a:Duke2000, a:WalcottEtAl2012}, depending on the focus
of the modeling approach. Here, we distinguish two bound conformations
of the motors and one unbound state. The essential mechanical elements
of \myo in our model are depicted schematically in \fig{Fig01}(a). The
motor head binds the motor to the substrate, which in the case of \myo
is an actin filament. The motor head also is the active domain of \myo
which binds \atp or the products of \atp hydrolysis---\adp and a
phosphate group \pho. Hinged to the motor head is the rigid lever arm
which can exist either in the primed (gray) or the stretched (black)
conformation. The lever arm amplifies small conformational changes in
the head domain of the motor so that the tip of the lever arm swings
forward by a distance $d$ in the transition from primed to stretched
conformation. This movement stretches the elastic neck linker, which is
modeled as a linear elastic element with spring constant $\km$. Elastic
forces in the neck linker are transmitted to the anchor, through which a
\myo motor can integrate firmly into \myo motor filaments such as
cytoskeletal minifilaments.

\begingroup
\squeezetable
\begin{table}
\begin{ruledtabular}
\begin{tabular}{ l | c | c | r }
name & symbol & values & model value \\
\hline
thermal energy         & $\kt$                      & ---                                                           & $4.14\pN\nm$\\
power-stroke distance  & $d$                        & $8\nm$            \cite{a:VilfanDuke2003b},
                                                      $10\nm$           \cite{a:WalcottEtAl2012}                    &  $8\nm$\\
motor elasticity       & $\km$                      & $0.3\pN\nm^{-1}$  \cite{a:Duke1999,a:WalcottEtAl2012},
                                                      $2.5\pN\nm^{-1}$  \cite{a:VilfanDuke2003b},
																		$3.0\pN\nm^{-1}$  \cite{a:ChenGao2011}                        & $2.5\pN\nm^{-1}$\\
transition rates       & $k_{01}$                   & $6\Hz$            \cite{a:WalcottEtAl2012, a:ChenGao2011},
																		$40\Hz$           \cite{a:VilfanDuke2003b, a:WalcottEtAl2012} & $40\Hz$\\
                       & $k_{10}$                   & $0\Hz$            \cite{a:WalcottEtAl2012},
																		$2\Hz$            \cite{a:VilfanDuke2003b}                    & $2\Hz$\\
                       & $k^0_{20}$                 & $\sim 18\Hz$      \cite{a:WalcottEtAl2012},
																		$80\Hz$           \cite{a:VilfanDuke2003b},
																	   $\sim 350\Hz$     \cite{a:WalcottEtAl2012},                   & $80\Hz$\\
							  & $k^0_{12} \simeq k^0_{21}$ & $10^3\Hz$         \cite{a:VilfanDuke2003b}                    & $10^3\Hz$\\
post-power-stroke bias & $\epp$                     & $-60\pN\nm$       \cite{a:VilfanDuke2003b}                    & $-60\pN\nm$\\
unbinding distance     & $\delta$                   & $0.328\nm$        \cite{a:VilfanDuke2003b},                   
																		$1.86\nm$         \cite{a:WalcottEtAl2012},                   
																		$2.60\nm$         \cite{a:WalcottEtAl2012}                    & $0.328\nm$\\
unbinding force        & $\fdel$                    & ---                                                           & $12.62\pN$\\
\end{tabular}
\end{ruledtabular}
\caption{Parameters determining the dynamics of the ensemble
non-processive motors. The third column list parameter values used in
previous models. The values used in this manuscript are taken from
\textcite{a:VilfanDuke2003b} and are listed in the last column.}
\label{tab:Tab01}
\end{table}
\endgroup

Driven by the hydrolysis of \atp, \myo cycles through a sequence of
mechanical and chemical conformations to generate force and directed
motion. The exact sequence of reaction steps, the rates of transitions
as well as the molecular parameters of \myo are subject of debate. The
basic sequence of conformations we use in our model, however, is well
supported by experimental observation. In particular, the reversal of
the power stroke under load has been observed for similar types of
single headed myosin molecules \cite{a:SellersVeigel2010}. In
\tab{Tab01} we list the values for the most important parameters as
determined experimentally or used in earlier models. These parameters
depend on the exact experimental conditions, e.g.~\atp concentration and
spatial arrangement of motors, and also on the exact type of \myo
\cite{a:VeigelEtAl2003, a:SoaresEtAl2011}. In the main body of our paper
we use the parameters of the last column of \tab{Tab01}. These are taken
from \textcite{a:VilfanDuke2003b}. As shown schematically in
\fig{Fig01}(b), we model the \myo motor cycle by three discrete
mechano-chemical states \cite{a:Duke2000} with stochastic transitions
between them. In the unbound state $(0)$, the motor head is loaded with
\adp and \pho and the lever arm is in its primed conformation. The
primed conformation is a high energy state, which stores part of the
approximately $80\pN\nm$ of energy released in \atp hydrolysis. The
motor then reversibly transitions to the weakly-bound state $(1)$ with
on-rate $k_{01}$ and off-rate $k_{10}$. Concomitant with the release of
\pho, the lever arm swings to the stretched conformation, thereby
releasing most of the energy stored on the primed neck linker, and the
motor enters the post-power-stroke state $(2)$. With the stretched lever
arm, the motor molecule is close to its conformational ground state and
there is a strong free energy bias $\epp$ favoring the post-power-stroke
state. Compared to the binding transitions, transitions between the
bound states are relatively fast, with unloaded transition rates
$k^0_{12} \simeq k^0_{21}$. Replacing \adp by \atp, unbinding from the
substrate and hydrolysis of \atp completes the motor cycle and brings
the motor back to the unbound state $(0)$ with primed lever arm. The
unloaded off-rate for this last step is $k^0_{20}$. Due to the energy
released in \atp hydrolysis this transition is considered as
irreversible, thus defining the direction of the motor cycle. Most
importantly in our context, both power stroke and unbinding from the
post-power-stroke state depend on load. The power stroke $(1) \to (2)$
moves the lever arm forward by the power-stroke distance $d$ and strains
the elastic neck linker of a motor. Replacing \adp by \atp and unbinding
from $(2)$ requires an additional movement of the lever arm in the same
direction as the power stroke, thus straining the neck linker further by
the unbinding distance $\delta$ and making unbinding slower under load.
The load dependent rates for these transition are denoted without the
superscript (see \fig{Fig02}).

\subsection{Parallel cluster model for ensembles of non-processive motors}

Because non-processive molecular motors are bound only during a small
fraction of the motor cycle they have to cooperate in groups to generate
sustained levels of force or persistent motion against an external load.
A sufficiently large number of motors ensures permanent attachment of
the group while individual motors continuously unbind and rebind as they
go through their motor cycle. \fig{Fig02}(a) illustrates the coupling of
\myo motors in an ensemble working against an external load. With their
anchors the motors are firmly integrated into the rigid backbone of the
motor filament whereas the motor heads bind to the substrate. In
\fig{Fig02}(a), the motors are oriented such that the lever arm swings
towards the right during the power stroke, so that the motors exert
force on the motor filament towards the right. The external load pulls
the motor filament towards the left, against the motor direction.
Because the motors are attached directly to the motor filament, they are
working effectively in parallel against the external load. Such parallel
arrangement was confirmed experimentally for the \myo motors in the
muscle sarcomere \cite{a:ColombiniEtAl2007, a:ThoresenEtAl2013}. We will
discuss two paradigmatic situations for the external load: $(i)$ a
constant external load, which is independent of the position of the
motor filament, and $(ii)$ an elastic external load, which increases
linearly with the displacement of the motor filament. For constant
external load, the ensemble will eventually reach a steady state of
motion with load dependent velocity. For a linear external load, an
isometric state with vanishing velocity is expected. Experimentally, the
unipolar ensemble of \myo motors in \fig{Fig02}(a) would represent one
half of a thick filament in the muscle sarcomere or of a minifilament in
the cytoskeleton. In this case, the external load is generated by the
motors in the other half of the bipolar motor filament or is due to the
tension in a surrounding actin network. In reconstituted assays, a
constant load could be realized through viscous forces in a flow chamber
or applying active feedback control; a linear load might be realized
using using elastic elements such as optical traps.

\FIG{Fig02}{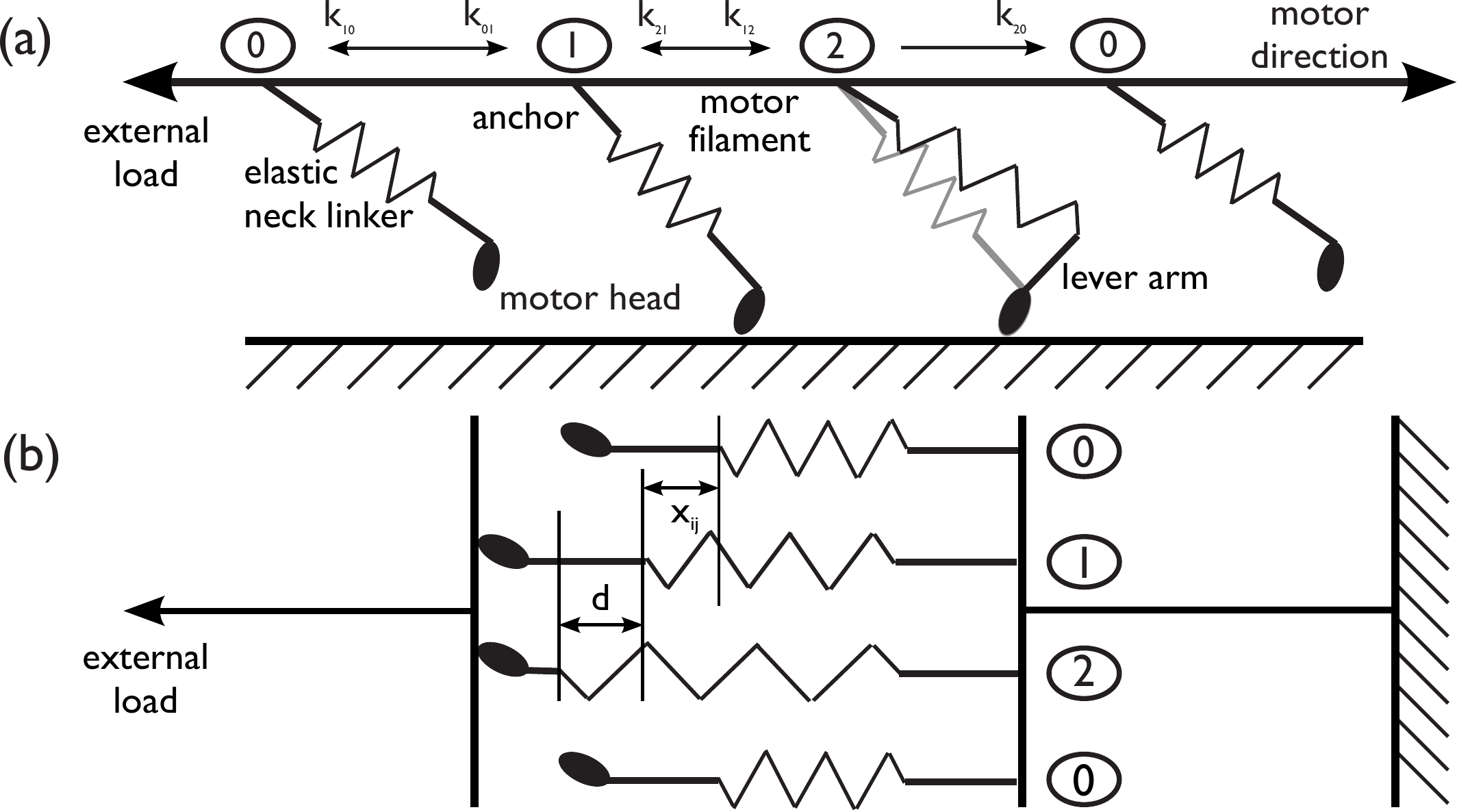}{(a) Mechanical coupling of an ensemble of \myo
motors. The motors pull on the motor filament towards the right against
the external load pulling to the left. The external load is balanced by
the elastic forces in the neck linkers of the bound motors. Typical
cases for the external load are ($i$) constant load and ($ii$) linear
load increasing with the displacement of the motor filament. The
mechanical coupling of the motors through the rigid motor filament
induces a dynamic coupling due to the strain dependence of the
transition rates. (b) In the \emph{parallel cluster model} (PCM) all
motors in a given mechano-chemical state are assumed to have the same
strain. Weakly-bound motors have the strain $\xij$. The power stroke
corresponds to a shortening of the bound molecular motor so that motors
in the post-power-stroke state have the strain $\xij+d$. The arrangement
of the motors is equivalent to adhesion clusters of parallel bonds,
where the closed bonds can be in two different conformations.}

For an ensemble of parallel motors in mechanical equilibrium, the
external load $\fext$ is balanced by the sum of elastic forces $\fn =
\km\xin$ in the neck linkers of all bound motors:
\begin{equation}\label{eq:ForceBalance}
\fext = \sum_{bound} \fn = \km \sum_{bound} \xin\,.
\end{equation}
In this expression, $\km$ is the elastic constant of the neck linkers,
$\xin$ their elongation (or strain) and the index $n$ runs over all
motors which are bound to the substrate. An imbalance of forces induces
a change of the position of the motor filament. This changes the strain
$\xin$ of all bound motors simultaneously until the balance of forces in
\eq{ForceBalance} is restored. In addition, a linear external load would
be changed by the displacement of the motor filament. In the following,
we assume that the relaxation time towards mechanical equilibrium is
negligible in comparison with the time scale for stochastic transitions,
so that the mechanical state of an ensemble always obeys the force
balance in \eq{ForceBalance}. Anchoring of the motors to the rigid motor
filament in combination with the condition of the force balance
introduces a tight mechanical coupling between motors: a stochastic
transition of one motor changes the force balance and hence the strain
of all bound motors. Thus, the strain $\xin$ of a bound motor does not
only depend on the state of the motor itself but results from the past
activity of the motor ensemble. Specifically, $\xin$ is determined by
the displacement of the motor filament after binding of a motor.
Therefore, \eq{ForceBalance} determines the sum but not the individual
strains of the motors: while the sum over the strain of the bound motors
vanishes at vanishing external load, the individual strains $\xin$ will
in general not. The distribution of the $\xin$ will be determined by the
randomly distributed times during which a motor remains bound to the
substrate and the displacement of the motor filament during these times.

Because the rates for the stochastic transitions $(1) \leftrightarrow
(2)$ and $(2) \to (0)$ depend on the strain of a motor, the mechanical
coupling leads to a dynamical coupling of the motors, as illustrated in
\fig{Fig02}(a). A stochastic description of the ensemble dynamics as a
Markov process would thus require not only the mechano-chemical state
but also the strain of every motor as state variables. Denoting the
total number of motors in an ensemble by $\nt$, the state space for an
ensemble with $\nt$ motors would then encompass $3^{\nt}$ discrete
states and $\nt$ independent, continuous variables ($\nt-1$ considering
\eq{ForceBalance}). This complexity prohibits analytical solutions and
previous approaches either used mean field models or computer
simulations (see e.g.~\cite{a:LeiblerHuse1993, a:Duke1999, a:Duke2000}).
Here, we use mean-field elements to arrive at an analytically tractable
model, which preserves the molecular details contained in the
crossbridge model and allows to study stochastic effects due to finite
ensemble size. We make the assumption that all motors in the same
mechano-chemical state have the same strain. This assumption is the
essence of our \emph{parallel cluster model} (PCM); its validity will be
discussed in \scn{PCMValidation} and demonstrated by comparison with
computer simulations. As illustrated in \fig{Fig02}(b), the PCM
effectively describes the ensemble of molecular motors as an adhesion
cluster of parallel bonds \cite{a:ErdmannSchwarz2004a,
a:ErdmannSchwarz2004c, a:ErdmannSchwarz2006}. In this picture, the power
stroke shortens the closed bonds by the power-stroke distance $d$.
Therefore, closed bonds can be in two conformations with different
lengths in which they carry different loads. All motors in a given
conformation, however, carry an equal share of the external load and
have the same strain. Thus, all motors in a given mechano-chemical state
are mechanically equivalent within the PCM so that the state of a motor
ensemble of $\nt$ motors can be characterized by the number of motors in
each of the mechano-chemical states. We use the number $i$ of bound
motors ($0 \leqslant i \leqslant \nt$) and the number $j$ ($0 \leqslant
j \leqslant i$) of motors in the post-power-stroke state. The number of
motors in the weakly-bound state then follows as $i-j$ and the number of
unbound motors is $\nt-i$. The strain of the motors in the weakly-bound
state $(1)$ is referred to as $\xij$, where the indices indicate the
dependence on the ensemble state $(i,j)$. Since the power stroke
stretches the neck linker by $d$, the strain of the motors in the
post-power-stroke state $(2)$ is given by $\xij+d$. With $i-j$ motors
with strain $\xij$ and $j$ motors with strain $\xij+d$, the force
balance in the PCM reads
\begin{equation}\label{eq:PCMForceBalance}
\fext = \km \left[(i-j) \xij + j (\xij+d)\right] = \km \left[i \xij + j d\right]\,.
\end{equation}
This expression can be solved for the strain $\xij$ of the weakly-bound
motors. For constant external load, $\fext=\const$, \eq{PCMForceBalance}
yields
\begin{equation}\label{eq:PCMConstStrain}
\xij = \frac{(\fext/\km) - j d}{i}\,.
\end{equation}
For linear external load, we have to introduce an external coordinate
describing the position of the motor ensemble. We define $z$ as the
average position of the bound motor heads. The position of the motor
filament then is given by $z-\xij$. The definition of ensemble position
is described in detail in \scn{EnsMov}. With the external elastic
constant $\kf$, the linear external load is $\fext = \kf (z - \xij)$.
Inserting this into \eq{PCMForceBalance} yields
\begin{equation}\label{eq:PCMLinStrain}
\xij = \frac{(\kf/\km) z - j d}{i + (\kf/\km)}
\end{equation}
for the strain of the weakly-bound motors. Here, we define the strain of
a weakly-bound motor to be positive when the neck linker is stretched in
the direction of the external load (towards the left in \fig{Fig02}(a))
whereas the average position $z$ of bound motor heads increases in the
motor direction (towards the right in \fig{Fig02}(a)) (see
\scn{EnsMov}). If all bound motors are in the weakly-bound state
($j=0$), the strain $\xij = x_{i0}$ is positive, that is, the neck
linkers pull on the motor filament against the external load. When the
external load is not too large, the strain of the weakly-bound motors
can become negative, if sufficiently many motors have gone through the
power stroke. In this case, the neck linkers of the weakly-bound motors
pull the motor filament against the motor direction, thereby supporting
the external load. Because $j \leqslant i$, the strain $\xij+d$ of
motors in the post-power-stroke state is always positive. It is this
pulling of post-power-stroke motors which eventually drives force
generation and motion by the ensemble.

The major benefit of the PCM lies in the fact, that it eliminates the
history dependence of the strain and introduces $\xij$ as a state
function. Within the PCM, the strain of all motors follows from the
current ensemble state $(i,j)$ and the external load $\fext$. For
constant external load, $\fext=\const$ takes the role of a parameter and
the ensemble dynamics is fully characterized by $(i,j)$. For linear
external load, $\fext = \kf (z-\xij)$ is changed through the activity of
motors so that additional rules for the position $z$ of the ensemble are
required to fully characterize the dynamics of an ensemble.

\subsection{Local thermal equilibrium of bound motors}

\FIG{Fig03}{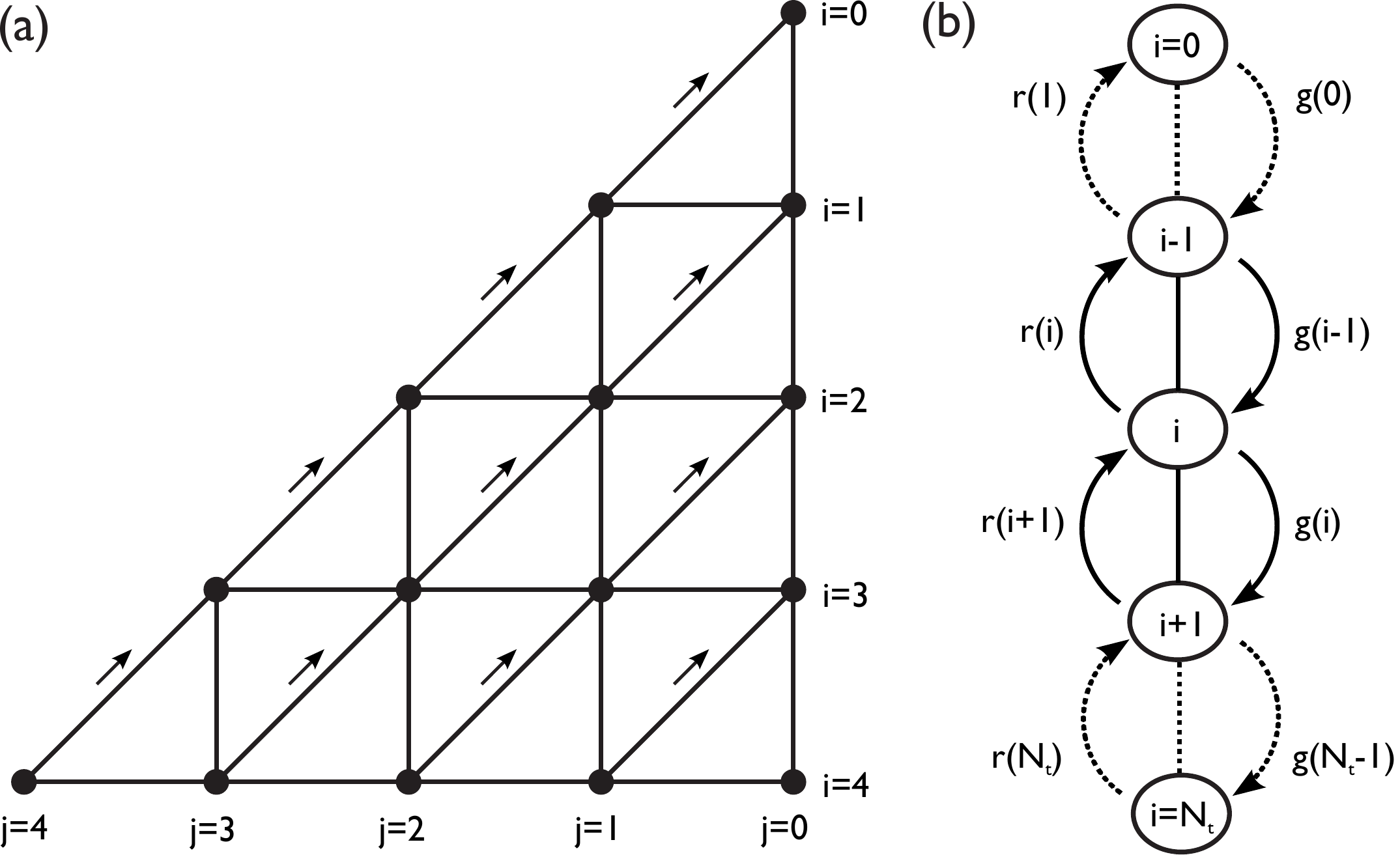}{(a) Two-dimensional stochastic reaction network for
transitions between the states $(i,j)$ of an ensemble within the PCM.
Binding to and unbinding from the weakly-bound state changes the number
of bound motors $i$ (vertical lines). The power stroke changes the
number of motors in the post-power-stroke state $j$ (vertical lines).
Unbinding from the post-power-stroke state changes both $i$ and $j$
(diagonal lines). This is the only irreversible transition and is marked
by an arrow. (b) \emph{Local thermal equilibrium} (LTE) projects all
states with different $j$ but given $i$ onto a single state so that the
state of an ensemble is described by the number $i$ of bound motors
alone. Transitions are described by effective reverse and forward rates.}

For an ensemble with $\nt$ \myo motors, the number $i$ of bound motors
ranges from $0$ to $\nt$. The number $j$ of motors in the
post-power-stroke state ranges from $0$ to $i$. \fig{Fig03}(a) shows the
corresponding network of states $(i,j)$ which are connected by the
possible stochastic transitions. Binding to and unbinding from the
weakly-bound state change $i$ by $\pm 1$ without changing $j$ (vertical
transitions). The power stroke and its reversal change $j$ by $\pm 1$
but leave $i$ constant (horizontal transitions). Unbinding from the
post-power-stroke state reduces $i$ and $j$ simultaneously (diagonal
transitions). This is the only irreversible transition in the network
and is marked by an arrow. In total there are $(\nt+1)(\nt+2)/2$ states
and $3\nt(\nt+1)/2$ transitions (of which $\nt(\nt+1)$ are reversible)
in this network.

To further reduce the complexity of the model, we take advantage of the
strong separation of time scales between the slow binding and unbinding
transitions and the transitions between the bound states, which are at
least an order of magnitude faster \cite{a:Duke1999, a:VilfanDuke2003b}
(see \tab{Tab01}). Following previous modeling approaches, we assume
that a \emph{local thermal equilibrium} (LTE) is maintained within the
bound states \cite{a:VilfanDuke2003b}. For a given number $i$ of bound
motors, the conditional probability to find $j$ motors in the
post-power-stroke state and $i-j$ in the weakly-bound state is given by
the Boltzmann distribution
\begin{equation}\label{eq:LTEDist}
p(j|i) = \frac{1}{Z_i} \Exp{-\eij/\kt}\,.
\end{equation}
Here, $Z_i$ is the appropriate partition sum for given $i$,
\begin{equation}
Z_i = \sum_{j=0}^i \Exp{-\eij/\kt}\,.
\end{equation}
The energy $\eij = j\epp + \eel_{ij} + \ext_{ij}$ of an ensemble in
state $(i,j)$ is the sum of the free energy bias $\epp \simeq -60\pN\nm
< 0$ towards the post-power-stroke state for $j$ motors, the elastic
energy $\eel_{ij}$ stored in the neck linkers and a possible external
contribution $\ext_{ij}$. The elastic energy of the neck linkers is
given by
\begin{equation}\label{eq:ConstEnergy}
\eel_{ij} = \frac{\km}{2} \left[(i-j)\xij^2 + j(\xij+d)^2\right]\,.
\end{equation}
The strain $\xij$ is given by \eq{PCMConstStrain} for constant and
\eq{PCMLinStrain} for linear external load. The external contribution to
the energy vanishes for constant external load, $\ext_{ij} = 0$. For
linear external load, $\ext_{ij}$ is given by the energy stored in the
external harmonic potential,
\begin{equation}\label{eq:ElasticEnergyLinear}
\ext_{ij} = \frac{\kf}{2}(z-\xij)^2\,,
\end{equation}
where $\xij$ is given by \eq{PCMLinStrain} and $z-\xij$ is the position
of the motor filament.

Combining \eqs{PCMConstStrain}{ConstEnergy} reveals that for constant
external load, the elastic energy of the neck linkers is identical in
states $(i,j)$ and $(i,i-j)$. $\eel_{ij}$ is minimal when all bound
motors are either weakly-bound ($j=0$) or in the post-power-stroke state
($j=i$). Intermediate states with $0<j<i$ have a larger elastic energy
because bound motors in opposite conformations are pulling against each
other. Because of this symmetry of $\eel_{ij}$, the free energy bias
$\epp$ translates directly into a strong bias of the LTE distribution
towards the post-power-stroke state so that almost all bound motors are
in the post-power-stroke state. The symmetry of the elastic energy is
not affected by the exact value of the constant external load, which
merely changes the absolute values of $\eel_{ij}$, so that the bias of
the LTE distribution persists for arbitrary values of a constant
external load.

A linear external load is increased by the power stroke of a motor.
Therefore, elastic energy $\eel_{ij}$ and external energy $\ext_{ij}$
tend to increase with an increasing number of post-power-stroke motors
and introduce a bias towards the weakly-bound state, opposite to the
free energy bias $\epp$. To demonstrate this, we compare the energy in
the extreme states $(i,0)$ and $(i,i)$. For a given $z$, $\eel_{ij}$ and
$\ext_{ij}$ take the smallest value in state $(i,0)$, in which all bound
motors are weakly-bound with strain $x_{i0} = (\kf/\km)z / (i+\kf/\km)$.
In state $(i,i)$, all bound motors are in the post-power-stroke state
with the strain $x_{ii}+d = (\kf/\km)(z+d) / (i+\kf/\km)$. Because
$x_{i0} < x_{ii}+d$, the elastic energy in state $(i,0)$ is smaller than
in state $(i,i)$, $\eel_{i0} < \eel_{ii}$. On the other hand, $x_{i0} >
x_{ii}$ and $z-x_{i0} < z-x_{ii}$, so that also the external energy
$\ext_{ij} = (\kf/2) (z-\xij)^2$ is smaller in state $(i,0)$ than in
state $(i,i)$, $\ext_{i0} < \ext_{ii}$. The total energy difference is
\begin{equation}\label{eq:LinDelE}
\left(\eel_{ii} + \ext_{ii}\right) - \left(\eel_{i0} + \ext_{i0}\right)
                = \frac{i \km \kf}{i\km + \kf}\frac{d (d + 2z)}{2}\,.
\end{equation}
It increases with increasing $\kf$ and $z$ so that the bias of the LTE
distribution will shift from the post-power-stroke state to the
weakly-bound state at large values of the external elastic constant
$\kf$ or the ensemble position $z$. This transition eventually will
stall ensemble movement, because movement is driven by post-power-stroke
motors and requires passage through the motor cycle.

\subsection{One-step master equation for binding dynamics}

As illustrated in \fig{Fig03}(b), the assumption of an LTE within the
bound motors effectively projects all ensemble states $(i,j)$ with
different $j$ but given $i$ onto a single variable. Thus, the state of
an ensemble is described by the number $i$ of bound motors alone. In
this effectively one-dimensional system, the probability $p_i(t)$ to
find $i$ motors bound to the substrate at time $t$ follows the one-step
master equation
\begin{equation}\label{eq:MasterEq}
\dd{p_i}{t} = r(i+1) p_{i+1} + g(i-1) p_{i-1} - \left[r(i) + g(i)\right] p_i\,.
\end{equation}
Once $p_i(t)$ is known, the probability $p_{ij}(t) = p(j|i) p_i(t)$ to
find the ensemble in the state $(i,j)$ at time $t$ is obtained as the
product of $p_i(t)$ with the time independent LTE distribution $p(j|i)$
from \eq{LTEDist}. In the one-step master equation, the effective
reverse rate $r(i)$ describes the rate at which bound motors unbind from
the substrate, that is, $r(i)$ is the rate of the transition $i \to
i-1$. The effective forward rate $g(i)$ describes the rate at which free
motors bind to the substrate, that is, $g(i)$ is the rate of the
transition $i \to i+1$. The effective transition rates $r(i)$ and $g(i)$
with their dependence on $i$ and $\fext$ define the stochastic dynamics
of binding and unbinding in the motor ensemble.

In the state $(i,j)$ of an ensemble, weakly-bound motors unbind with
off-rate $k_{10}(i,j)$ and post-power-stroke motors unbind with off-rate
$k_{20}(i,j)$. Following previous modeling approaches \cite{a:Duke1999,
a:Duke2000}, we assume that the off-rate from the weakly-bound state is
independent of the load on a motor and therefore independent of the
state of an ensemble, $k_{10}(i,j) = k_{10} = \const$. Unbinding from
the post-power-stroke state requires the lever arm to work against the
external load over the unbinding distance $\delta$. Assuming a
Kramers-type load dependence, the off-rate from the post-power-stroke
state decreases exponentially with the load $\fij = \km (\xij+d)$ on the
neck linker:
\begin{equation}\label{eq:k20}
k_{20}(i,j) = k^0_{20} \Exp{-\fij/\fdel}\,.
\end{equation}
The unbinding force scale $\fdel = \kt/\delta \simeq 12.6\pN$ is set by
the thermal energy $\kt$ and the unbinding distance $\delta$. In state
$(i,j)$, there are $i-j$ weakly-bound motors and $j$ post-power-stroke
motors. Because all stochastic transitions proceed independently, the
rate for unbinding of a motor in state $(i,j)$ is the sum over the
single-motor transition rates:
\begin{equation}\label{eq:ReverseRateIJ}
r(i,j) = (i-j) k_{10}(i,j) + j k_{20}(i,j)\,.
\end{equation}
The effective reverse rate $r(i)$ for the transition $i \to i-1$ in the
one-dimensional system is then obtained by averaging $r(i,j)$ over $j$
with the LTE distribution from \eq{LTEDist}:
\begin{equation}\label{eq:ReverseRate}
r(i) = \sum_{j=0}^i r(i,j) p(j|i)\,.
\end{equation}
Considering the exponential dependence of $k_{20}(i,j)$ on strain and
the strain dependence of the LTE distribution, $r(i)$ is a strongly
non-linear function of the external load and the number of bound motors.
Since the off-rate $k_{20}(i,j)$ decreases under load (see \eq{k20}),
the post-power-stroke state of \myo behaves as a catch bond. Due to the
strong bias of the LTE distribution towards the post-power-stroke state,
\myo motors predominantly unbind from the post-power-stroke state. This
implies that also the effective reverse rate $r(i)$ decreases under load
and the \myo ensemble as a whole behaves as a catch bond. For constant
external load, the bias persists for arbitrary values of $\fext$, so
that the catch bond character of \myo motors is found for all values of
the external load. For elastic external load, the bias of the LTE
distribution passes over to the weakly-bound state for very stiff
external springs. This means that \myo behaves as a catch bond at small
values of $\kf$ and $z$, but unbinds with load-independent rate at large
values of $\kf$ or $z$. For very large loads, it is expected that
unbinding of motors is accelerated under load as for slip bonds
\cite{a:ColombiniEtAl2007}, but such large loads will not be considered
here.

The only pathway for binding is the transition $(0) \to (1)$ from the
unbound state to the weakly-bound state of a motor. Because unbound
motors are not subject to any load, the on-rate is assumed to be
constant, $k_{01} = \const$. With $\nt-i$ motors binding independently,
the effective forward rate is given by
\begin{equation}\label{eq:ForwardRate}
g(i) = (\nt-i) k_{01}\,.
\end{equation}
The forward rate $g(i)$ increases linearly with the number $\nt-i$ of
unbound motors but is otherwise independent of the state of the
ensemble. Because there is no dependence on $j$, averaging with the LTE
distribution is not required.

With the definition of the effective transition rates in
\eqs{ReverseRate}{ForwardRate}, the one-step master equation of
\eq{MasterEq} for the binding dynamics in an ensemble of \myo motors is
fully characterized for the case of constant external load. If the
external load depends on the position of the ensemble, as for a linear
load, \eq{MasterEq} has to be solved together with additional rules for
the movement of the motor ensemble, which will be introduced in the next
section.

\subsection{Ensemble movement}\label{scn:EnsMov}

With the introduction of the PCM, we have focused on modeling the
dynamics of binding and unbinding of molecular motors in an ensemble.
The displacement of the ensemble, on the other hand, seems to be
eliminated by the analogy to a cluster of parallel adhesion bonds.
Nevertheless, there are clear prescriptions for the transformation of
binding and unbinding of motors to a displacement of the ensemble. In
order to derive these prescriptions and to elucidate the inherent
approximations, we take a step back and consider the general case of an
ensemble of molecular motors without the approximation of the PCM, that
is, we consider an ensemble in which every motor is characterized by an
individual value of the strain. The spatial coordination of the motors
is schematically depicted in \fig{Fig02}(a). The anchors are integrated
into the motor filament at fixed positions. Because the motor filament
is rigid, the relative positions of the anchors are constant and we can
assume that all anchors are at the same position $\zfil$, which is
identified with the position of the motor filament. In the following, we
consider a reference state in which $i$ motors are bound to the
substrate and $\nt-i$ are unbound. The neck linkers of the bound motors
have the strains $\xin$, where the index $n \in \{1, \dots, i\}$ labels
the bound motors. The position $\zn$ of a bound motor head on the
substrate is related to the position $\zfil$ of the motor filament via
its strain $\xin$ as $\zn = \zfil + \xin$ for weakly-bound and $\zn =
\zfil+\xin-d$ for post-power-stroke motors. To abbreviate notation, we
define the offset of a motor head from its anchor as $\xn := \xin$ for
weakly-bound and $\xn := \xin-d$ for post-power-stroke motors. Using
this definition, the position of a bound motor head can be written as
\begin{equation}\label{eq:MotorHeadPos}
\zn = \zfil + \xn
\end{equation}
for all bound motors with $n \in \{1,\dots,i\}$. We now define the
position of an ensemble as the average position of the bound motor
heads:
\begin{equation}\label{eq:EnsemblePos}
\zavg := \frac{1}{i} \sum_{n=1}^i \zn\,.
\end{equation}
With this definition, the ensemble position $\zavg$ can only change
through binding or unbinding of motors, because motor heads are bound at
fixed positions on the substrate. By contrast, the position $\zfil$ of
the motor filament also changes through transitions within the bound
states which change the balance of forces in \eq{ForceBalance}. Only for
completely detached ensembles, in which no motor is bound to the
substrate, we have to use the position $\zfil$ of the motor filament as
the ensemble position. Because unbound motors have vanishing strain and
unbound motor heads are at the same position as the anchors, $\zfil$ is
identical to the average position of the unbound motor heads. Inserting
$\zn$ from \eq{MotorHeadPos} into \eq{EnsemblePos}, the average position
of the bound motor heads is
\begin{equation}
\zavg = \zfil + \xavg_{ij}\,.
\end{equation}
The average offset $\xavg_{ij}$ between anchors and motor heads is
related to the external load $\fext$ via the balance of forces in
\eq{ForceBalance} as
\begin{equation}
\xavg_{ij} := \frac{1}{i} \sum_{n=1}^i \xn = \frac{(\fext/\km) - j d}{i}\,.
\end{equation}
Unlike the absolute position $\zavg$ and the individual values of $\xn$
or $\xin$, which all result from the history of the ensemble,
$\xavg_{ij}$ follows from the current state of the ensemble alone. In
particular, it depends on the external load $\fext$, the number $i$ of
bound motors and the number $j$ of post-power-stroke motors.

We now calculate how the average position of bound motor heads changes
through binding of one additional motor. Assuming that $i$ motors are
bound initially with average position $\zavg$ and that the new motor
binds with vanishing strain at $\zfil$ to the substrate, the new average
position $\zavg'$ of $i+1$ bound motor heads is
\begin{equation}
\zavg' = \frac{i\zavg + \zfil}{i+1} = \zfil + \frac{i}{i+1} \xavg_{ij}\,.
\end{equation}
Thus, binding of a motor changes ensemble position by
\begin{equation}
\dzavgon_{ij} := \zavg' - \zavg
               = \left[\frac{i}{i+1} - 1\right] \xavg_{ij}
               = -\frac{\xavg_{ij}}{i+1}\,.
\end{equation}
Like $\xavg_{ij}$, the binding step $\dzavgon_{ij}$ is a function of the
ensemble state before binding. After binding, the average offset of the
motors is adjusted to
\begin{equation}\label{eq:DZOnAvgIJ}
\xavg_{i+1j}' = \frac{i}{i+1} \xavg_{ij}\,.
\end{equation}
Combining this with $\zavg'$ confirms that the position of the motor
filament remains unchanged, $\zfil' = \zavg' - \xavg_{i+1j}' = \zavg -
\xavg_{ij} = \zfil$. This is required for consistency, because the
balance of forces is not affected by a motor binding with vanishing
strain. In a more general description, motors could be allowed to bind
with a finite value of the strain chosen from a random distribution with
vanishing mean. In this case, \eq{DZOnAvgIJ} for $\dzavgon_{ij}$ would
remain valid in the ensemble average.

Unlike the binding step $\dzavgon_{ij}$, the change of the ensemble
position upon unbinding depends on which of the motors unbinds. Assuming
that a motor head with offset $\xn$ unbinds from the position $\zn =
\zfil+\xn$ on the substrate, the average position $\zavg''$ of the $i-1$
remaining motor heads is
\begin{equation}
\zavg'' = \frac{i\zavg - \zn}{i-1}
        = \frac{i\left(\zfil + \xavg_{ij}\right) - \left(\zfil + \xn\right)}{i-1}
        = \zfil + \frac{i\xavg_{ij} - \xn}{i-1}\,.
\end{equation}
Thus, unbinding of a motor from $\zn$ changes the position of the
ensemble by
\begin{equation}\label{eq:DZOffN}
\dzavgoff_{ij,n} = \zavg'' - \zavg
                 = \frac{i\xavg_{ij}-\xn}{i-1} - \xavg_{ij}
                 = \frac{\xavg_{ij}-\xn}{i-1}\,.
\end{equation}
Assuming that all motors are equally likely to unbind, the average of
the unbinding step over all bound motors vanishes:
\begin{equation}\label{eq:DZOffAvg}
\dzavgoff_{ij} = \frac{1}{i} \sum_{n=1}^{i} \dzavgoff_{ij,n}
               = \frac{\xavg_{ij} - \xavg_{ij}}{i-1} = 0\,.
\end{equation}
For weakly-bound motors this assumption is valid, because the off-rate
$k_{10} = \const$ is independent of strain. Post-power-stroke motors, on
the other hand, are catch bonds with an off-rate $k_{20}(i,j)$
decreasing exponentially with increasing strain. Therefore,
post-power-stroke motors unbind preferentially with small strain and
small offset, $\xn < \xavg_{ij}$. Averaging the unbinding step
$\dzavgoff_{ij,n}$ with the actual off-rates would then lead to a
positive displacement $\dzavgoff_{ij} \geqslant 0$. The size of the
unbinding step depends on the distribution of strains of the bound motor
heads and vanishes when all motors have the same offset. In addition,
unbinding of a motor with non-zero strain changes the balance of forces
so that the position $\zfil$ of the motor filament is changed.

When the last bound motor unbinds and the ensemble detaches completely
from the substrate, the motor head relaxes instantaneously from its
position $\zavg = z_1 = \zfil-x_1$ on the substrate to the position
$\zfil = \zavg-x_1$ of the motor filament. Because the position of the
detached ensemble is described by the position $\zfil$ of the motor
filament, unbinding of the last motor changes the ensemble position by
\begin{equation}\label{eq:DZOffDet}
\dzavgoff_{1j} = - x_1 \quad\text{for}\quad j = 0,1\,.
\end{equation}
The dependence on the state of the unbinding motor is included in the
definition of the offset $x_1$.

\FIG{Fig04}{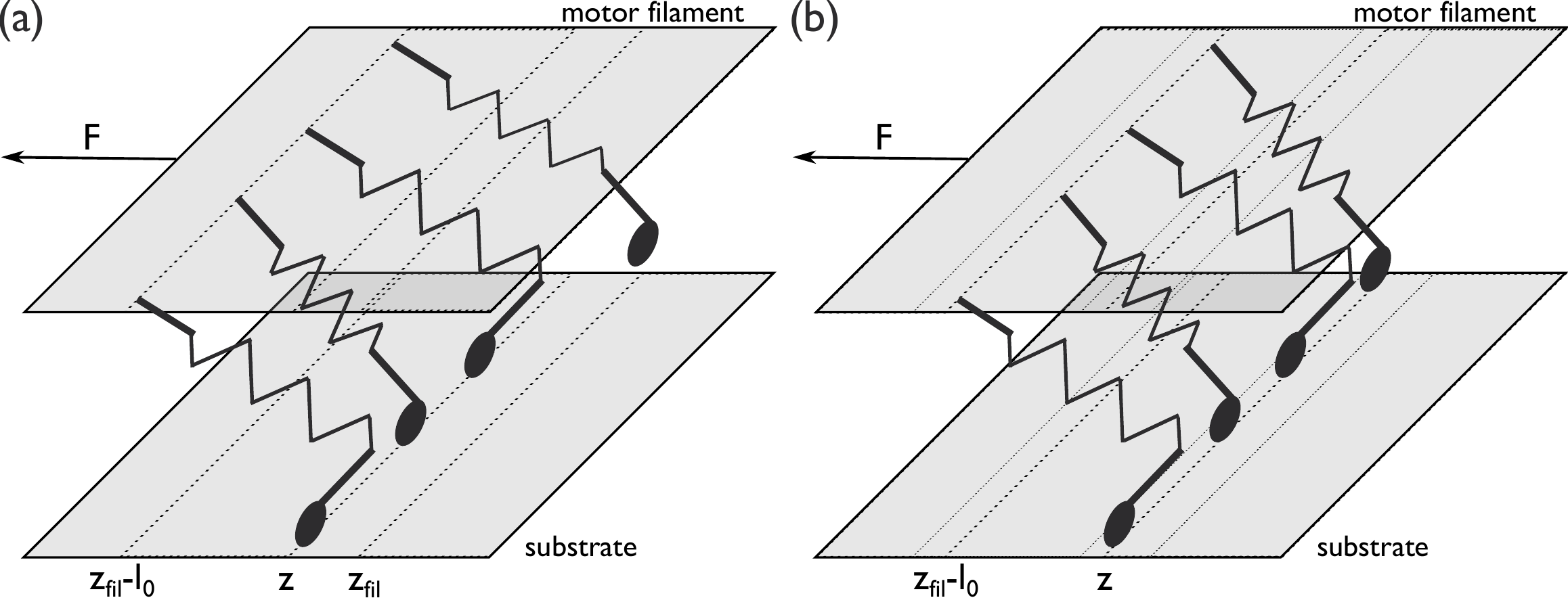}{Change of ensemble position $z$ upon binding of a
motor within the PCM. (a) Before binding, all bound motor heads are at
the ensemble position $z$. In the illustrated case, the strain of the
weakly-bound motor is negative, $\xij < 0$, so that motor heads will
bind at $\zfil = z-\xij > z$ ahead of the current ensemble position. The
post-power-stroke motors have positive strain, $\xij+d > 0$, and work
against the external load and the elastic force from the weakly-bound
motor. For the illustration, motors are depicted with length $\ell_0$ so
that the anchors are at $\zfil-\ell_0$. (b) After a motor has bound, all
bound motor heads are shifted to the new ensemble position $z$ to
implement the PCM assumption of equal $\xij$ of all bound motors.
Because the external force is distributed over a larger number of bound
motors, the position $\zfil$ of the motor filament also shifts to larger
values.}

Within the PCM, weakly-bound motors have the strain $\xij$ and the
strain of post-power-stroke motors is $\xij+d$, so that all bound motors
are characterized by the same offset $\xij$. Therefore, all bound motor
heads are at the same position $z = \zfil+\xij$ on the substrate. To
apply the general expressions for the displacement to the PCM, the
averages $\zavg$ and $\xavg_{ij}$ are replaced by the quantities $z$ and
$\xij$ which are the same for all motors in the PCM ensemble. In a given
state $(i,j)$, binding of a new motor changes the position $z$ of the
ensemble by
\begin{equation}\label{eq:dzonij}
\dzon_{ij} = -\frac{\xij}{i+1}\,.
\end{equation}
This is the actual change of the average position of the bound motor
heads assuming that the new motor has bound with vanishing strain at the
position $\zfil$. As illustrated in \fig{Fig04}, in order to implement
the assumption of the PCM, all $i+1$ bound motor heads have to be
shifted to the new common position $z' = z+\dzon_{ij}$ on the substrate
after binding. This shift is not meant to correspond to an actual
physical process but is a theoretical procedure required to maintain the
PCM assumption of identical strains of bound motors. As in the general
case, the position of the motor filament does not change upon binding of
a motor because the balance of forces is unchanged. Because all bound
motor heads have the same offset $\xij$ and are at the same position $z$
on the substrate, the position $z$ of the ensemble is unchanged by the
unbinding of a motor:
\begin{equation}
\dzoff_{ij} = 0 \quad\text{for}\quad i \geqslant 2\,.
\end{equation}
This is the same result as for the average in \eq{DZOffAvg}. Comparison
with the general case reveals that the PCM predicts too small a
displacement upon unbinding and will underestimate the velocity of an
ensemble when there is a wide distribution of the strains of motors.
When the last motor unbinds from the substrate, according to
\eq{DZOffDet} the position of the motor ensemble changes by the
unbinding step
\begin{equation}\label{eq:dzoffij}
\dzoff_{1j} = -x_{1j} \quad\text{for}\quad j = 0,1\,.
\end{equation}
\eqr{dzonij}{dzoffij} completely specify the rules for ensemble movement
resulting from the binding dynamics in an ensemble. The ensemble moves
forward, when a motor binds while the strain $\xij$ of the weakly-bound
motors is negative and moves backwards when $\xij$ is positive.
Unbinding of a motor does not change the position unless the last motor
unbinds. In this case, the position of the unbinding motor head relaxes
to the position of the motor filament.

The velocity $v_{ij}$ of an ensemble in state $(i,j)$ is given by the
product of the displacement step induced by a binding or unbinding
transition with the rate at which this transition proceeds. Unbound
motors bind with the effective forward rate $g(i)$ defined in
\eq{ForwardRate}. Unbinding only changes ensemble position when the last
motor unbinds. Weakly-bound motor unbind with the constant off-rate
$k_{10} = \const$; post-power-stroke motors unbind with the off-rate
$k_{20}(1,1)$. Thus, the velocity of the ensemble in state $(i,j)$ is
\begin{align}\label{eq:vij}
v_{ij} &= g(i) \dzon_{ij} + \left[k_{10} \dzoff_{10} \delta_{j0} + k_{20}(1,1) \dzoff_{11} \delta_{j1}\right] \delta_{i1}\\
		 &= -g(i) \frac{\xij}{i+1} - \left[k_{10} x_{10} \delta_{j0} + k_{20}(1,1) x_{11} \delta_{j1}\right] \delta_{i1}\,.
\end{align}
The last term applies only to unbinding from the state $i=1$ and
distinguishes between weakly-bound motors ($j=0$) and post-power-stroke
motors ($j=1$).

To combine the expressions for ensemble displacement and velocity with
the solutions of the one-step master equation for the binding dynamics,
we have to average over the variable $j$ using the LTE distribution. The
offset of the bound motors in state $i$ is given by
\begin{equation}
x_i = \sum_{j=0}^i \xij p(j|i) = \frac{1}{Z_i} \sum_{j=0}^i \xij \Exp{-\eij/\kt}\,.
\end{equation}
The binding step of the ensemble due to the transition $i \to i+1$ then
becomes
\begin{equation}
\dzon_i = -\sum_{j=0}^i \dzon_{ij} p(j|i)
        = -\frac{1}{i+1} \sum_{j=0}^i \xij p(j|i)
        = -\frac{x_i}{i+1}
\end{equation}
for $i \geqslant 1$. The unbinding step is $\dzoff_i = 0$ for $i
\geqslant 2$ and $\dzoff_1 = -x_1$ for $i=1$. Averaging the velocity
$v_{ij}$ from \eq{vij} over $j$ yields the velocity in state $i$:
\begin{equation}\label{eq:BoundVeloI}
v_i = g(i) \dzon_i - \left[k_{10} x_{10} p(0|1) + k_{20}(1,1) x_{11} p(1|1)\right] \delta_{i1}\,.
\end{equation}
The expression for the velocity applies to attached ensembles with at
least one bound motor, that is, $i \geqslant 1$. To complete the
description of ensemble movement, the velocity of detached ensembles
with $i=0$ has to be defined. The position of the detached ensemble is
described by the position of the motor filament, $\zfil$, which is
identical to the position of the unbound motor heads. We assume that the
external load $\fext$ moves the motor filament through the viscous
environment with effective mobility $\eta$. For constant external load,
detached ensembles move with constant velocity
\begin{equation}\label{eq:UnboundVeloConst}
\vu = - \eta \fext\,.
\end{equation}
The negative sign follows from the definition of the direction of the
external load opposite to the working direction of the motors. Detached
ensembles attach to the substrate with forward rate $g(0) = \nt k_{01}$
so that the random attachment times follow an exponential distribution
with average $g^{-1}(0)$. For constant velocity $v_0 = \const$, this
implies an exponential distribution also for the size of the backsteps.
The average backstep size is then given by
\begin{equation}\label{eq:DZOnAtt}
\dzon_0 = \frac{v_0}{g(0)} = -\frac{\eta\kf}{\nt k_{01}}\,.
\end{equation}
For a linear external load, $\fext = \kf\zfil$, the velocity of the
detached ensemble depends on the position $\zfil = z$ of the ensemble:
\begin{equation}\label{eq:UnboundVeloLin}
\vu(z) = \dzfil = -\eta\kf\zfil\,.
\end{equation}
Therefore, the average backstep size for linear external load depends on
the position $\zfil^0$ of the motor filament at detachment:
\begin{equation}
\dzon_0 = -\frac{\eta\kf\zfil^0}{\nt k_{01} + \eta\kf}\,.
\end{equation}
For a large mobility with $\eta\kf \gg \nt k_{01}$, the average
backsteps size is $\dzon_0 \approx \zfil^0$, that is, the ensemble is
effectively reset to the initial position $z=0$.

Together, the master equation of \eq{MasterEq} for the stochastic
binding dynamics and the rules for the displacement upon binding and
unbinding of motors fully characterize dynamics and movement of an
ensemble of molecular motors. For constant external load, ensemble
movement is slaved to binding and unbinding of motors, because on- and
off-rates are independent of the position $z$ of the ensemble. In this
case, the master equation can be solved independently for the
probability distribution $p_i(t)$ and the average velocity of an
ensemble can be inferred from this solution. The average bound velocity,
that is, the average velocity of ensembles with at least one bound
motor, is given by
\begin{equation}\label{eq:AvgBoundVelo}
\vb(t) = \sum_{i=1}^{\nt} v_i \pbnd_i(t)
       = \sum_{i=1}^{\nt} v_i \frac{p_i(t)}{1-p_0(t)}\,.
\end{equation}
The probability distribution $\pbnd_i(t)$ is normalized over the
attached states $i \in \{1, \dots, \nt\}$ of the ensemble. The effective
velocity of an ensemble, which includes the backward motion (slips) of
the unbound ensemble, is given by the average
\begin{equation}\label{eq:AvgEffVelo}
\veff(t) = \sum_{i=0}^{\nt} v_i p_i(t)
\end{equation}
over all states $i \in \{0, \dots, \nt\}$. Because $\vu = -\eta\fext
\leqslant 0$, the effective velocity is smaller than the bound velocity,
$\veff(t) \leqslant \vb(t)$. For linear load, also a probability
distribution for the unbound velocity would have to be determined. From
the average ensemble velocity as function of time, the position $z(t)$
can be calculated as
\begin{equation}
z(t) = z_0 + \int_0^t \veff(t') dt'\,.
\end{equation}
For linear external load, the transition rates characterizing the master
equation depend on the position of the motor ensemble, so that the
master equation has to be solved together with the displacement of the
ensemble, which usually has to be done numerically.

\section{Results}

\subsection{Constant load}

\subsubsection{Analytical solutions of the one-step master equation}

Mathematically, the reduction of the stochastic binding dynamics on the
two-dimensional network of states of \fig{Fig03}(a) to the
one-dimensional system of \fig{Fig03}(b) described by \eq{MasterEq} is a
dramatic advance, because many general results are known for one-step
master equations \cite{b:VanKampen2003}. For constant external load, the
transition rates are independent of the position $z$ of the ensemble and
stationary solutions of the one-step master equation can be derived
analytically. For a single variable and in the absence of sources and
sinks, stationarity implies detailed balance, that is, $r(i+1) p_{i+1} =
g(i) p_i$. Iterating this condition yields the stationary probability
$p_i(\infty)$ to find $i$ bound motors in an ensemble:
\begin{equation}\label{eq:StationaryDist}
p_i(\infty) = \frac{\prod_{j=0}^{i-1} \frac{g(j)}{r(j+1)}}{1 + \sum_{k=1}^{\nt} \prod_{j=0}^{k-1} \frac{g(j)}{r(j+1)}}\,.
\end{equation}
This distribution immediately allows to calculate the average number of
bound motors as
\begin{equation}\label{eq:AvgNBound}
\nb = \Avg{i} = \sum_{i=0}^{\nt} i p_i(\infty)\,.
\end{equation}
In order to calculate averages restricted to attached ensembles with $i
\geqslant 1$, the stationary probability distribution $p_i(\infty)$ has
to be re-normalized for the $\nt$ attached states:
\begin{equation}\label{eq:StationaryDistBound}
\pbnd_i(\infty) = \frac{p_i(\infty)}{1 - p_0(\infty)}
                = \frac{\prod_{j=0}^{i-1} \frac{g(j)}{r(j+1)}}{\sum_{k=1}^{\nt} \prod_{j=0}^{k-1} \frac{g(j)}{r(j+1)}}\,.
\end{equation}

The average detachment time $\tdet$ of an ensemble is defined as the
mean first passage time of the ensemble from the initial attached state,
in which only a single motor is bound ($i=1$), to complete detachment of
the ensemble, where all motors have dissociated ($i=0$). The mean first
passage time $\tdet$ can be calculated analytically using the adjoint
master equation \cite{b:VanKampen2003}:
\begin{equation}\label{eq:BoundTime}
\tdet = \sum_{j=1}^{\nt} \frac{1}{r(j)} \prod_{k=1}^{j-1} \frac{g(k)}{r(k)}\,.
\end{equation}
The average attachment time of an ensemble is defined as the mean first
passage time $\tatt$ from the detached state ($i=0$) to the initial
attached state ($i=1$). This transition involves only a single binding
step so that the mean first passage time is given by the inverse of the
forward rate $g(0)$:
\begin{equation}\label{eq:UnboundTime}
\tatt = \frac{1}{g(0)} = \frac{1}{\nt k_{01}}\,.
\end{equation}
A measure for the ability of ensembles of non-processive motors to
generate force and directed motion is the duty ratio of an ensemble.
For a single molecular motor, the duty ratio is defined as the fraction
of time in the motor cycle, during which the motor is attached to its
substrate. Processive motors usually are characterized by large duty
ratios close to unity, which allows them to walk along the substrate for
many motor cycles. Non-processive motors, on the other hand, are
characterized by small duty ratios, which reduces the interference
between cooperating motors. For ensembles of molecular motors, we define
the ensemble duty ratio $\rd$ as
\begin{equation}\label{eq:DutyRatio}
\rd = \frac{\tdet}{\tdet + \tatt}\,.
\end{equation}
This is the ratio of detachment time $\tdet$, which is the average time
during which an ensemble remains attached to the substrate before
detaching again, to the average time it takes to complete one
attachment-detachment cycle of an ensemble, which is the sum $\tdet +
\tatt$ of detachment and attachment time. To allow for efficient motion
and force generation, the duty ratio of an ensemble of non-processive
motors should be close to unity, comparable to that of processive motors.

The average bound velocity of an ensemble in the stationary state is
given by \eq{AvgBoundVelo} with the stationary probability distribution
from \eq{StationaryDistBound} replacing $\pbnd_i(t)$, that is,
\begin{equation}\label{eq:StatAvgBoundVelo}
\vb = \sum_{i=1}^{\nt} v_i \pbnd_i(\infty)\,.
\end{equation}
Correspondingly, the average effective velocity in the stationary state
is found by inserting $p_i(\infty)$ from \eq{StationaryDist} in
\eq{AvgEffVelo} as
\begin{equation}\label{eq:StatAvgEffVelo}
\veff = \sum_{i=0}^{\nt} v_i p_i(\infty)\,.
\end{equation}
The effective velocity can be also expressed using the ensemble duty
ratio:
\begin{equation}
\veff = \rd\vb + (1-\rd)\vu = \frac{\tdet\vb + \tatt\vu}{\tdet + \tatt}\,.
\end{equation}
Because $\vu \leqslant 0$, the effective velocity of an ensemble is
always smaller than the bound velocity. The closer the ensemble duty
ratio is to unity, the closer is the effective velocity to the bound
velocity.

Processive motors can be characterized by their processivity, that is,
the average number of steps a motor takes on a substrate before
unbinding. For ensembles of non-processive motors, we can use the
average walk length $\dw$ between attachment and complete detachment as
a measure for the effective processivity. Assuming that relaxation to
the stationary distribution $\pbnd_i(\infty)$ is fast compared to the
detachment time $\tdet$, the ensemble moves with constant bound velocity
$\vb$ from \eq{StatAvgBoundVelo} over the detachment time $\tdet$ so
that the average walk length is given by the product
\begin{equation}\label{eq:WalkLength}
\dw = \vb \tdet\,.
\end{equation}
Unlike processive motors, $\dw$ does not correspond to a fixed number of
binding and unbinding steps of motors because the displacement steps
depend on the state of the ensemble.

\subsubsection{Binding dynamics}

To demonstrate the stochastic effects resulting from the finite number
of motors in an ensemble and the influence of the catch bond character
of the post-power-stroke state, we first study the dependence of the
binding dynamics on ensemble size $\nt$ and external load $\fext$.

\FIG{Fig05}{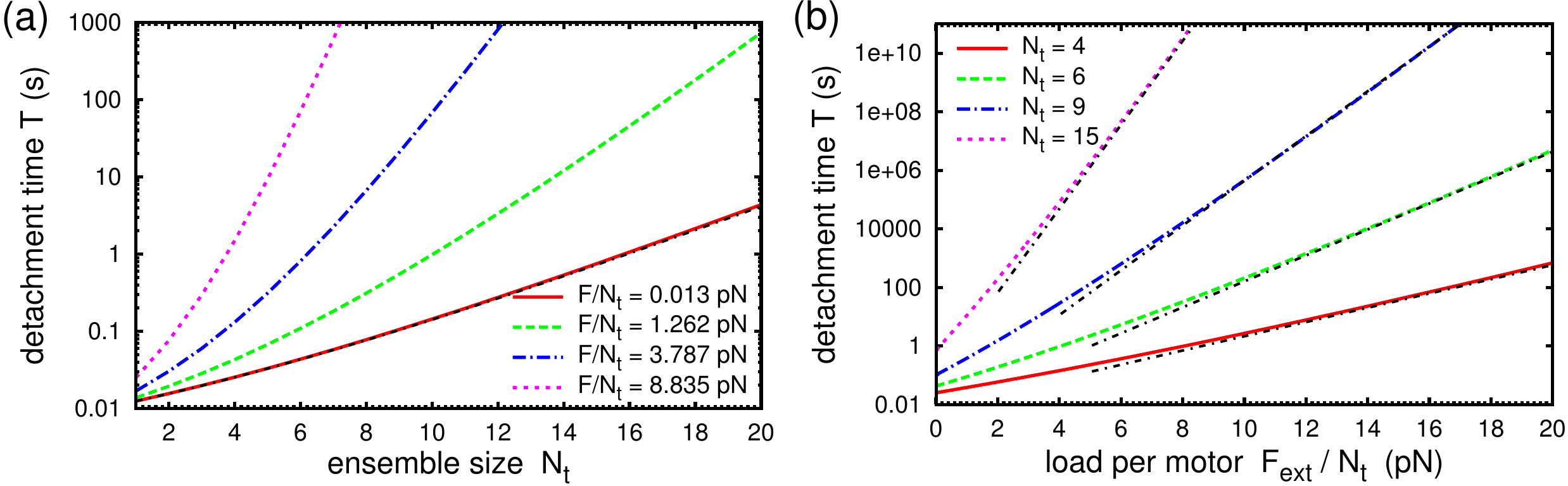}{Analytical results for the parallel cluster model
with constant external load: average detachment time $\tdet$. (a)
$\tdet$ as function of ensemble size $\nt$ for the values $\fext/\nt =
0.0126\pN$, $1.262\pN$, $3.787\pN$ and $8.835\pN$ of the external load
per motor. The black, dash-dotted curve is the approximation of
\eq{BoundTimeApprx} for $\fext/\nt = 0$. (b) $\tdet$ as function of the
external load per motor $\fext/\nt$ for ensemble sizes $\nt=4$, $6$, $9$
and $15$. Black, dash-dotted curves are exponential approximations.
Constant parameters are listed in \tab{Tab01}.}

\fig{Fig05}(a) shows the average detachment time $\tdet$ (see
\eq{BoundTime}) of an ensemble of \myo motors as function of ensemble
size $\nt$ for different values of the external load per motor
$\fext/\nt$. $\tdet$ appears to increases exponentially with $\nt$,
where prefactor and scale of the exponential increase with $\fext/\nt$.
An approximation for $\tdet$ can be derived for vanishing external load
under the assumption that all bound motors are in the post-power-stroke
state, which is justified by the strong bias of the LTE distribution
towards the post-power-stroke state. For $\fext=0$, the dynamics of the
bound motors is not coupled so that not only the on-rate $k_{01}$ but
also the off-rate $k_{20} \simeq k_{20}^0$ is independent of the
ensemble state $(i,j)$. A series expansion of \eq{BoundTime} for
$\tdet$ then leads to
\begin{equation}\label{eq:BoundTimeApprx}
\tdet \approx \frac{1}{k_{01} \nt} \left[\Exp{\ln\left[\frac{k_{20}^0 + k_{01}}{k_{20}^0}\right]\nt} - 1\right]
      = \tatt \left[\Exp{\ln\left[\frac{k_{20}^0 + k_{01}}{k_{20}^0}\right]\nt} - 1\right]\,.
\end{equation}
Comparison with the exact results for $\fext/\nt = 0.013 \pN$ in
\fig{Fig05}(a) shows excellent agreement. For finite load, no closed
form can be found, because the off-rate $k_{20}(i,j)$ is a strongly
non-linear function of the ensemble state $(i,j)$. A fit of $\tdet$ to a
function of the type of \eq{BoundTimeApprx} with adapted prefactor and
scale of the exponential yields a better approximation than a pure
exponential but deviations at small $\nt$ are still observed, in
particular for large external load (not shown).

\fig{Fig05}(b) shows the average detachment time $\tdet$ for different
ensemble sizes as function of the external load per motor. $\tdet$
increases exponentially for not too small values of $\fext/\nt$ where
the scale of the exponential increases with $\nt$. This exponential
increase is a consequence of the catch bond character of the
post-power-stroke state: for all values of a constant external load, the
LTE distribution is strongly biased towards the post-power-stroke state,
that is, $p(i|i) \lesssim 1$ and $0 \lesssim p(j\neq i|i)$. Hence,
unbinding will occur predominantly from the post-power-stroke state so
that the effective reverse rate $r(i) \approx p(i|i) k_{20}(i,i) \propto
\Exp{\fext/i\fdel}$ decreases exponentially under load. This induces the
exponential increase of $\tdet$ observed in \fig{Fig05}(b). Only for
very large loads beyond $\fext/\nt \simeq 20\fdel$, unbinding from the
post-power-stroke state would become slow enough to make unbinding from
the weakly-bound state significant so that $\tdet$ would reach a
plateau. At this level of force, however, forced unbinding would have to
be taken into account \cite{a:ColombiniEtAl2007} and we do not consider
such large forces in our model. The average attachment time $\tatt$ of a
\myo ensemble involves only a single binding step, so that the
dependence on $\fext$ and $\nt$ is rather weak: $\tatt = \left(k_{01}
\nt\right)^{-1}$ is independent of external load and decreases inversely
with ensemble size.

\FIG{Fig06}{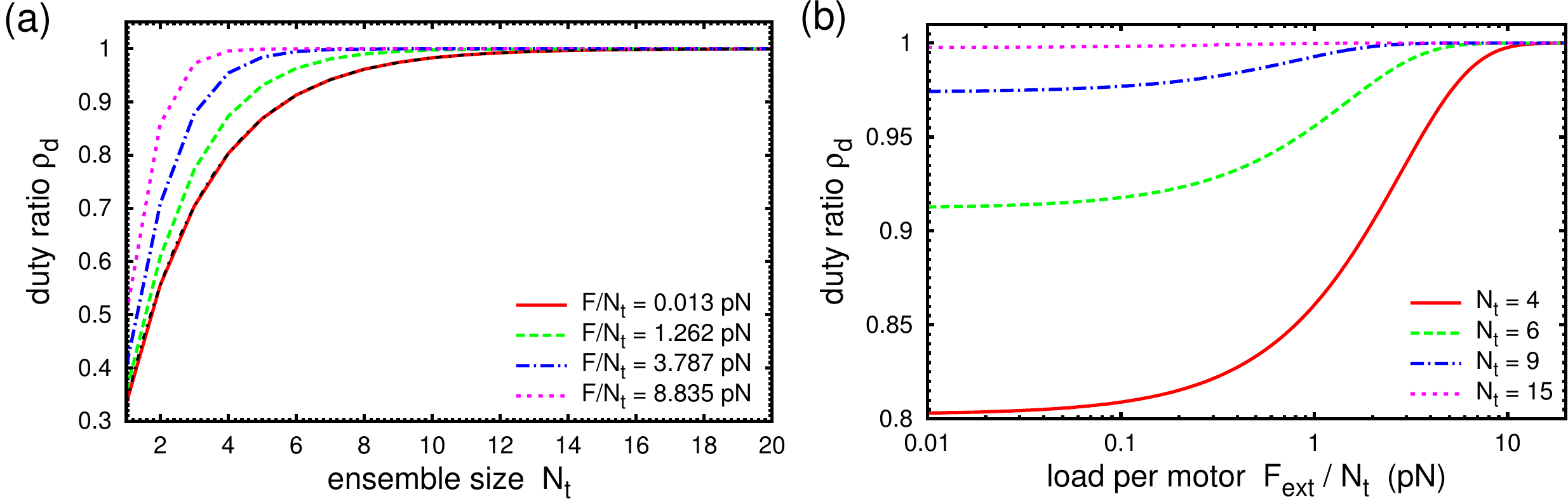}{Analytical results for the parallel cluster model
with constant external load: ensemble duty ratio $\rd$. (a) $\rd$ as
function of ensemble size $\nt$ for the values $\fext/\nt = 0.0126\pN$,
$1.262\pN$, $3.787\pN$ and $8.835\pN$ of the external load per motor.
The black, dash-dotted curve is the approximation of \eq{DutyRatioApprx}.
(b) $\rd$ as function of the external load per motor $\fext/\nt $ for
ensemble sizes $\nt=4$, $6$, $9$ and $15$. Constant parameters are
listed in \tab{Tab01}.}

The ability of a molecular motor to generate sustained levels of force
or continuous motion depends crucially on its duty ratio. Due to the
increase of the detachment time $\tdet$ and the decrease of the
attachment time $\tatt$ with $\nt$, the ensemble duty ratio can be
adjusted via the ensemble size $\nt$. The minimal number of motors,
which could allow for a duty ratio close to unity and almost continuous
attachment of an ensemble, is determined by the inverse of the duty
ratio of a single motor. For $\fext=0$, the off-rate of \myo can be
approximated as $k_{20} \simeq k_{20}^0 \simeq 80\Hz$ because \myo
unbinds almost exclusively from the post-power-stroke state. With the
on-rate $k_{01} = 40\Hz$, the duty ratio of a single \myo is
\begin{equation}\label{eq:DutyRatioSingle}
\rds = \frac{\tdet}{\tatt+\tdet} = \frac{k_{01}}{k_{01}+k_{20}^0} \simeq 0.33\,.
\end{equation}
This value is significantly larger than the observed duty ratio of
skeletal and smooth muscle \myo but comparable to the duty ratio
($\simeq 0.23$) of non muscle \myo \cite{a:SoaresEtAl2011}. For $\rds
\simeq 0.33$, a minimum of $\nt \simeq 3$ motors is required for
continuous attachment. However, due to the stochastic binding dynamics
and the lack of coordination of individual motors, the ensemble duty
ratio will be smaller than the sum over the single motor duty ratios and
a larger number of motors will be required to ensure continuous
attachment. For vanishing external load, the approximation of
\eq{BoundTimeApprx} for the average detachment time $\tdet$ can be used
to derive an approximation for the duty ratio. Inserting
\eq{BoundTimeApprx} for $\tdet$ in \eq{DutyRatio} for the ensemble duty
ratio yields
\begin{equation}\label{eq:DutyRatioApprx}
\rd \approx 1 - \Exp{-\ln\left[\frac{k_{20}^0+k_{01}}{k_{20}^0}\right]\nt}\,.
\end{equation}
With increasing $\nt$, the duty ratio saturates exponentially from the
single motor value $\rd = 1 - k_{20}^0/(k_{20}^0+k_{01})$ for $\nt=1$
towards $\rd \simeq 1$ for large $\nt$. \fig{Fig06}(a) shows $\rd$ as
function of $\nt$ for different values of the external load per motor.
\eq{DutyRatioApprx} provides an excellent approximation for
near-vanishing load. For $\fext/\nt = 0.013\pN$, the duty ratio is $\rd
\simeq 0.33$ for a single motor. For $\nt=3$, the duty ratio is $\rd
\simeq 0.7 < 1$ and reaches unity for ensemble sizes beyond $\nt \simeq
15$. With increasing external load, the duty ratio is elevated already
for $\nt=1$ and a smaller number of motors is required to reach a duty
ratio of $\rd \simeq 1$. \fig{Fig06}(b) shows the ensemble duty ratio
(see \eq{DutyRatio}) as function of the external load per motor. For
$\nt=4$, $\tdet \simeq 0.025\s$ and $\tatt \simeq 0.006\s$ so that $\rd
\simeq 0.8$ at $\fext/\nt = 0$. Due to the exponential increase of
$\tdet$ with $\fext/\nt$, the duty ratio increases quickly and saturates
at $\rd \simeq 1$ above $\fext/\nt \simeq \fdel \simeq 12.6\pN$. With
increasing ensemble size $\nt$ the duty ratio increases and the limiting
value $\rd \simeq 1$ is reached at smaller values of $\fext/\nt$. For
$\nt=15$ the duty ratio is practically unity for all values of the
external load, because $\tdet \simeq 0.73\s$ and $\tatt \simeq 0.0017\s$
so that $\rd \simeq 0.998$ at $\fext=0$.

\FIG{Fig07}{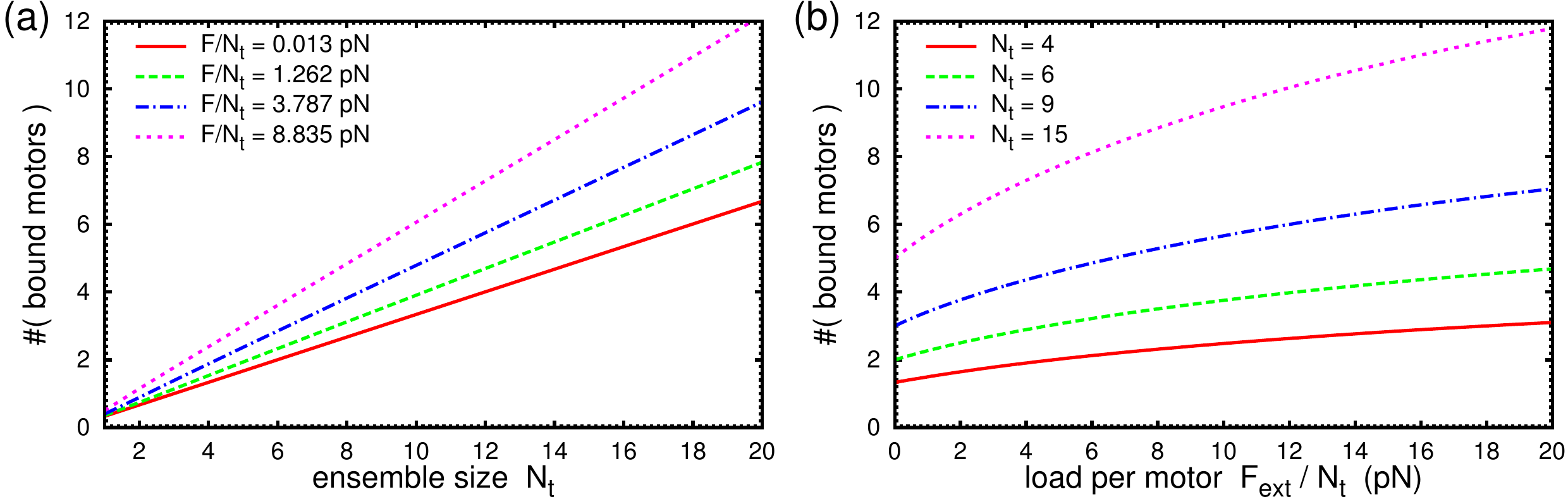}{Analytical results for the parallel cluster model
with constant external load: average number of bound motors $\nb$. (a)
$\nb$ as function of ensemble size $\nt$ for the values $\fext/\nt =
0.0126\pN$, $1.262\pN$, $3.787\pN$ and $8.835\pN$ of the external load
per motor. (b) $\nb$ as function of external load per motor $\fext/\nt $
for ensemble sizes $\nt=4$, $6$, $9$ and $15$. Constant parameters are
listed in \tab{Tab01}.}

\fig{Fig07}(a) shows the average number of bound motors $\nb$ as
function of ensemble size $\nt$ for different values of the external
load. For all values of the load, $\nb$ increases linearly with $\nt$,
where the increase becomes steeper under larger external load.
\fig{Fig07}(b) shows $\nb$ as function of the external load per motor
for different ensemble sizes. At vanishing load, $\nb \simeq \rds \nt
\simeq 0.33 \nt$. For $\fext=0$, the motors bind independently, because
the PCM assumes that motors in equivalent states have equal strains and
because most bound motors are in the post-power-stroke state, so that no
internal stress is built up between motors in different states. With
increasing $\fext/\nt$, the average number of bound motors increases
sub-linearly and plateaus towards $\nt$ for large loads. The recruitment
of additional bound motors under increasing load was described
theoretically \cite{a:Duke1999} and has been observed experimentally for
\myo in muscle \cite{a:PiazzesiEtAl2007}. The increase of the average
number of bound motors under load observed in \fig{Fig07}(b), as well as
the increase of $\tdet$ and $\rd$ under load, confirms that \myo as a
whole behaves as a catch bond over a large range of values of a constant
external load. Thus, the efficiency of an ensemble of non-processive
motors for the generation of motion and force, which is determined by
detachment time, duty ratio and number of bound motors, can be adjusted
by changing the ensemble size $\nt$ or by using the force sensitivity of
the motors.

\subsubsection{Stochastic trajectories}

To gain more insight into the movement of an ensemble of non-processive
molecular motors and the relation between binding and movement, it is
instructive to look at single, stochastic trajectories. We use the
Gillespie algorithm \cite{a:Gillespie1976} to simulate stochastic
binding and unbinding trajectories according to the one-step master
equation of \eq{MasterEq} and apply the rules for the displacement upon
binding and unbinding to implement ensemble movement. Within the
Gillespie algorithm, the transition rates $r(i)$ and $g(i)$ are used to
choose time and type of the next stochastic transition from an
exponential probability distribution. After a binding transition $i \to
i+1$ with $i \geqslant 1$, the position $z$ of the ensemble is changed
by $\dzon_i = -x_i/(i+1)$. For $i=0$, the position of the detached
ensemble is changed by $\dzon_0 = -\eta\fext\tau$ before attachment.
Here, $\tau$ is a random attachment time with average $\tatt$. After an
unbinding transition $i \to i-1$, $z$ is unchanged for $i \geqslant 2$.
If the last motor unbinds, that is for $i=1$, ensemble position is
changed by $\dzoff_1 = -x_1$. After adjusting ensemble position, the new
value of the strain $\xij$ of weakly-bound motors is determined from the
balance of forces in \eq{ForceBalance} and the position $\zfil$ of the
motor filament is set to $\zfil = z-\xij$. With the updated LTE
distribution $p(j|i)$, the average strain of weakly-bound motors, $x_i =
\sum_{j=0}^i \xij p(j|i)$, and the transition rates $r(i)$ and $g(i)$
are calculated and new random time and type of the next reaction are
chosen.

\FIG{Fig08}{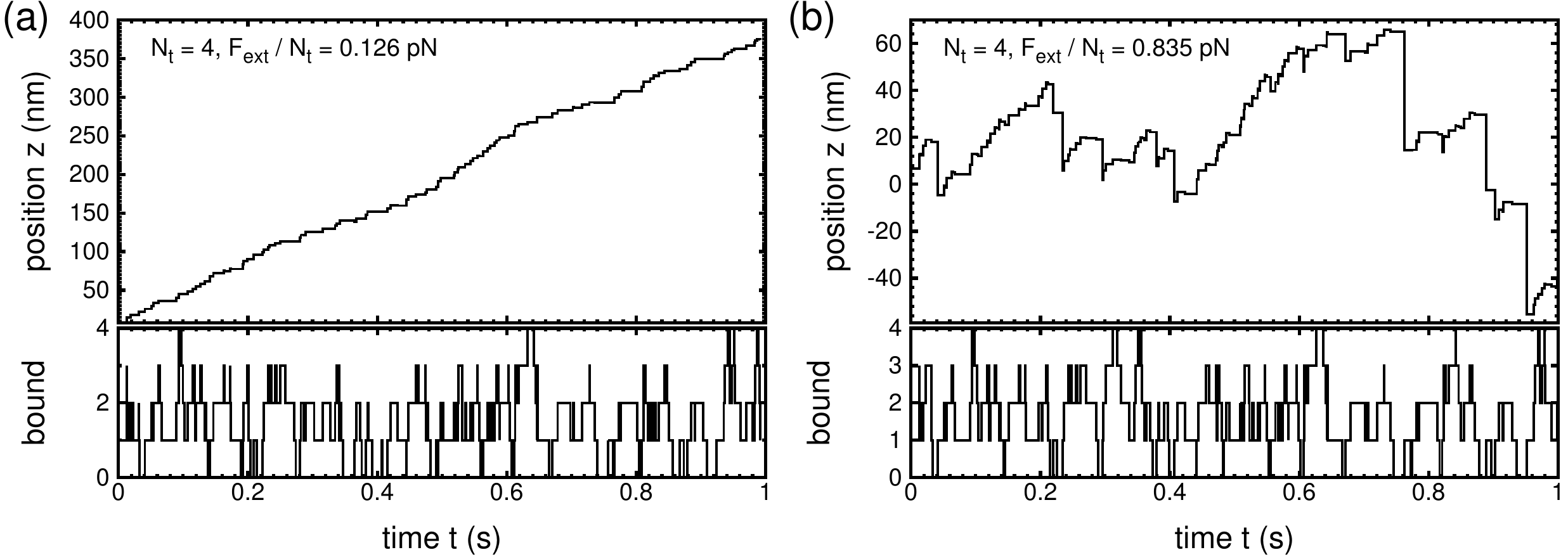}{Stochastic trajectories for constant external load.
Ensemble position $z$ (upper panel) and number $i$ of bound motors
(lower panel) as function of time $t$ for ensemble size $\nt=4$ and
external load per motor (a) $\fext/\nt = 0.126\pN$ and (b) $\fext/\nt =
0.835\pN$. In (a) and (b) the detached ensemble slides backwards with
mobility $\eta = 10^3 \mob$. Constant parameters are listed in \tab{Tab01}.}

\fig{Fig08}(a) shows a stochastic trajectory of an ensemble with $\nt=4$
motors working against the constant external load $\fext/\nt = 0.126
\pN$. The lower panel shows the fluctuating number of bound motors $i$,
the upper panel the ensemble position $z$. The external load is below
the stall force so that the attached ensemble moves forward with a
velocity fluctuating around the average bound velocity $\vb > 0$. The
number of bound motors fluctuates strongly and the ensemble frequently
detaches completely from the substrate. Unlike $z$, the position $\zfil
= z-x_i$ of the motor filament is also changed by a change of the strain
$x_i$ of the motors. When a motor binds, the external load is
distributed over a larger number of motors, so that the strain is
reduced and the motor filament slides forward in addition to the change
of $z$. When a motor unbinds, the strain $x_i$ of the remaining bound
motors increases, so that the motor filament slides backwards while $z$
remains constant. Trajectories of $\zfil = z-x_i$ therefore are rather
close to $z$ but show stronger fluctuations (not shown). Complete
detachment of the ensemble leads to backward steps of average size
$\dzon_0 = -\vu\tatt$. For a small load as in \fig{Fig08}(a), however,
backsteps are too small to be resolved so that detachment events appears
as pauses in the trajectory. \fig{Fig08}(b) shows a stochastic
trajectory for an ensemble with $\nt=4$ motors but at larger external
load. At this load, the average bound velocity is positive so that the
net movement of the attached ensemble is forward, although the strain
$x_1$ is positive for a single bound motor so that binding of the second
motor leads to a backward step in $z$. Detachment of the ensemble has
become only marginally less frequent under the larger load, but the size
of the backward steps has increased (note the larger $z$ scale in (b)
compared to (a)) such that the effective velocity is close to zero.
Hence the value of the external load is close to the effective stall
force, at which the forward movement of the attached ensemble balances
the backward slips of the detached ensemble.

\FIG{Fig09}{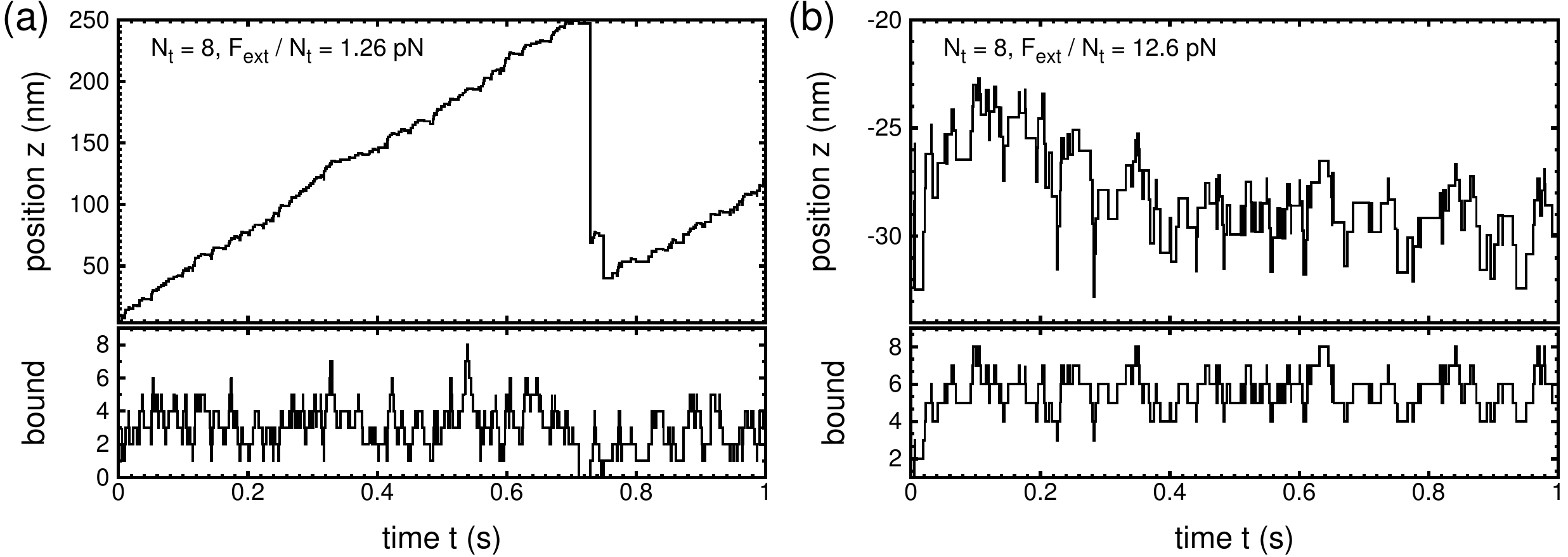}{Stochastic trajectories for constant external load.
Ensemble position $z$ (upper panel) and number $i$ of bound motors
(lower panel) as function of time $t$ for ensemble size $\nt=8$ and
external load per motor (a) $\fext/\nt = 1.26\pN$ and (b) $\fext/\nt =
12.6\pN$. In (a) and (b) the unbound ensemble slides backwards with
mobility $\eta = 10^3 \mob$. Constant parameters are listed in \tab{Tab01}.}

\fig{Fig09}(a) shows a stochastic trajectory of an ensemble at larger
ensemble size $\nt=8$. The external load is below the stall force, so
that the attached ensemble moves forward with slightly fluctuating
velocity. Due to the larger ensemble size, complete detachment occurs
less frequently but the large external load leads to large backsteps. In
\fig{Fig09}(b), the ensemble size is the same as in (a) but $\fext$ is
increased to a value close to the stall force with $\vb \simeq 0$. Due
to the catch bond character of \myo, the typical number of bound motors
is increased and complete detachment is rare. Nevertheless, the ensemble
position $z$ fluctuates strongly because binding at small $i$ leads to
backsteps $\dzon_i \leqslant 0$ whereas binding at larger $i$ leads to
forward steps $\dzon_i \geqslant 0$ of the ensemble. At the stall force
these two effects balance so that the average bound velocity vanishes,
$\vb=0$.

\subsubsection{Velocity and walk length}

\FIG{Fig10}{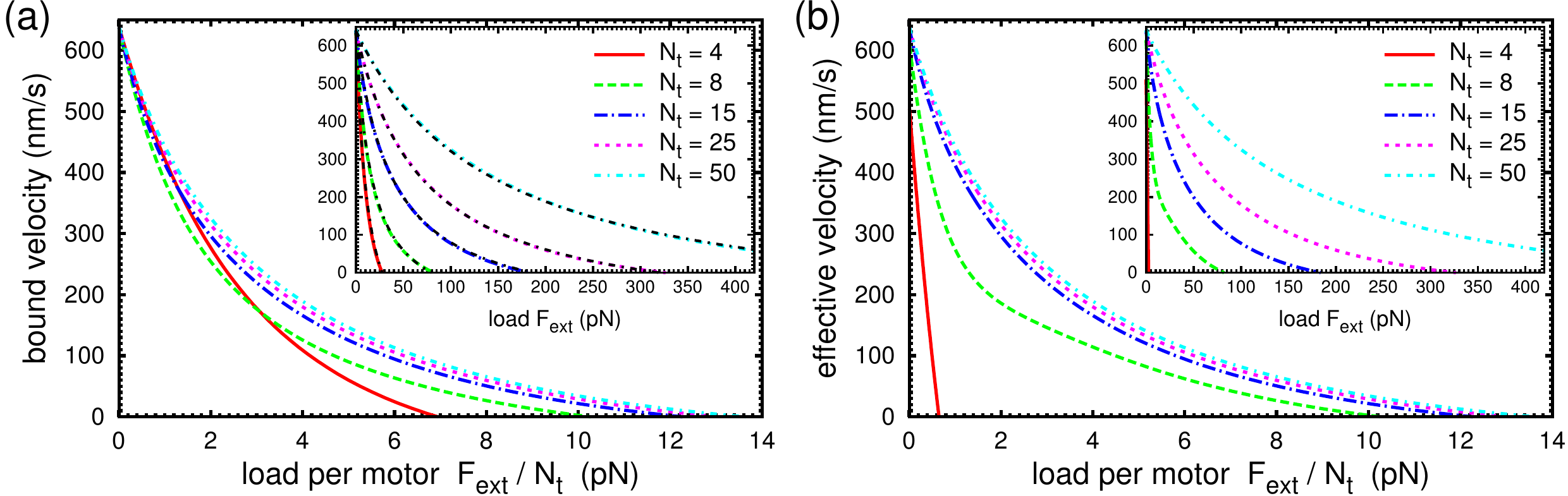}{Analytical results for the parallel cluster model
with constant external load. (a) Average bound velocity $\vb$ and (b)
average effective velocity $\veff$ as function of the external load per
motor $\fext/\nt$ for ensemble sizes $\nt=4$, $8$, $15$, $25$ and $50$.
For $\veff$ the viscous mobility of detached ensembles is $\eta = 10^3
\mob$. The insets show $\vb$ and $\veff$ as function of external load
$\fext$. Black, dash-dotted curves in the inset of (a) show the
Hill-relation from \eq{HillRelation} with $\alpha = 0.46\pN$,
$0.205\pN$, $0.2\pN$, $0.215\pN$ and $0.215\pN$ for $\nt=4$, $8$, $15$,
$25$ and $50$. Constant parameters are listed in \tab{Tab01}.}

\fig{Fig10}(a) shows analytical results for the average bound velocity
$\vb$ (see \eq{StatAvgBoundVelo}) as function of the external load per
motor for different ensemble sizes. For $\fext=0$, the average bound
velocity is $\vb(\fext=0) \simeq 640\nm\s^{-1}$ and is independent of
ensemble size $\nt$. From its maximal value, $\vb$ decreases with
increasing $\fext/\nt$ in a concave fashion and becomes negative for
loads above the stall force $\fs$. The concave shape of the
force-velocity relation has been explained before by the increase of the
average number of bound motors under load \cite{a:Duke1999, a:Duke2000}
which is a consequence of the catch bond character of \myo. Compared to
a system with constant $\nb$, the increase of $\nb$ reduces the load on
the individual bound motors and reduces their strain $\xij$. According
to \eq{BoundVeloI} for $v_i$, this increases the velocity at given
external load and hence the stall force of the ensemble. Beyond
sufficiently large ensemble sizes of $\nt \geqslant 15$ the
force-velocity relation $\vb(\fext/\nt)$ hardly changes with further
increasing $\nt$. At $\nt=15$, the stall force per motor is $\fs/\nt
\simeq 12.4\pN$. With increasing ensemble size, it increases slightly to
$\fs/\nt \simeq 13.5\pN$ for $\nt=50$ because the force-velocity
relation becomes very shallow near $\fs$. Smaller ensembles show a more
rapid decrease of $\vb$ from the value at $\fext=0$ as well as a smaller
$\fs/\nt$. The inset in \fig{Fig10}(a) shows $\vb$ as function of the
absolute load $\fext$ for the same values of $\nt$ as in the main panel.
The analytical results from the model are compared to the effective,
Hill-type force-velocity relation \cite{a:Hill1939b}
\begin{equation}\label{eq:HillRelation}
v_{\rm hill}(\fext) = \vb(\fext=0) \frac{\fs - \fext}{\fs + (\fext/\alpha)}\,.
\end{equation}
For the comparison, the values of $\vb(\fext=0)$ and $\fs$ are taken
from the model results. The parameter $\alpha$ has units of force and is
used to fit the curvature of the force-velocity relations. The
comparison shows that \eq{HillRelation} describes the force-velocity
relation for all $\nt$ extremely well. The parameter $\alpha$ is almost
identical for ensemble sizes $\nt=8$ to $50$ with a typical value of
$\alpha \simeq 0.21\pN$ and differs significantly only for $\nt=4$. For
$\nt=4$, the ensemble detaches frequently so that the off-step
$\dzoff_{1j}$ from \eq{dzoffij}, in which the strain of the last
unbinding motor is released and changes the ensemble position,
contributes significantly to the bound ensemble velocity. This causes
the markedly different dependence on $\fext/\nt$ with smaller curvature
of the force-velocity curve for $\nt=4$. For $\nt=1$, this off-step upon
unbinding can still generate forward movement.

The good fit of $\vb(\fext)$ by $v_{\rm hill}(\fext)$ demonstrates the
qualitative agreement of the PCM with the experimental force-velocity
curve of muscle, for which $v_{\rm hill}$ was originally derived. Our
model can now be used to estimate the values for load free velocity and
stall force and to elucidate their dependence on the model parameters.
For vanishing force, the number of bound motors on average is $\nb
\simeq \rd\nt \simeq \nt/3$. Since almost all of these motors are in the
post-power-stroke state, we find $x_i = -d$ and the binding step of the
ensemble is $\dzon_{\nb} = -x_i/(\nb+1) \simeq d/\nb $. Together with
the forward rate $g(\nb) = (\nt-\nb) k_{01} = (2\nt/3) k_{01}$ we obtain
the bound velocity $\vb(\fext=0) \simeq g(\nb) \dzon_{\nb} \simeq d
\left[(\nt - \nb) / \nb\right] k_{01} = 2 d k_{01} = 640 \nm\s^{-1}$. To
estimate the stall force, we again assume that all $\nb$ bound motors
are in the post-power-stroke state. At the stall force, the strain
$x_{\nb} = (\fs/\km\nb) - d$ should vanish so that $\dzon_{\nb} = 0$.
Thus, the stall force follows the relation $\fs/\nb = \km d \simeq
20\pN$. For $\nt=15$ at $\fs/\nt \simeq 12.4\pN$ the number of bound
motors can be read from \fig{Fig07}(b) as $\nb \simeq 0.68\nt$. This
yields the ratio $\fs/\nb \simeq 18.2\pN$, which is consistent with the
estimate. The estimate of the stall force shows that $\fs$ increases
linearly with the number of bound motors. If the number of bound motors
remained constant at $\nb(\fext=0) \simeq 0.33\nt$, the stall force
would be reduced to one half the actual value. On the other hand, the
velocity at vanishing external load decreases with the number of bound
motors. If $\nb$ at $\fext=0$ had the same value $\nb = 0.68\nt$ found
at $\fext=\fs$, the unloaded velocity would be reduced to $\vb(\fext=0)
\simeq 160 \nm\s^{-1}$. In this way, the catch bond character of \myo
motor allows ensembles to adapt the typical number of bound motors to
the environmental conditions and to increase the dynamic range of an
ensemble. At small external load, a small number of bound motors is able
to generate fast movement of the ensemble. At large external load, a
large number of bound motors is needed to overcome the external load and
to generate slow forward movement.

\fig{Fig10}(b) shows the average effective velocity $\veff$ (see
\eq{StatAvgEffVelo}) as function of the external load per motor for
different ensemble sizes. For large ensembles with $\nt \geqslant 15$,
the effective velocity is essentially identical to the bound velocity
because large ensembles rarely detach from the substrate ($\rd \simeq
1$). The difference between $\vb$ and $\veff \leqslant \vb$, which is
observed for smaller ensemble sizes, is determined by the frequency
$\tdet^{-1}$ of detachment and the size $\dzon_0$ of the backsteps. As
shown in \fig{Fig05}, $\tdet^{-1}$ decreases exponentially with $\fext$,
while $\dzon_0$ increases linearly with $\fext$ (see \eq{DZOnAtt}). For
$\fext=0$, detached ensembles do not move so that even very small
ensembles (including single motors) effectively move forward. Because
the frequency of detachment as well as the average duration $\tatt =
(k_{01}\nt)^{-1}$ of detachment events increases with decreasing $\nt$,
$\veff(\fext=0)$ is smaller for smaller $\nt$. For $\nt=4$ it is
$\veff(\fext=0) \simeq 513\nm\s^{-1}$ and $\veff$ decreases very rapidly
with increasing external load. The stall force is reduced to $\fs/\nt
\simeq 1\pN$ compared with the value $\fs/\nt \simeq 7\pN$ for the bound
ensemble. For $\nt=8$, the effective velocity initially decreases
quickly under load. Once the detachment frequency has decreased
sufficiently, the force-velocity relation becomes rather shallow. As
detachment becomes very rare under further increasing load, the stall
force for $\nt=8$ is almost identical to the stall force of the bound
ensemble with $\vb=0$. The interplay of the linear increase of the size
of the backsteps and the exponential decrease of the detachment
frequency under load can also lead to a non-monotonous force-velocity
relation.

\FIG{Fig11}{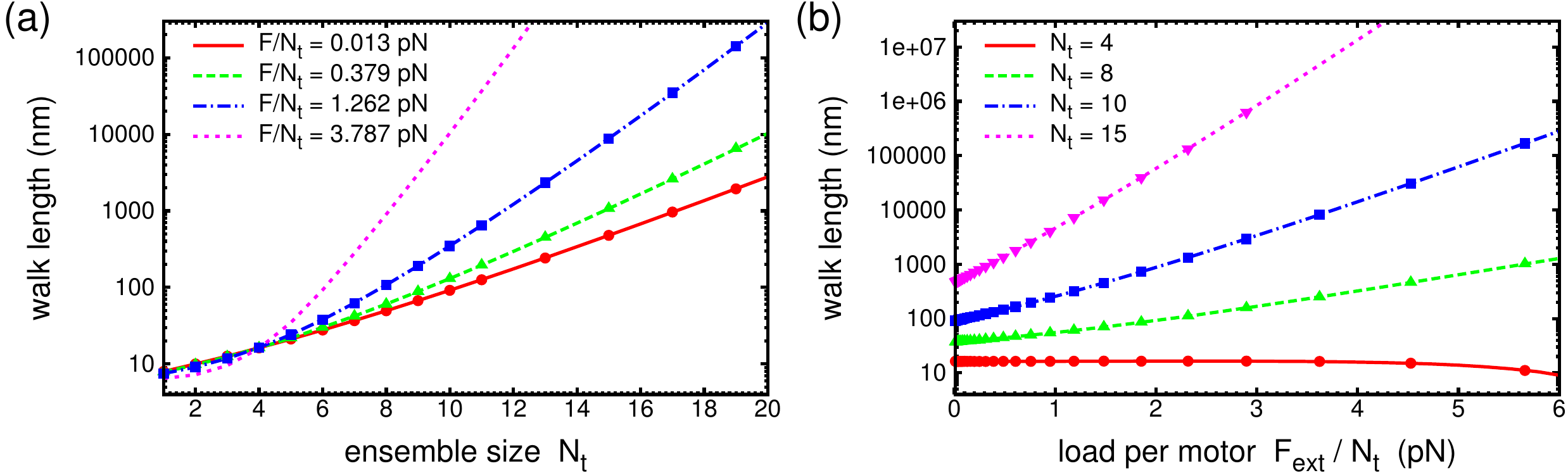}{Comparison of analytical and numerical results for
the parallel cluster model with constant external load: Average walk
length $\dw$. (a) $\dw$ as function of $\nt$ for $\fext/\nt = 0.013\pN$,
$0.379\pN$, $1.26\pN$ and $3.79\pN$. (b) $\dw$ as function of
$\fext/\nt$ for ensemble sizes $\nt=4$, $8$, $10$ and $15$. The
approximation $\dw = \vb T_{10}$ (see \eq{WalkLength}) (curves) is
compared to results from stochastic simulations (symbols). Constant
model parameters are listed in \tab{Tab01}.}

\fig{Fig11}(a) shows the average walk length of an ensemble as function
of ensemble size for different values of the external load $\fext/\nt$.
The stationary approximation $\dw = \vb\tdet$ is compared to numerical
results from stochastic simulations. After an initial transient, the
average walk length increases exponentially with $\nt$. This increase
reflects the exponential increase of the detachment time $\tdet$ with
$\nt$. The initial transient is due to the variation of $\vb$ for small
$\nt$ at given $\fext/\nt$. With increasing external load, the walk
length increases as long as $\fext/\nt$ is below the stall force. For
the smallest external load, the walk length reaches $\dw \simeq 500\nm$
for $\nt=15$. For $\fext/\nt \simeq 1.3\pN \simeq 0.1\fs$, the walk
length increases to $10^4\nm$ because the exponential increase of
$\tdet$ outruns the decrease of the velocity. \fig{Fig11}(b) shows $\dw$
as function of the external load per motor for different ensemble sizes.
For $\nt=4$, the walk length decreases slowly because the quick decrease
of $\vb$ under load compensates the increase of the detachment time. For
larger $\nt$, $\dw$ increases with $\fext/\nt$ over the range of force
shown in the figure. Only when the stall force is reached, the walk
length plummets to negative values. The stationary approximation
describes the exact numerical results remarkably well over the whole
range of ensemble size and external load. This confirms that bound
ensembles are characterized by the stationary values for bound velocity
$\vb$ and processivity $\dw$, in analogy to processive motors.

\subsubsection{Validation of the parallel cluster model}\label{scn:PCMValidation}

The assumption underlying the parallel cluster model---motors in
equivalent mechano-chemical states have identical strains---is not
justified a priori. In fact, a finite distribution of strains is
expected because motors remain bound at fixed positions on the substrate
for random time intervals while the ensemble moves with fluctuating
velocity. When the ensemble moves forward, the bound motor which was
bound for the longest time should have the largest strain while the
motor which has bound most recently should have the smallest strain. For
groups of processive motors (see Refs.~\cite{a:KlumppLipowsky2005a,
a:BergerEtAl2012, a:KunwarMogilner2010}) the load dependence of the
velocity seems to provide a natural mechanism for equalizing the load on
the motors: as a motor moves ahead of the group it will be subject to a
large load; this reduces the velocity of the motor so that the group
will catch up with the advancing motor and the loads will be equalized.
On the other hand, if a motor trails behind the group it will be subject
to the smallest load and will have the largest velocity so that the
trailing motor catches up with the group. Such a mechanism does not seem
to be at work for ensembles of non-processive motors, which cannot move
along the substrate. However, movement of an ensemble of non-processive
motors requires continuous unbinding and binding of the motors. Thereby,
the motors remain bound to the substrate for short time intervals and
the strain, which was build up while the motor was bound, is released
completely before the motor binds again with vanishing strain. Frequent
binding and unbinding of motors as a prerequisite of movement in
combination with the release of strain after unbinding should result in
a narrow distribution of strains of the bound motors, which is the basis
for the PCM. Interestingly, equalizing the load in a group of processive
motors requires several step of the motors along the substrate, that is,
several unbinding and binding events of the individual motor heads,
whereas the strain of a non-processive motors is released in a single
step. Due to the catch bond character of \myo it might occur that motors
remain bound for long time intervals and build up excessive strains.
This, however, should only possible for a small fraction of bound
motors, because the majority of motors in the ensemble is required to
displace the ensemble and build up the strain. Therefore, most bound
motors will still have a narrow distribution of strains.

To validate the assumptions of the PCM, we compare analytical results
obtained within the PCM for constant external load to results from
computer simulations which do not use the PCM assumption of equal motor
strains. Moreover, these simulations do not apply the LTE of bound
states but include stochastic transitions between weakly-bound state and
post-power-stroke state explicitly. In the simulations, the motor cycle
is described by the three distinct mechano-chemical states depicted in
\fig{Fig01}. Without LTE, the stochastic transitions in an ensemble with
$\nt$ motors proceed on the two-dimensional network of mechano-chemical
state shown in \fig{Fig03}(a). The mechano-chemical state of the
ensemble has to be complemented by the strain $\xin$ of every bound
motor in order to calculate strain dependent transition rates. Every
bound motor head is assigned an individual position $\zn$ on the
substrate. For given external load $\fext$ and positions $\zn$, the
strain $\xin$ for every bound motor is calculated from the balance of
forces in \eq{ForceBalance}. The position $z$ of the ensemble is defined
as the average position of the bound motor heads (see \eq{EnsemblePos}).
As in the PCM, detached ensembles slip backwards in the direction of the
external load with mobility $\eta$.

Transitions between unbound and weakly-bound state are independent of
strain so that the transitions $(0) \to (1)$ and $(1) \to (0)$ proceed
with constant transition rates $k_{01} = \const$ and $k_{10} = \const$,
respectively. The values are listed in \tab{Tab01}. For transitions $(1)
\to (2)$ and $(2) \to (1)$ between weakly-bound and post-power-stroke
state we assume constant transition rates $k_{12} = k^0_{12} \exp(-\epp
/ 2\kt)$ and $k_{21} = k_{21}^0 \exp(+\epp/2\kt)$, respectively. For our
simulations we use $k_{12}^0 = k_{21}^0 = 10^3\Hz$ but the actual value
does not affect the results as long as the forward rate $k_{12}$ is not
smaller than the off-rate $k_{10}$ from the weakly-bound state. Because
$k_{21}/k_{12} = \exp(\epp/\kt)$, a Boltzmann distribution for two
states with free energy difference $\epp < 0$ will establish in
equilibrium. Compared to the LTE distribution of \eq{LTEDist}, the
elastic energy of the motors has been omitted. For constant external
load, however, the LTE distribution is dominated by the strong bias
$\epp$ towards the post-power-stroke state so that the omission will
have little effect on results. Unbinding from the post-power-stroke
state is irreversible. The transition $(2) \to (0)$ proceeds with strain
dependent transition rate
\begin{equation}\label{eq:OffStrain}
k_{20}(\xin) = k_{20}^0 \exp(-\kf\xin/\fdel)
\end{equation}
for $\xin \geqslant 0$ and $k_{20}(\xin) = k_{20}^0$ for $\xin < 0$,
where $k_{20}^0$ and $\fdel$ from \tab{Tab01} are used. Unbinding is
slowed down when the neck linker is stretched in the direction of the
external force but remains constant when the neck linker is compressed
in the opposite direction. The latter case does not occur in the PCM
model (as long as the external load is positive) so that the distinction
was not necessary. Results of simulations using the Kramers' type
off-rate for positive and negative strain are discussed in \app{AppB}.
As in the simulations with the PCM, we used the Gillespie algorithm
\cite{a:Gillespie1976} to simulate the stochastic reactions: the
reaction rates for all the motors are used to choose the waiting times
between transition and the kind of reaction from the appropriate
probability distributions. After every transition, strains $\xin$ and
transition rates are updated and the next reaction is determined.

Neglecting the change of the elastic energy in the kinetic description
of the power stroke seems necessary because of the large value of the
stiffness $\km$ of the neck linkers assumed in our model. The neck
linker of a motor is stretched when it goes through the power stroke.
This step is favorable as long as the increase of the elastic energy of
the neck linker is smaller than the (negative) free energy bias $-\epp$
towards the post-power-stroke state. For a larger number of bound
motors, $i \gg 1$, the power stroke stretches the neck linker of a motor
approximately by the power-stroke distance $d \simeq 8\nm$ and the
decrease of the elastic energy of the other bound motors can be
neglected. For $\km \simeq 2.5 \pN\nm^{-1}$, the elastic energy
increases by $\km d^2/2 \simeq 80\pN\nm > 60\pN\nm \simeq -\epp$. For
the given elastic constant of the neck linkers, individual motors are
effectively unable to perform the power stroke when a large number of
bound motors holds the motor filament in place. Thus, the motor ensemble
is stuck kinetically with most motors in the weakly-bound state although
the energy of the ensemble as a whole would be reduced by a transition
of all bound motors to the post-power-stroke state. To overcome this
problem and to make the power stroke favorable also for individual
motors in a kinetic description, significantly smaller values of $\km
\simeq 0.3 \pN\nm^{-1}$ have been used \cite{a:Duke1999, a:Duke2000}.
For larger values $\km \simeq 2.5 \pN\nm^{-1}$ as in our model, the LTE
assumption has been used before \cite{a:VilfanDuke2003b}. For a kinetic
description with a large value of the neck linker stiffness, variants of
the motor cycle have been used, which do not require an explicit load
dependence of the power stroke \cite{a:ChenGao2011, a:WalcottEtAl2012}.
This approach, however, does not allow to describe the transition of the
LTE distribution to the weakly-bound state when working against very
stiff external springs (see \scn{LinLTE}) or the synchronization of the
power stroke against large forces \cite{a:Duke1999}.

\FIG{Fig12}{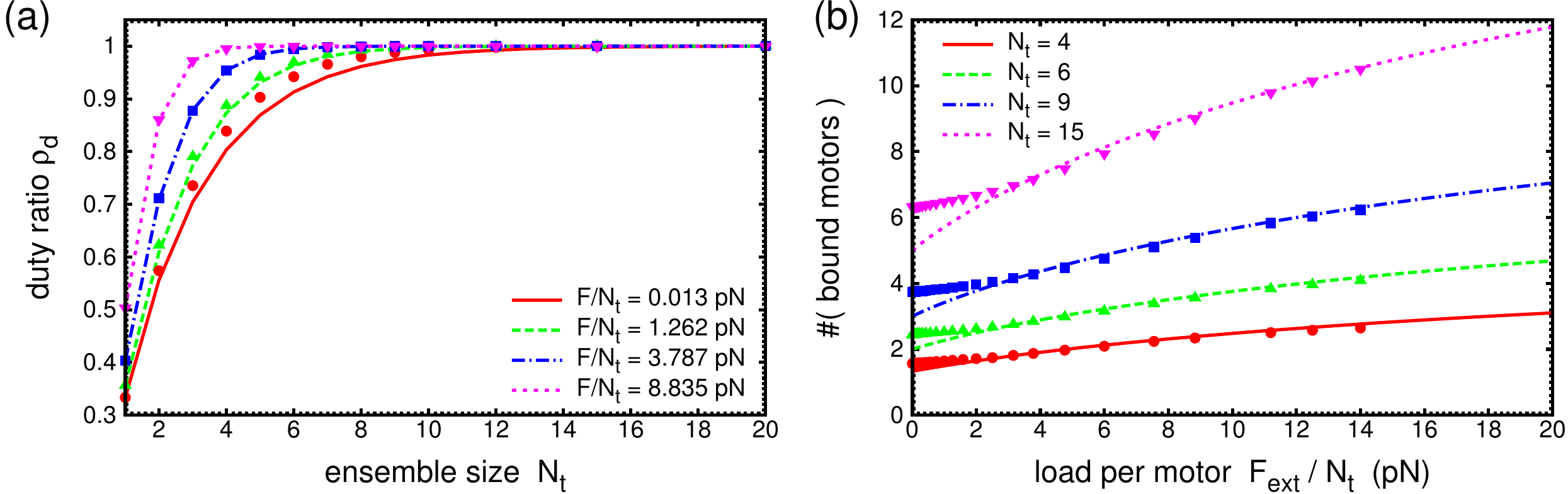}{Validation of the parallel cluster model:
comparison of analytical results using the PCM (lines) with computer
simulations with individual motor strains (symbols). (a) Duty ratio
$\rd$ as function of ensemble size $\nt$ for the values $\fext/\nt =
0.0126\pN$, $1.262\pN$, $3.787\pN$ and $8.835\pN$ of the external load
per motor. (b) Average number of bound motors $\nb$ as function of
external load per motor $\fext/\nt $ for ensemble sizes $\nt = 4$, $6$,
$9$ and $15$. Constant parameters are listed in \tab{Tab01}.}

In the following, we compare analytical results from the PCM with
numerical results from simulations with individual motor strains.
\fig{Fig12}(a) shows the ensemble duty ratio $\rd$ as function of
ensemble size for different values of the external load per motor;
\fig{Fig12}(b) shows the average number of bound motors $\nb$ as
function of $\fext/\nt$ for different values of $\nt$. For all ensemble
sizes, analytical results from the PCM agree very well with numerical
results with individual motor strains. Significant deviations are only
observed for small external load where the analytical results
underestimate the duty ratio as well as the number of bound motors.
These deviations are caused by the distribution of strains among bound
motors. For vanishing and small external load, unbinding of those motors
with positive strain will be slowed down while unbinding of those motors
with negative strain is unaffected. With increasing external load, the
strain of the motors is dominated by the external load and the effects
of the distribution of strains become negligible.

\FIG{Fig13}{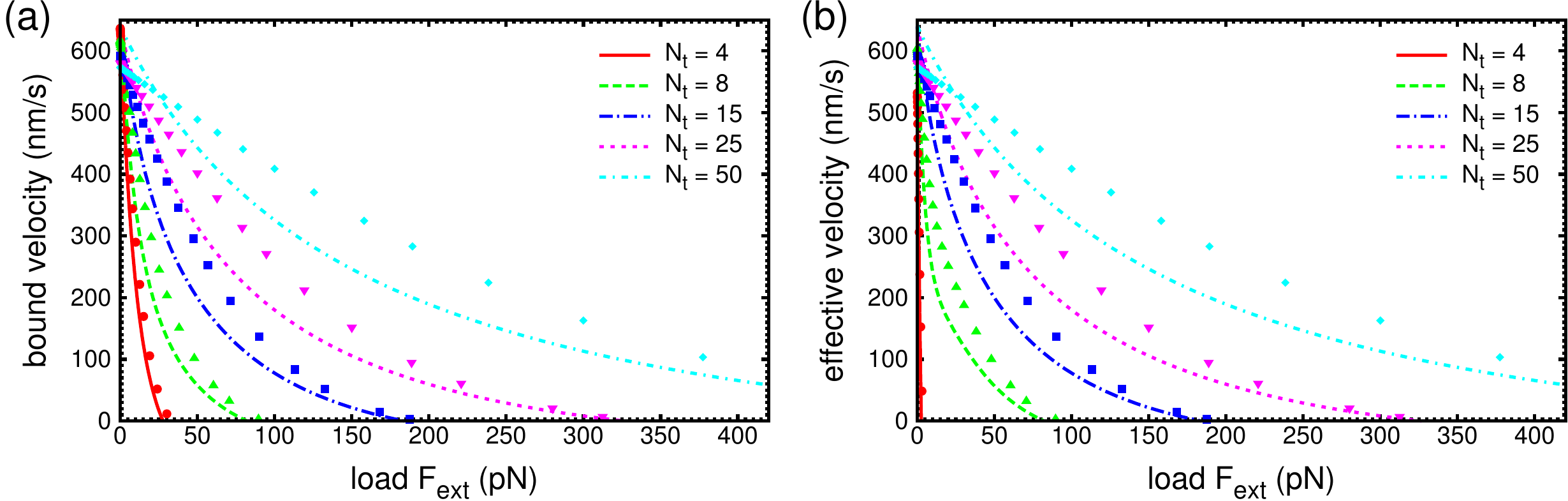}{Validation of the parallel cluster model:
comparison of analytical results using the PCM (lines) with computer
simulations with individual motor strains (symbols). (a) Average bound
velocity $\vb$ and (b) average effective velocity $\veff$ as function of
the external load per motor $\fext/\nt$ for ensemble sizes $\nt=4$, $8$,
$15$, $25$ and $50$. The mobility of the free ensemble entering $\veff$
is $\eta = 10^3 \mob$. Constant parameters are listed in \tab{Tab01}.}

\fig{Fig13} compares analytical and simulation results for the bound
velocity $\vb$ in (a) and for the effective velocity $\veff$ in (b) as
function of external load per motor for different ensemble sizes. The
agreement is quite good for small $\nt$ but clear deviations are
observed for large $\nt$ at small and intermediate values of $\fext$. At
vanishing and small load, the simulations show a decrease of the bound
velocity $\vb$ with increasing $\nt$. In the PCM, $\vb(\fext=0)$ was
independent of $\nt$. The $\nt$ dependence of $\vb$ is caused by the
increase of the average number of bound motors at small external load
observed in \fig{Fig12}(b). As demonstrated in the previous section, an
increasing number of bound motors reduces the bound velocity at
vanishing external load. At intermediate values of $\fext$, the
numerical results for $\vb$ and $\veff$ are larger than predicted by the
PCM because the numerical force-velocity relation is less concave than
the analytical one. This underestimation of the velocity has been
predicted in \scn{EnsMov} and is due to the preferential unbinding of
post-power-stroke motors with small strain in the presence of a
distribution of internal strains. Close to the stall force, the
analytical and numerical results again agree very well even for large
values of $\nt$.

\subsection{Linear load}

\subsubsection{LTE distribution and effective reverse rate} \label{scn:LinLTE}

For constant external load, the elastic energy stored in the neck
linkers was symmetric against exchanging weakly-bound and
post-power-stroke motors, $j \leftrightarrow i-j$. This symmetry was not
affected by the value of $\fext$, so that the LTE distribution remained
strongly biased towards the post-power-stroke state for all values of a
constant external load. For a linear external load, on the other hand,
the elastic energy stored in the neck linkers of bound motors and in the
external spring favor the weakly-bound state (see \eq{LinDelE}). Because
this contribution to the elastic energy of the ensemble increases with
increasing stiffness of the external spring, the bias of the LTE
distribution will shift towards the weakly-bound state for large values
of $\kf$. This transition between post-power-stroke and weakly-bound
state has been described as a possible basis for unconventional elastic
behavior of muscle fibers \cite{a:CaruelEtAl2013}.

\FIG{Fig14}{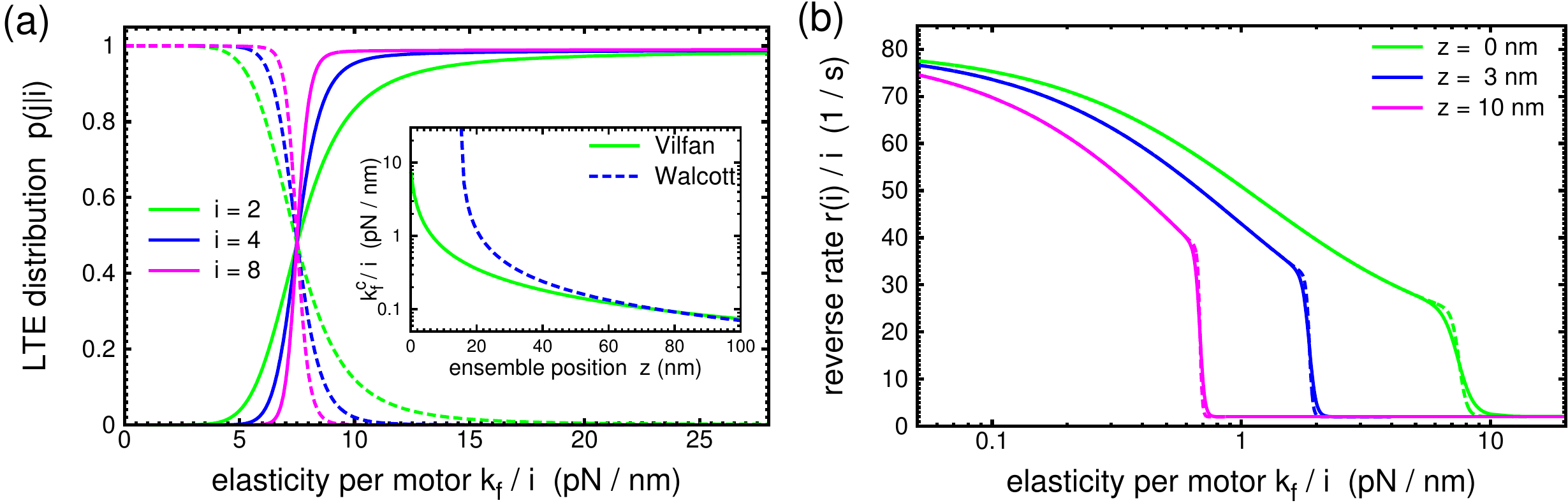}{Transition of LTE distribution: (a) Conditional
probabilities $p(0|i)$ (solid lines) and $p(i|i)$ (dashed lines) from
the LTE distribution (see \eq{LTEDist}) as function of the external
spring constant per bound motor, $\kf/i$, for $i=2$, $4$ and $8$ bound
motors and ensemble position $z=0$. \emph{Inset:} Critical external
spring constant per motor, $\kfc/i$, as function of ensemble position
$z$. The solid curve uses the model parameters listed in \tab{Tab01}.
For the dashed curve, the parameters $\km = 0.3 \pN\nm^{-1}$ and $d =
10\nm$ were used as in \textcite{a:WalcottEtAl2012}. (b) Effective
reverse rate per bound motor, $r(i)/i$, (see \eq{ReverseRate}) as
function of the external spring constant per bound motor, $\kf/i$, for
$i = 4$ (dashed lines) and $8$ (solid lines) bound motors and ensemble
positions $z = 0\nm$, $3\nm$ and $10\nm$. Constant parameters are listed
in \tab{Tab01}.}

\fig{Fig14}(a) plots the conditional probabilities $p(i|i)$ (all
bound motors in the post-power-stroke state) and $p(0|i)$ (all bound
motors in the weakly-bound state) from the LTE distribution of
\eq{LTEDist} as function of the external spring constant per bound
motor, $\kf/i$, for the ensemble position $z=0$ and different numbers of
bound motors, $i$. The probabilities $p(j|i)$ for the intermediate
states $0 < j < i$ are negligible due to internal strains built up by
bound motors in opposite states working against each other. For small
$\kf/i$, most bound motors are in the post-power-stroke state with
$p(i|i) \lesssim 1$. At a critical value of the external spring
constant, $\kfc/i \simeq 7.5 \pN\nm^{-1}$, the bias of the LTE
distribution shifts rapidly from the post-power-stroke state to the
weakly-bound state. The ratio $\kfc/i$ is independent of the number of
bound motors but the transition becomes sharper with increasing $i$.
The power stroke of a bound motor is driven by the free energy bias
$\epp < 0$ towards the post-power-stroke state. Thus, the transition of
the LTE distribution from post-power-stroke to weakly-bound state
occurs, when the increase of the elastic energy of an ensemble upon the
transition of $i$ bound motors from weakly-bound to post-power-stroke
state (see \eq{LinDelE}) exceeds the free energy gain $-j\epp$ upon this
transition. Solving the condition
\begin{equation}
\frac{i \km \kfc}{i\km + \kfc} \frac{d(2z + d)}{2} = -i\epp = i\left|\epp\right|
\end{equation}
for the critical spring constant yields
\begin{equation}\label{eq:kfc}
\frac{\kfc}{i} = \km \left[\frac{\km d(2z+d)}{2\left|\epp\right|}-1\right]^{-1}\,.
\end{equation}
The ratio $\kfc(z)/i$ is independent of $i$ for all values of $z$. This
is due to the parallel arrangement of the motors under the external
load. The inset in \fig{Fig14}(a) plots $\kfc(z)/i$ as function of $z$.
Because $\kf \geqslant 0$, a finite critical elastic constant exists for
all $z \geqslant 0$ only if $\km d^2 > 2 \left| \epp \right|$. This is
the case for our model parameters listed in \tab{Tab01}. In
\textcite{a:WalcottEtAl2012}, the parameters $d = 10\nm$ and $\km =
0.3\pN\nm^{-1}$ are used to characterize the power stroke. For these
values, $\km d^2 < 2 \left|\epp\right|$ so that the transition of the
LTE distribution can only occur above a finite ensemble position $z
\geqslant (\left| \epp \right| / \km d) - d/2$. The corresponding
$\kfc(z)/i$ is also plotted in the inset in \fig{Fig14}(a). For both
parameter sets, the critical spring constant decreases as $\kfc/i
\propto z^{-1}$ for large $z$ so that the external load $\kfc z/i$ at
the LTE transition becomes independent of $z$.

The LTE transition from post-power-stroke to weakly-bound state affects
the binding dynamics of an ensemble quantitatively and qualitatively,
because the off-rate from the weakly-bound state is significantly
smaller than the unloaded off-rate from the post-power-stroke state,
$k_{10} \ll k_{20}^0$, and because $k_{10}$ is independent of the load
on a motor. \fig{Fig14}(b) plots the effective reverse rate divided
by the number of bound motors, $r(i)/i$, as function of $\kf/i$ for
different values of $i$ and $z$. For $\kf/i \to 0$, motors unbind
predominantly from the post-power-stroke state ($p(i|i) \lesssim 1$) and
$r(i)/i$ approaches the value of the unloaded off-rate from the
post-power-stroke state, $r(i)/i \to k_{20}^0 \simeq 80 \s^{-1}$. From
this limit, $r(i)/i$ decreases exponentially with $\kf/i$, because the
load on a post-power-stroke motor increases linearly with $\kf/i$ and
$k_{20}$ decreases exponentially under load. Below the LTE distribution,
it is $r(i)/i \gg k_{10}$. Thus, the effective reverse rate decreases
strongly as the LTE distribution shifts towards the weakly-bound state
and approaches the off-rate from the weakly-bound state, $r(i)/i \to
k_{10} \simeq 2\s^{-1} = \const$ for large $\kf > \kfc$. The exponential
decrease of $r(i)/i$ and the critical elastic constant are independent
of $i$ but the transition becomes sharper with increasing $i$. With
increasing $z$, the initial exponential decrease of $r(i)/i$ becomes
faster. Nevertheless, the drop of $r(i)/i$ at the LTE becomes more
pronounced because the LTE transition occurs at smaller values of
$\kf/i$, as predicted by \eq{kfc}.

In an ensemble with $\nt$ molecular motors, the number $i$ of bound
motors fluctuates continuously. Because the critical elastic constant
$\kfc(z)$ is proportional to $i$, the LTE distribution follows the
fluctuations of $i$ and alternates between post-power-stroke state (for
large $i$ with $\kf/i < \kfc(z)/i$) and weakly-bound state (for small
$i$ with $\kf/i > \kfc(z)/i$) when the ensemble is in the transition
region. Because the transition to the weakly-bound state occurs first
for the smallest $i$ and unbinding from the weakly-bound state is
significantly slowed down, onset of the LTE distribution will stabilize
the ensemble against unbinding. Furthermore, the critical elastic
constant is itself a dynamic quantity, because $\kfc(z)/i$ reduces with
increasing ensemble position. Thus, for linear external load two
different mechanisms can stabilize an ensemble as it moves to larger
$z$: $(i)$ for ensemble positions below the LTE transition, unbinding is
slowed down by the catch bond character of motors in the
post-power-stroke state and $(ii)$ above the LTE transition threshold,
unbinding is slowed down by the transition to the weakly-bound state.
Moreover, because ensemble movement relies on the presence of motors in
the post-power-stroke state, ensemble movement feeds back negatively on
itself and the ensemble will stall as the LTE transition threshold is
reached.

\subsubsection{Stochastic trajectories}

As for the case of constant external load, it is instructive to study
individual stochastic trajectories in order to gain more insight into
the interplay of ensemble movement and binding dynamics. The stochastic
trajectories are generated using the Gillespie algorithm as described
for the case of constant load. The upper panel of the stochastic
trajectories, is now used to display the the external load $\kf z$,
which is proportional to ensemble position but omits the strongly
fluctuating contribution of the strain of the motors. The actual
external load depends on the number of bound motors and their states: if
all bound motors are in the post-power-stroke state, the external load
is larger than $\kf z$. If all bound motors are in the weakly-bound
state, the external load is smaller than $\kf z$.

\FIG{Fig15}{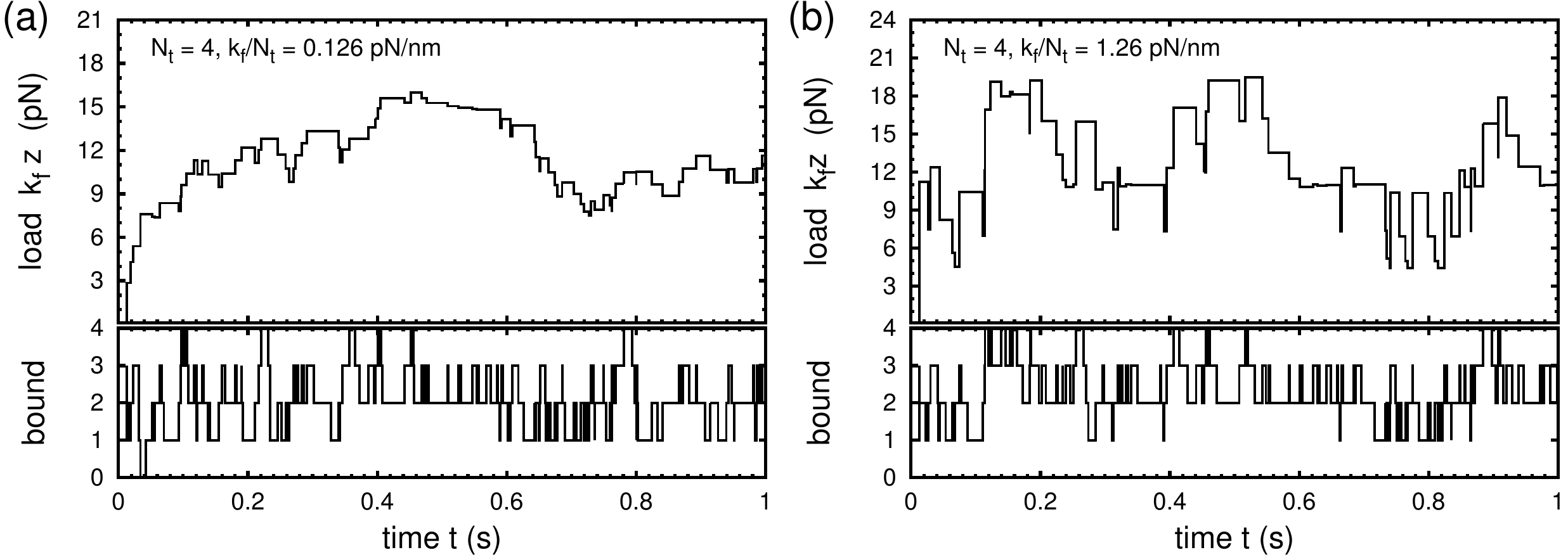}{Stochastic trajectories for linear external load.
Elastic load $\kf z$ on the ensemble (upper panel) and number $i$ of
bound motors (lower panel) as function of time $t$ for ensemble size
$\nt=4$ and external elastic constant per motor (a) $\kf/\nt = 0.126
\pN\nm^{-1}$ and (b) $\kf/\nt = 1.26 \pN\nm^{-1}$. In (a) and (b)
detached ensembles are stationary with mobility $\eta=0$. Constant
parameters are listed in \tab{Tab01}.}

\fig{Fig15}(a) shows a stochastic trajectory of an ensemble with $\nt=4$
motors working against a linear external load with elastic constant
$\kf/\nt = 0.126 \pN\nm^{-1}$ ($\kf = 0.504 \pN\nm^{-1}$). At $z=0$, the
ratio $\kf/i$ is below the critical elastic constant $\kfc/i \simeq 7.5
\pN\nm^{-1}$ for all $i \geqslant 1$. Therefore, ensemble position
initially increases gradually. During this transient movement, the
ensemble occasionally detaches completely, because for all $i \geqslant
1$ unbinding occurs from the post-power-stroke state with large
intrinsic off-rate against small external load. After the initial
transient, the ensemble reaches a stationary state in which $\kf z$
fluctuates around a constant average. The typical external load $\kf z
\simeq 12\pN$ in this isometric state corresponds to the ensemble
position $z \simeq 24\nm$. For this value of $z$, the critical elastic
constant is lowered to $\kfc(z)/i \simeq 0.3 \pN\nm^{-1}$. With $\kf =
0.504 \pN\nm^{-1}$, the LTE distribution shifts to the weakly-bound
state for $i=1$ but remains in the post-power-stroke state for $i > 1$.
Due to the very slow unbinding of the last ($i=1$) bound motor from the
weakly-bound state, complete detachment of the ensemble is no longer
observed in the isometric state, although $i$ continues to fluctuate
between $i=1$ and $\nt$. This demonstrates the stabilization of the
ensemble due to the LTE transition. On the other hand, the LTE
transition for $i=1$ also causes stalling of the ensemble and prevents
movement beyond the isometric state. Assuming that all bound motors are
in the post-power-stroke state, even for the largest values of $z$ the
strain $\xij$ is negative for all $i > 1$ so that the ensemble still
steps forward in these states. (Although the size of the steps may be
reduced due to the increased probability for weakly-bound motors in the
proximity of the LTE transition). Because the ensemble can only step
backwards when all bound motors are in the weakly-bound state, the
ensemble alternates between forward stepping for large $i$ ($\kf/i$
below the threshold) and backward stepping for small $i$ ($\kf/i$ above
the threshold). The isometric state is reached, when these two
contributions balance and the ensemble fluctuates around a constant
average position. Excursion to large $z$ will shift the LTE to the
weakly bound state for higher values of $i$ and induce quick backward
movement and restoring of the the isometric state. Excursions to small
$z$, on the other hand, will shift the LTE distribution towards the
post-power-stroke state for $i=1$, thus inducing quick forward movement
towards the isometric state. This stalling mechanism is different from
the case of constant external load, where the LTE distribution was
always in the post-power-stroke state and the stall force was determined
by vanishing $\xij$ under large load. \fig{Fig15}(b) shows a stochastic
trajectory of an ensemble with $\nt=4$ motors working against the larger
external elastic constant $\kf/\nt = 1.26 \pN\nm^{-1}$. The isometric
load $\kf z \simeq 12\pN$ is comparable to the trajectory from (a). Due
to the larger elastic constant, however, this corresponds to the smaller
ensemble position $2.4 \nm$. Moreover, small fluctuations in $z$ induce
strong fluctuations of $\kf z$. The critical elastic constant for the
LTE transition is $\kfc/i \simeq 2.5 \pN\nm^{-1}$. Again, the value of
$\kf \simeq 5.04 \pN\nm^{-1}$ is below the critical elastic constant for
$i > 1$ and above for $i=1$ so that complete detachment is prevented by
the LTE transition to the weakly-bound state in the lowest bound state
$i=1$. As in (a), the strain $\xij$ of the bound motors is negative for
$i>1$ (assuming they are all in the post-power-stroke state) so that the
stalling of the ensemble is caused by the LTE transition.

For the trajectories with $\nt = 4$ in \fig{Fig15}, the LTE transition
occurred only in the lowest bound state $i=1$. For increasing $\nt$,
more states with $i \geqslant 1$ undergo the LTE transition and
fluctuations of $i$ in the isometric state are effectively restricted to
values above the threshold (compare \fig{FigS04}). For large ensembles
as in \fig{FigS04}, the alternating forward and backward motion in the
isometric state displays a characteristic pattern of rapid increase of
$\kf z$ concomitant with an increase of $i$, followed by a gradual
decrease of $\kf z$ accompanied by a decrease of $i$. This pattern,
which becomes more pronounced for larger values of the external spring
constant, is reminiscent of oscillation pattern for ensembles of motors
working against an elastic element \cite{a:JuelicherProst1997}. A
Fourier analysis, however, has not shown any characteristic time scale
for the fluctuations.

\FIG{Fig16}{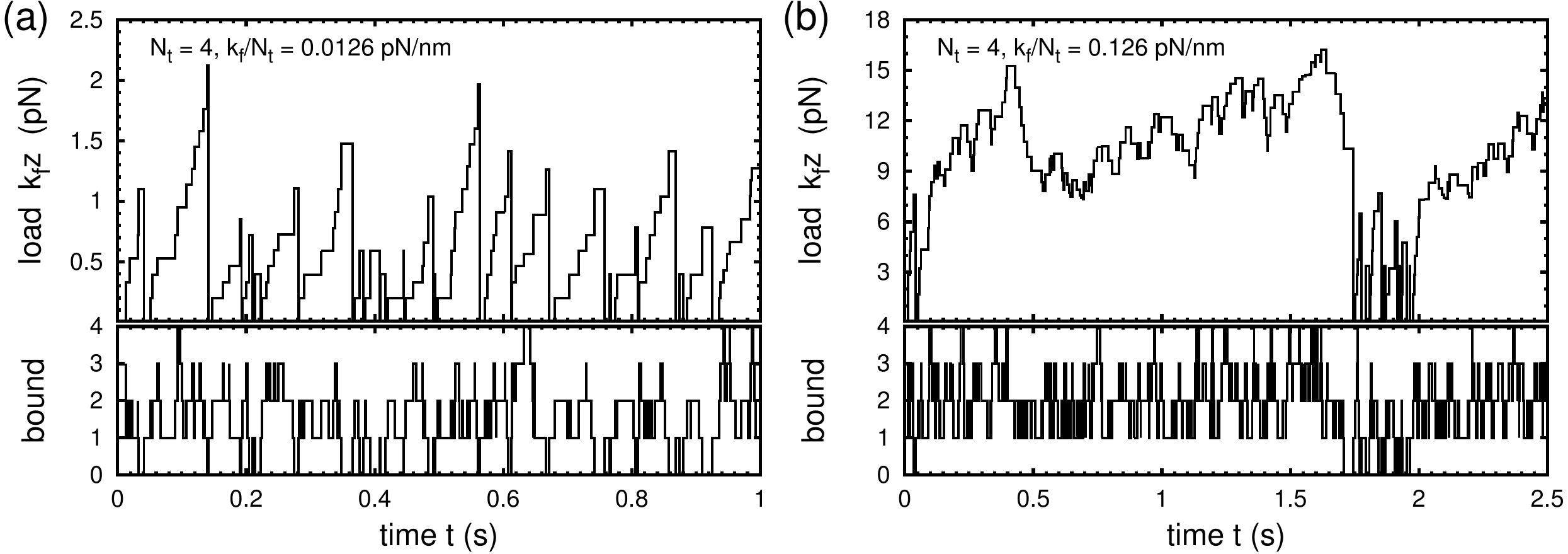}{Stochastic trajectories for linear external load.
Elastic load $\kf z$ on the ensemble (upper panel) and number $i$ of
bound motors (lower panel) as function of time $t$ for ensemble size
$\nt=4$ and external elastic constant per motor (a) $\kf/\nt = 0.126
\pN\nm^{-1}$ and (b) $\kf/\nt = 1.26 \pN\nm^{-1}$. In (a) and (b)
detached ensembles are reset to $z=0$ with infinite mobility $\eta =
\infty$. Constant parameters are listed in \tab{Tab01}.}

In the above trajectories, detached ensembles are stationary with
mobility $\eta=0$. \fig{Fig16} demonstrates the effect of resetting an
ensemble to $z=0$ after detachment. \fig{Fig16}(a) shows a trajectory
for $\nt=4$ with the small external elastic constant $\kf/\nt = 0.0126
\pN\nm^{-1}$. The small ensemble size and the small external elastic
constant allows frequent detachment during the initial, transient
movement. Because the position is reset to $z=0$ after detachment, the
ensemble cannot reach the isometric state which would stabilize the
ensemble and prevent detachment. Therefore, the stochastic trajectories
show a characteristic pattern of gradual linear buildup of load,
followed by rapid release upon detachment. This pattern resembles
trajectories observed experimentally in three bead assays
\cite{a:FinerEtAl1994, a:VeigelEtAl2003, a:DeboldEtAl2005}, active gels
\cite{a:MizunoEtAl2007, a:SoaresEtAl2011} and motility assays
\cite{a:PlacaisEtAl2009}. \fig{Fig16}(b) shows a trajectory with larger
external elastic constant. During the transient increase of $\kf z$, the
ensemble detaches occasionally and is reset to $z=0$. Because the
isometric ensemble position is smaller for larger $\kf/\nt$, the
isometric state can eventually be reached. This isometric state is not
affected by the movement of the detached ensemble. As observed above,
complete detachment from the isometric state is rare. As detachment
occurs in a rare fluctuation, however, the ensemble detaches several
times on its trajectory until the isometric state is again reached.

\subsubsection{Binding dynamics}

To study the interplay of ensemble movement and binding dynamics, we
first analyze the dependence of detachment time $\tdet$ on ensemble size
$\nt$ and external elastic constant $\kf$. Because the one-step master
equation \eq{MasterEq} cannot be solved with position dependent
transition rates, the detachment time is calculated numerically by
averaging the first passage time from $i=1$ to $i=0$ over repeated,
stochastic trajectories. The trajectories all start at $z=0$ and are
terminated as soon as the ensemble detaches, so that the mobility $\eta$
has no influence on the result.

\FIG{Fig17}{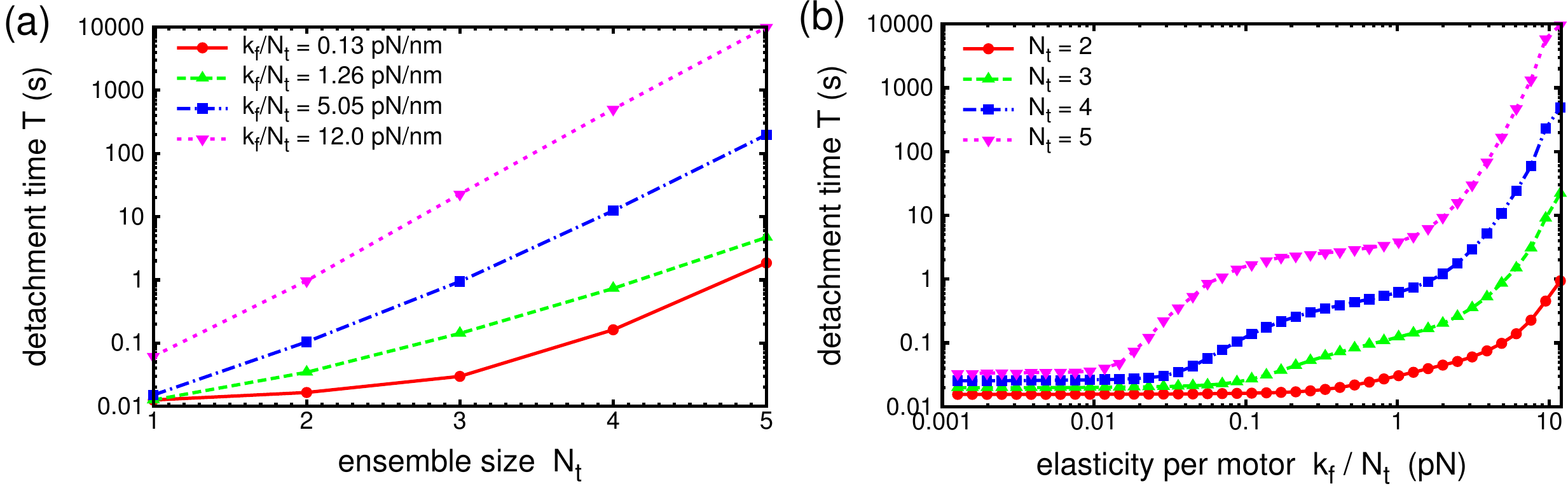}{Numerical results for the parallel cluster model
with linear external load: average detachment time $\tdet$. (a) $\tdet$
as function of ensemble size $\nt$ for the values $\kf/\nt = 0.126
\pN\nm^{-1}$, $1.262 \pN\nm^{-1}$, $3.787 \pN\nm^{-1}$ and $12.0
\pN\nm^{-1}$ of the external elastic constant per motor. (b) $\tdet$ as
function of the external elastic constant per motor $\kf/\nt$ for
ensemble sizes $\nt=2$, $3$, $4$ and $5$. Constant parameters are listed
in \tab{Tab01}.}

\fig{Fig17}(a) plots simulation results for the average detachment time
$\tdet$ of an ensemble as function of ensemble size $\nt$ for different
values of the external elastic constant $\kf/\nt$. The largest value of
$\kf/\nt$ is above the critical external elastic constant at $z=0$ so
that unbinding proceeds predominantly from the weakly-bound state for
all values of $\nt$ and is independent of load. Thus, $\tdet$ increases
approximately exponentially with increasing $\nt$ and reaches $\tdet
\simeq 10^4 \s$ already for $\nt \simeq 5$. For constant load, such
detachment times required external loads well beyond the stall force.
For the smallest value of $\kf/\nt$, which is well below $\kfc(z=0)/i$,
a transient regime with a very slow increase of $\tdet$ is observed.
Here, the average detachment time results from a combination of fast
detachment from the post-power-stroke state during the transient
increase of $z$ and slow unbinding from the weakly-bound state (for
small values of $i$) once the isometric state is reached. For
sufficiently large $\nt$, detachment before reaching the isometric state
becomes unlikely and $\tdet$ is determined by slow unbinding from the
weakly-bound state. Thus, the increase of $\tdet$ with $\nt$ becomes
similar to the exponential increase observed for large $\kf/\nt$. For
increasing values of $\kf/\nt$, the transient regime of slow increase of
$\nt$ becomes less pronounced and $\tdet$ increases exponentially for
most values of $\nt$.

\fig{Fig17}(b) plots $\tdet$ as function of $\kf/\nt$ for different
$\nt$. The plot reveals three different regimes of the detachment time
which corresponds to different detachment mechanisms. At very small
$\kf/\nt$, ensembles detach during the initial increase of $z$. Here,
unbinding of motors proceeds predominantly from the post-power-stroke
state under small external load, so that the detachment time is almost
independent of $\kf/\nt$. In an intermediate regime, the detachment time
increases significantly. Here, the ensembles can reach the isometric
state so that the contribution of slow unbinding from the weakly-bound
state becomes more prominent. Because the ensembles adjust themselves
dynamically to the isometric state, large ensembles display a plateau
region with constant $\tdet$. When the external elastic constant
approaches the absolute threshold $\kfc/\nt \simeq 7.5\pN\nm^{-1}$ for
$z=0$, the LTE distribution shifts towards the weakly-bound state for an
increasing number of states $i$. This increases the typical number of
bound motors and the detachment time increases rapidly (see \fig{FigS04}).

\FIG{Fig18}{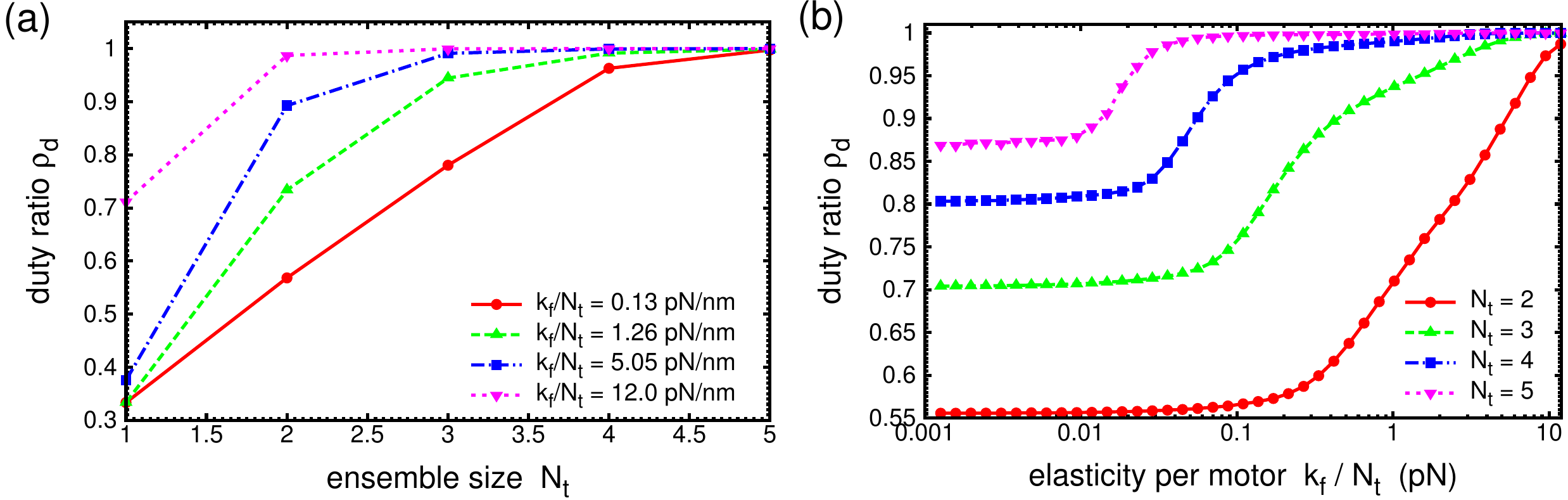}{Numerical results for the parallel cluster model
with linear external load: ensemble duty ratio $\rd$. (a) $\rd$ as
function of ensemble size $\nt$ for the values $\fext/\nt = 0.13
\pN\nm^{-1}$, $1.26 \pN\nm^{-1}$, $5.05 \pN\nm^{-1}$ and $12.0
\pN\nm^{-1}$ of the external elastic constant per motor. (b) $\rd$ as
function of the external elastic constant per motor $\kf/\nt $ for
ensemble sizes $\nt=2$, $3$, $4$ and $5$. Detached ensembles are
stationary with vanishing mobility, $\eta=0$. Constant parameters are
listed in \tab{Tab01}.}

\fig{Fig18}(a) plots the ensemble duty ratio $\rd$ as function of
ensemble size $\nt$ for different values of the external elastic
constant $\kf/\nt$. Because the attachment time $\tatt$ is independent
of $\kf$ and decreases with $\nt^{-1}$, the dependence of $\rd$ on $\kf$
and $\nt$ mainly reflects the corresponding dependence of $\tdet$. For
the largest value with $\kf/\nt > \kfc/\nt$, the duty ratio at $\nt=1$
is increased significantly with respect to the duty ratio of a single
free motor, $\rds \simeq 0.33$. Due to the exponential increase of the
detachment time, $\rd \simeq 1$ is reached already for $\nt \geqslant
2$. For smaller $\kf/\nt < \kfc/\nt$, the duty ratio at $\nt=1$ is close
to the duty ratio of a single free motor. This indicates that unbinding
proceeds predominantly from the post-power-stroke state. Due to the
rapid increase of $\tdet$ with $\nt$, permanent attachment with $\rd
\simeq 1$ is achieved for $\nt \geqslant 5$. This is significantly
smaller than the ensemble size $\nt=15$ required for permanent
attachment under constant external load. \fig{Fig18}(b) plots $\rd$ as
function of $\kf/\nt$ for different $\nt$. In analogy to the detachment
time, $\rd$ displays two regimes: a constant duty ratio at small
$\kf/\nt$ followed by a rapid increase at intermediate values of
$\kf/\nt$. Because $\rd \simeq 1$ is already reached here, the strong
increase of $\tdet$ for $\kf/\nt \gtrsim \kfc/\nt$ cannot be resolved.

\FIG{Fig19}{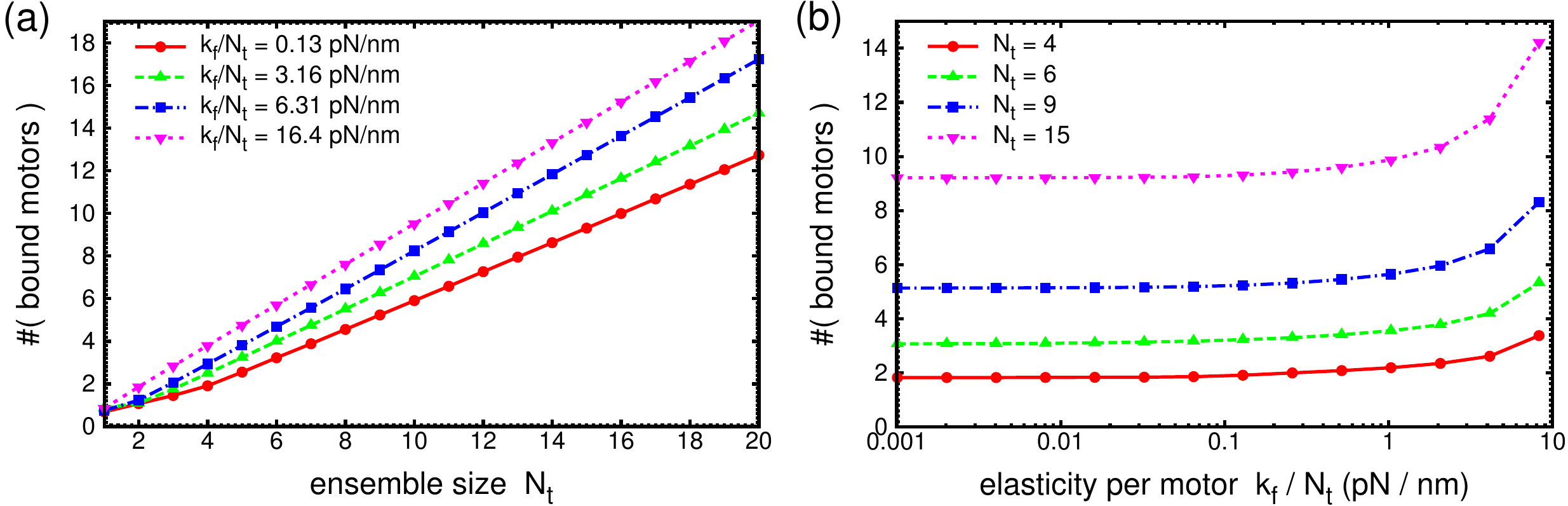}{(a) Numerical results for the parallel cluster
model with linear external load: average number $\nb$ of bound motors.
(a) $\nb$ as function of ensemble size $\nt$ for the values $\kf/\nt =
0.13 \pN\nm^{-1}$, $3.16 \pN\nm^{-1}$, $6.31 \pN\nm^{-1}$ and $16.4
\pN\nm^{-1}$ of the external elastic constant per motor. (b) $\nb$ as
function of external elastic constant per motor $\kf/\nt $ for the
ensemble sizes $\nt=4$, $6$, $9$ and $15$. Detached ensembles at
stationary with $\eta=0$. Constant parameters are listed in \tab{Tab01}.}

To calculate the average number of bound motors $\nb = \Avg{i}$, we
average $i$ over long trajectories in which the ensembles are allowed to
detach from the substrate. The detached ensembles are stationary with
vanishing mobility, $\eta=0$, so that the isometric state of the
ensembles is probed. \fig{Fig19}(a) plots the average number $\nb$ of
bound motors as function of ensemble size $\nt$ for different values of
the external elastic constant per motor $\kf/\nt$. For all the values of
$\kf/\nt$ the average number of bound motors increases linearly with
$\nt$. The slope becomes steeper with increasing $\kf/\nt$ but saturates
for very large $\kf/\nt$. Only for small values of $\kf/\nt$ there is a
short transient with a slower increase of $\nb$. \fig{Fig19}(b) plots
$\nb$ as function of the external elastic constant per motor $\kf/\nt$
for different values of $\nt$. For $\kf/\nt < \kfc(z=0)/\nt$, the
average number of bound motors is constant. Here, the ensembles adjust
themselves to an isometric state at finite ensemble position $z>0$. As
the value of $\kf/\nt$ exceeds the critical values $\kfc(z=0)/\nt$, the
average number of bound motors increases steeply. Here, the LTE
transition to the weakly-bound state occurs already at $z=0$ and for an
increasing number of states $i$ (see \fig{FigS04}).

\subsubsection{Average external load}

For linear external load, analogous to the force-velocity relation as a
characteristic for the dynamic properties of an ensemble under constant
external load, is the average external load $\favg = \Avg{\kf z}$ of the
ensemble in the stationary state. As explained in the context of the
stochastic trajectories, this quantity differs from the actual load in
the external elastic element by leaving out the strain $\xij$ of the
motors.

\FIG{Fig20}{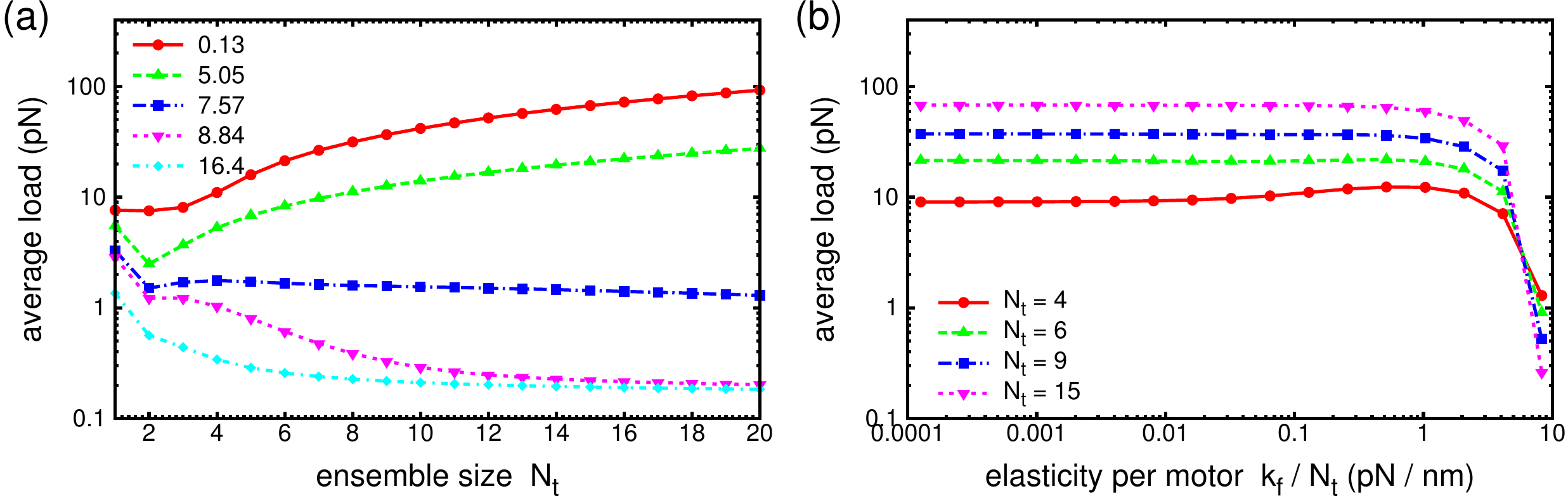}{Numerical results for the parallel cluster model
with linear external load: average external load $\favg$ for $\eta=0$.
(a) $\favg$ as function of ensemble size $\nt$ for values $\kf/\nt =
0.13 \pN\nm^{-1}$, $5.05 \pN\nm^{-1}$, $7.57 \pN\nm^{-1}$, $8.84
\pN\nm^{-1}$ and $16.4 \pN\nm^{-1}$ of the external elastic constant per
motor. (b) $\favg$ as function of external elastic constant per motor
$\kf/\nt $ for ensemble size $\nt=4$, $6$, $9$ and $15$. Constant
parameters are listed in \tab{Tab01}.}

As for the average number of bound motors, the average external load is
determined by averaging over long trajectories with multiple unbinding
events. In \fig{Fig20}, the detached ensembles are stationary with
mobility $\eta=0$, so that the average load in the isometric state is
probed. \fig{Fig20}(a) plots the average external load $\favg$ in the
stationary state of an ensemble as function of ensemble size $\nt$ for
different values of the external elastic constant per motor, $\kf/\nt$.
Below the critical value for the LTE transition, $\kf/\nt < \kfc/\nt$,
the average external load increases with $\nt$. For very small external
elastic constants, $\kf/\nt < 1 \pN\nm^{-1}$, the curves are almost
identical for different $\kf$, but begin to decrease for $\kf/\nt > 1
\pN\nm^{-1}$. This decrease of $\favg$ for given $\nt$ is observed in
the trajectories of \fig{FigS04}. At the critical external elastic
constant, $\kf/\nt \simeq \kfc/\nt$, the average external load is
independent of $\nt$. Above the critical threshold, the LTE transition
towards the weakly-bound state occurs already at $z=0$ so that ensemble
movement is severely reduced. Here, $\favg$ decreases with increasing
$\nt$. \fig{Fig20}(b) plots the average external load as function of
$\kf/\nt$ for different $\nt$. For small $\kf/\nt < \kfc/\nt$, the
average external load is independent of $\kf/\nt$. Close to the critical
value, $\favg$ breaks down because ensemble movement is effectively
impossible.

\FIG{Fig21}{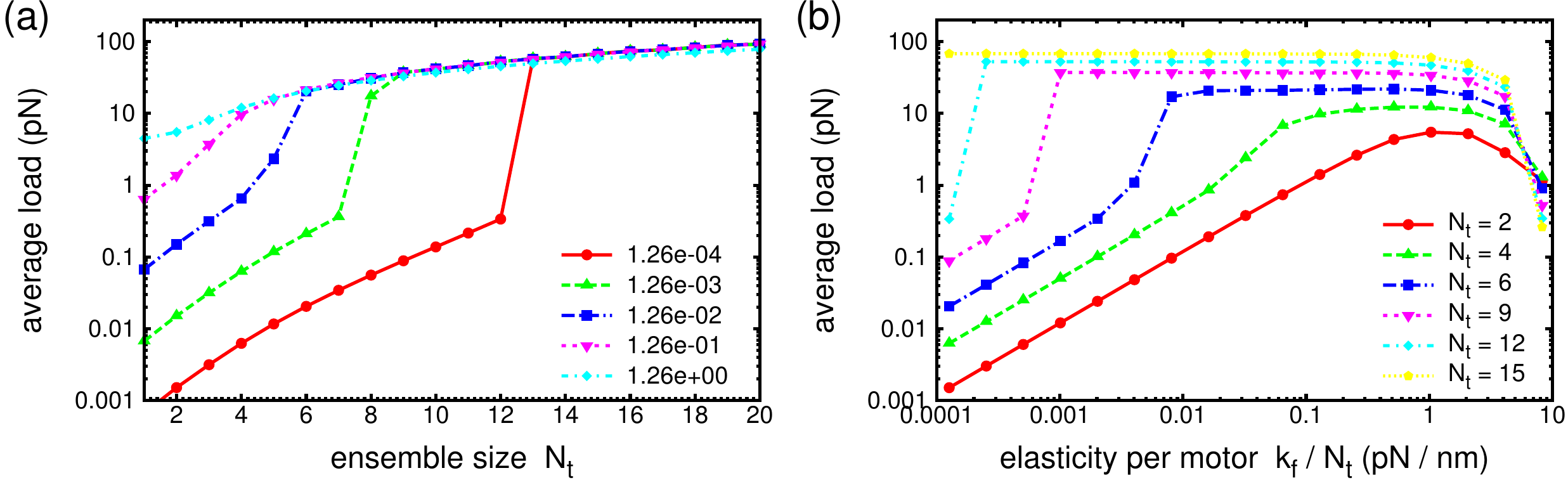}{Numerical results for the parallel cluster model
with linear external load: average isometric load $\favg$ for $\eta \to
\infty$. (a) $\favg$ as function of ensemble size $\nt$ for values
$\kf/\nt = 1.26 \times 10^{-4} \pN\nm^{-1}$, $0.00126 \pN\nm^{-1}$,
$0.0126 \pN\nm^{-1}$, $0.126 \pN\nm^{-1}$ and $1.26 \pN\nm^{-1}$. (b)
$\favg$ as function of external elastic constant per motor $\kf/\nt$ for
ensemble size $\nt=4$, $6$, $9$ and $15$. Constant parameters are listed
in \tab{Tab01}.}

\fig{Fig21}(a) plots the average external load as function of ensemble
size $\nt$ for different values of $\kf/\nt$ for the case that the
ensemble is reset to its initial position $z=0$ after complete
detachment. This corresponds to the limit of large mobility $\eta \to
\infty$. The values of $\kf/\nt $ are below the critical value
$\kfc(z=0)/\nt$ and from a range in which $\favg$ at $\eta=0$ was
independent of $\kf/\nt$. For small external elastic constant, the
isometric state corresponds to a large ensemble position $z$. Therefore,
ensembles detach frequently during the long transient movement towards
the isometric state. Because detached ensembles are reset to $z=0$, this
reduces the average external load, as observed in \fig{Fig16}(a).
Because detachment becomes less likely with increasing $\nt$ and
ensembles are stabilized in the isometric state, $\favg$ increases with
$\nt$ and eventually jumps discontinuously to the average load for
$\eta=0$. With increasing $\kf/\nt$, the average external load grows in
proportion to $\kf/\nt$ at given $\nt$. This means that the movement of
the ensemble during the initial transient is hardly affected by the
small external load. Because the isometric state corresponds to smaller
ensemble position $z$, however, the discontinuous jump to the curve for
$\eta=0$ occurs at smaller values of $\nt$ and becomes less pronounced.
\fig{Fig21}(b) plots the average external load for $\eta \to \infty$ as
function of $\kf/\nt$ for different $\nt$. For small $\kf/\nt $, $\favg$
increases linearly towards the average isometric load. For larger $\nt$,
$\favg$ for given $\kf/\nt$ increases and jumps discontinuously towards
isometric load. The discontinuity becomes more pronounced with
increasing $\nt$ and the position of the jump decreases.

\section{Discussion and summary}

In this paper we have introduced and analyzed a stochastic model for
ensembles of non-processive motors such as \myo working against an
external load. The model allows us to investigate in detail the effect
of a finite number of motors in an ensemble, most importantly the
stochastic binding dynamics of the motors. Introducing the parallel
cluster model and using the local thermal equilibrium approximation
allowed us to reduce the complexity of the model significantly,
eventually leading to efficient numerical simulations and analytical
results for stationary properties. In detail, we have analyzed two
paradigmatic situations in which the motor-ensemble works against either
a constant or a linear external load. Both situations are highly
relevant for a large class of experiments.

For constant external load, our results for large ensemble sizes are in
good qualitative agreement with previous model results for large
assemblies of \myo \cite{a:Duke1999, a:Duke2000}. Due to the local
thermal equilibrium assumption, however, our model is not able to
describe, e.g., the synchronization of motors under large external load
which is due to the kinetic hindrance of the power stroke. Average
quantities, however, are well represented. In particular the
force-velocity relation of an ensemble follows the characteristic
concave shape which is described by a Hill relation and is found
experimentally for muscle fibers \cite{a:Hill1939b} as well as for small
ensembles of \myo \cite{a:WalcottEtAl2012}. The parallel cluster model
makes it easy to identify the relevant quantities determining the
force-velocity curve. In particular the role of the load-sensitivity of
unbinding from the post-power-stroke state and the increase of the
number of bound motors under load for the adaption of the dynamic range
of an ensemble becomes clear. Due to the strong bias of the LTE
distribution towards the post-power-stroke state, \myo as a whole
behaves as a catch bond for the whole range of constant external load
considered in this manuscript. This induces the increase of the number
of bound motors under load, which is the basis for the concave shape of
the force-velocity relation. At small external load, a small number of
bound motors is able to work against the external load and the large
number of unbound motors generates fast ensemble movement with little
resistance from the bound motors. At large external load, on the other
hand, the increase of the number of bound motors allows to increase the
stall force of the ensemble relative to the case of a constant number of
bound motors. Thus, the mechanosensitive response of \myo to a constant
external load greatly increases the dynamic range over which \myo
ensembles can operate and the robustness of ensemble movement. As
demonstrated in \app{small}, the apparent load-sensitivity of \myo
becomes stronger for small duty ratios, for which a stronger increase of
the number of bound motors is observed. The relevance of this mechanism
of mechanosensitivity for the efficiency of motor ensembles is
underlined by the experimental observation of the increase of the number
of bound motors in muscle filaments which use \myo
\cite{a:PiazzesiEtAl2007} and the recent identification of a similar
load-sensitive step in myosin {I} \cite{a:LaaksoEtAl2008}. This last
result indicates that the mechanisms described here might have wider
applications. In addition to reproducing previous results for large
ensembles, our model in particular allows to identify the range of
ensemble sizes in which stochastic detachment of ensembles is relevant
and reduces the efficiency of ensembles. As demonstrated in \app{AppA},
this range depends strongly on the duty ratio of the single motor. The
smaller the single motor duty ratio, the larger is the number of motors
in an ensemble that is needed to ensure practically permanent
attachment. For the model parameters used in the main part of the
manuscript, the single motor duty ratio is similar but slightly larger
than reported for non-muscle \myo \cite{a:SoaresEtAl2011}. Here, it
turns out that roughly $\nt=15$ motors are needed for almost permanent
attachment. This number is in the range of the size of minifilaments, in
particular if one considers that due to spatial restrictions, not all
motors which are contained in a minifilament can actually bind to a
substrate at the same time. Thus, unbinding will be relevant for
cytoskeletal \myo minifilaments.  Interestingly, due to the
load-sensitivity of \myo minifilaments, detachment of the ensembles will
occur most frequently for ensemble under small load whereas ensembles
working against large external load are stabilized by the catch bond
character of \myo so that they can form efficient crosslinkers of actin
fibers.

For linear external load, it has been shown that the load-sensitivity of
the unbinding step is less relevant for the behavior of a motor
ensemble. Rather, the reverse transition from the post-power-stroke
state to the weakly-bound state, which is induced by the stiffness of
the external spring or by large external load, eventually stalls the
ensemble in the isometric state. As long as the stiffness of the
external spring is below the critical threshold, ensembles can move
forward until the isometric state is reached. Because ensembles adapt
their position dynamically towards the isometric state, characteristic
dynamic properties such as the detachment time or the typical number of
bound motors in the isometric state are independent of the external
elastic constant state and increases significantly only when the
critical stiffness is reached. Thus, measuring the change of the number
of bound motors with the stiffness of the external force is a
characteristic sign of the relative size of the unbinding rates from the
different bound states of the motors. The critical value of $\kf$ above
which ensemble are no longer able to generate force allows to determine
the free energy bias towards the post-power-stroke state. The
qualitative change of behavior of the ensemble should allow to observe
this experimentally even in noisy data, e.g.\ in extensions of the three
bead assay with better control over the number of motors. Typical
stochastic trajectories reveal a behavior which is qualitatively similar
to previously predicted or experimentally observed types of behavior.
For relatively large ensembles which do not unbind, a sort of irregular
oscillation pattern has been observed, similar to the oscillation
predicted using a ratchet model for large ensembles of motors
\cite{a:JuelicherProst1997}. The linear increase of stress followed by a
fast stress relaxation which is observed in three bead assays
\cite{a:FinerEtAl1994} and in active gels \cite{a:MizunoEtAl2007} has
been reproduced for the case that unbinding from the substrate occurs
before the ensemble reaches its equilibrium position.

The sequence of reaction in our basic crossbridge model is compatible
with experiments \cite{a:VeigelEtAl2003, a:SellersVeigel2010}. Compared
to other models, we have neglected one additional post-power-stroke
state, which will be important for detailed descriptions of the
force-velocity relation in skeletal muscle \cite{a:Duke1999} or
inclusion of the \atp dependence of the motor cycle
\cite{a:WalcottEtAl2012}. The \atp dependence, however, could also be
included in our model via an \atp dependent off-rate from the
post-power-stroke state. On the other hand, our model explicitly
includes the weakly-bound state (or pre-power-stroke state) as a load
bearing state, which is omitted by \textcite{a:WalcottEtAl2012}. Because for
constant external load, the weakly-bound state is hardly occupied so
that motors bind effectively directly to the post-power-stroke state as
in \textcite{a:WalcottEtAl2012}. For linear external load, however, the
isometric state of the ensemble is determined by the transition from the
post-power-stroke to the weakly-bound state, which could not be
described without a model for the power stroke. Without the weakly-bound
state, which is stable under load, this transition would destabilize the
ensemble instead of stabilizing it. The choice of the set of parameters
used in our model was motivated by previous modeling approaches and did
not aim at a description of a specific molecular motor. However, the
basic conclusions of the model do not depend on the exact choice of
parameters. The only prerequisite for the application of the model is
that transitions between the bound states are fast so that the local
thermal equilibrium can establish. Comparison with simulations without
PCM or LTE have shown that this requires a forward rate $k_{12}$ for the
power stroke which is faster than the off-rate $k_{10}$ from the
weakly-bound state. Thus, our model can be used with different choices
of parameters. This is demonstrated in \app{AppA} for different choices
of the single motor duty ratio. The discussion of the result has
demonstrated the dependence of experimentally measurable quantities such
as the number of bound motors, the load free velocity or the stall force
on the model parameters so that the parameters can be adapted to
describe a desired behavior of the motor ensemble.

Ensembles of non-processive motors behave very similarly to processive
motors with a specific force-velocity relation and a typical walk
length. The analytical expressions allow to integrate such ensembles in
larger systems in a similar manner as it has been done for processive
motors \cite{a:KlumppLipowsky2005a, a:KosterEtAl2003}. In such a
systems, ensembles of ensembles would be multiply coupled either in
series or in parallel through forces that they generate. Such model
could be used to describe, e.g.\ tension generation in a stress fibers
or the cell actin cortex. Since the relaxation to the stationary values
is relatively fast, slow external stimuli, changing for example the
motor activity, could be included to mimic signaling events which a cell
experiences, e.g.\ during cell migration.


\begin{thebibliography}{60}
\expandafter\ifx\csname natexlab\endcsname\relax\def\natexlab#1{#1}\fi
\expandafter\ifx\csname bibnamefont\endcsname\relax
  \def\bibnamefont#1{#1}\fi
\expandafter\ifx\csname bibfnamefont\endcsname\relax
  \def\bibfnamefont#1{#1}\fi
\expandafter\ifx\csname citenamefont\endcsname\relax
  \def\citenamefont#1{#1}\fi
\expandafter\ifx\csname url\endcsname\relax
  \def\url#1{\texttt{#1}}\fi
\expandafter\ifx\csname urlprefix\endcsname\relax\def\urlprefix{URL }\fi
\providecommand{\bibinfo}[2]{#2}
\providecommand{\eprint}[2][]{\url{#2}}

\bibitem[{\citenamefont{Lipowsky and Klumpp}(2005)}]{a:LipowskyKlumpp2005}
\bibinfo{author}{\bibfnamefont{R.}~\bibnamefont{Lipowsky}} \bibnamefont{and}
  \bibinfo{author}{\bibfnamefont{S.}~\bibnamefont{Klumpp}},
  \bibinfo{journal}{Physica A} \textbf{\bibinfo{volume}{352}},
  \bibinfo{pages}{53} (\bibinfo{year}{2005}).

\bibitem[{\citenamefont{Howard}(1997)}]{rv:Howard1997}
\bibinfo{author}{\bibfnamefont{J.}~\bibnamefont{Howard}},
  \bibinfo{journal}{Nature} \textbf{\bibinfo{volume}{389}},
  \bibinfo{pages}{561} (\bibinfo{year}{1997}).

\bibitem[{\citenamefont{Vale and Milligan}(2000)}]{rv:ValeMilligan2000}
\bibinfo{author}{\bibfnamefont{R.~D.} \bibnamefont{Vale}} \bibnamefont{and}
  \bibinfo{author}{\bibfnamefont{R.~A.} \bibnamefont{Milligan}},
  \bibinfo{journal}{Science} \textbf{\bibinfo{volume}{288}},
  \bibinfo{pages}{88} (\bibinfo{year}{2000}).

\bibitem[{\citenamefont{Gu{\'e}rin et~al.}(2010)\citenamefont{Gu{\'e}rin,
  Prost, Martin, and Joanny}}]{a:GuerinEtAl2010}
\bibinfo{author}{\bibfnamefont{T.}~\bibnamefont{Gu{\'e}rin}},
  \bibinfo{author}{\bibfnamefont{J.}~\bibnamefont{Prost}},
  \bibinfo{author}{\bibfnamefont{P.}~\bibnamefont{Martin}}, \bibnamefont{and}
  \bibinfo{author}{\bibfnamefont{J.-F.} \bibnamefont{Joanny}},
  \bibinfo{journal}{Curr. Opin. Cell Biol.} \textbf{\bibinfo{volume}{22}},
  \bibinfo{pages}{14} (\bibinfo{year}{2010}).

\bibitem[{\citenamefont{Vale}(2003)}]{rv:Vale2003}
\bibinfo{author}{\bibfnamefont{R.~D.} \bibnamefont{Vale}},
  \bibinfo{journal}{Cell} \textbf{\bibinfo{volume}{112}}, \bibinfo{pages}{467}
  (\bibinfo{year}{2003}).

\bibitem[{\citenamefont{Svoboda et~al.}(1993)\citenamefont{Svoboda, Schmidt,
  Schnapp, and Block}}]{a:SvobodaEtAl1993}
\bibinfo{author}{\bibfnamefont{K.}~\bibnamefont{Svoboda}},
  \bibinfo{author}{\bibfnamefont{C.~F.} \bibnamefont{Schmidt}},
  \bibinfo{author}{\bibfnamefont{B.~J.} \bibnamefont{Schnapp}},
  \bibnamefont{and} \bibinfo{author}{\bibfnamefont{S.~M.} \bibnamefont{Block}},
  \bibinfo{journal}{Nature (London)} \textbf{\bibinfo{volume}{365}},
  \bibinfo{pages}{721} (\bibinfo{year}{1993}).

\bibitem[{\citenamefont{Vale et~al.}(1996)\citenamefont{Vale, Funatsu, Pierce,
  Romberg, Harada, and Yanagida}}]{a:ValeEtAl1996}
\bibinfo{author}{\bibfnamefont{R.~D.} \bibnamefont{Vale}},
  \bibinfo{author}{\bibfnamefont{T.}~\bibnamefont{Funatsu}},
  \bibinfo{author}{\bibfnamefont{D.~W.} \bibnamefont{Pierce}},
  \bibinfo{author}{\bibfnamefont{L.}~\bibnamefont{Romberg}},
  \bibinfo{author}{\bibfnamefont{Y.}~\bibnamefont{Harada}}, \bibnamefont{and}
  \bibinfo{author}{\bibfnamefont{T.}~\bibnamefont{Yanagida}},
  \bibinfo{journal}{Nature (London)} \textbf{\bibinfo{volume}{380}},
  \bibinfo{pages}{451} (\bibinfo{year}{1996}).

\bibitem[{\citenamefont{Ashkin et~al.}(1990)\citenamefont{Ashkin, Sch{\"u}tze,
  Dziedzic, Euteneuer, and Schliwa}}]{a:AshkinEtAl1990}
\bibinfo{author}{\bibfnamefont{A.}~\bibnamefont{Ashkin}},
  \bibinfo{author}{\bibfnamefont{K.}~\bibnamefont{Sch{\"u}tze}},
  \bibinfo{author}{\bibfnamefont{J.~M.} \bibnamefont{Dziedzic}},
  \bibinfo{author}{\bibfnamefont{U.}~\bibnamefont{Euteneuer}},
  \bibnamefont{and} \bibinfo{author}{\bibfnamefont{M.}~\bibnamefont{Schliwa}},
  \bibinfo{journal}{Nature (London)} \textbf{\bibinfo{volume}{348}},
  \bibinfo{pages}{346} (\bibinfo{year}{1990}).

\bibitem[{\citenamefont{Klumpp and Lipowsky}(2005)}]{a:KlumppLipowsky2005a}
\bibinfo{author}{\bibfnamefont{S.}~\bibnamefont{Klumpp}} \bibnamefont{and}
  \bibinfo{author}{\bibfnamefont{R.}~\bibnamefont{Lipowsky}},
  \bibinfo{journal}{Proc. Natl. Acad. Sci. U.S.A.}
  \textbf{\bibinfo{volume}{102}}, \bibinfo{pages}{17284}
  (\bibinfo{year}{2005}).

\bibitem[{\citenamefont{Chai et~al.}(2009)\citenamefont{Chai, Lipowsky, and
  Klumpp}}]{a:ChaiEtAl2009}
\bibinfo{author}{\bibfnamefont{Y.}~\bibnamefont{Chai}},
  \bibinfo{author}{\bibfnamefont{R.}~\bibnamefont{Lipowsky}}, \bibnamefont{and}
  \bibinfo{author}{\bibfnamefont{S.}~\bibnamefont{Klumpp}},
  \bibinfo{journal}{Journal of Statistical Physics}
  \textbf{\bibinfo{volume}{135}}, \bibinfo{pages}{241} (\bibinfo{year}{2009}).

\bibitem[{\citenamefont{Koster et~al.}(2003)\citenamefont{Koster, van Guijn,
  Hofs, and Dogterom}}]{a:KosterEtAl2003}
\bibinfo{author}{\bibfnamefont{G.}~\bibnamefont{Koster}},
  \bibinfo{author}{\bibfnamefont{M.}~\bibnamefont{van Guijn}},
  \bibinfo{author}{\bibfnamefont{B.}~\bibnamefont{Hofs}}, \bibnamefont{and}
  \bibinfo{author}{\bibfnamefont{M.}~\bibnamefont{Dogterom}},
  \bibinfo{journal}{Proc. Natl. Acad. Sci. U.S.A.}
  \textbf{\bibinfo{volume}{100}}, \bibinfo{pages}{15583}
  (\bibinfo{year}{2003}).

\bibitem[{\citenamefont{Leduc et~al.}(2004)\citenamefont{Leduc, Camp{\'a}s,
  Zeldovich, Roux, Jolimaitre, Bourel-Bonnet, Goud, Joanny, Bassereau, and
  Prost}}]{a:LeducEtAl2004}
\bibinfo{author}{\bibfnamefont{C.}~\bibnamefont{Leduc}},
  \bibinfo{author}{\bibfnamefont{O.}~\bibnamefont{Camp{\'a}s}},
  \bibinfo{author}{\bibfnamefont{K.~B.} \bibnamefont{Zeldovich}},
  \bibinfo{author}{\bibfnamefont{A.}~\bibnamefont{Roux}},
  \bibinfo{author}{\bibfnamefont{P.}~\bibnamefont{Jolimaitre}},
  \bibinfo{author}{\bibfnamefont{L.}~\bibnamefont{Bourel-Bonnet}},
  \bibinfo{author}{\bibfnamefont{B.}~\bibnamefont{Goud}},
  \bibinfo{author}{\bibfnamefont{J.-F.} \bibnamefont{Joanny}},
  \bibinfo{author}{\bibfnamefont{P.}~\bibnamefont{Bassereau}},
  \bibnamefont{and} \bibinfo{author}{\bibfnamefont{J.}~\bibnamefont{Prost}},
  \bibinfo{journal}{Proc. Natl. Acad. Sci. U.S.A.}
  \textbf{\bibinfo{volume}{101}}, \bibinfo{pages}{17096}
  (\bibinfo{year}{2004}).

\bibitem[{\citenamefont{McMahon}(1984)}]{b:McMahon1984}
\bibinfo{author}{\bibfnamefont{T.~A.} \bibnamefont{McMahon}},
  \emph{\bibinfo{title}{{Muscles, Reflexes, and Locomotion}}}
  (\bibinfo{publisher}{Princeton University Press},
  \bibinfo{address}{Princeton, New Jersey}, \bibinfo{year}{1984}).

\bibitem[{\citenamefont{Piazzesi et~al.}(2007)\citenamefont{Piazzesi,
  Reconditi, Linari, Lucii, Bianco, Brunello, Decostre, Stewart, Gore, Irving
  et~al.}}]{a:PiazzesiEtAl2007}
\bibinfo{author}{\bibfnamefont{G.}~\bibnamefont{Piazzesi}},
  \bibinfo{author}{\bibfnamefont{M.}~\bibnamefont{Reconditi}},
  \bibinfo{author}{\bibfnamefont{M.}~\bibnamefont{Linari}},
  \bibinfo{author}{\bibfnamefont{L.}~\bibnamefont{Lucii}},
  \bibinfo{author}{\bibfnamefont{P.}~\bibnamefont{Bianco}},
  \bibinfo{author}{\bibfnamefont{E.}~\bibnamefont{Brunello}},
  \bibinfo{author}{\bibfnamefont{V.}~\bibnamefont{Decostre}},
  \bibinfo{author}{\bibfnamefont{A.}~\bibnamefont{Stewart}},
  \bibinfo{author}{\bibfnamefont{D.~B.} \bibnamefont{Gore}},
  \bibinfo{author}{\bibfnamefont{T.~C.} \bibnamefont{Irving}},
  \bibinfo{author}{\bibfnamefont{M.} \bibnamefont{Irving}},
  \bibinfo{author}{\bibfnamefont{V.} \bibnamefont{Lombardi}},
  \bibinfo{journal}{Cell} \textbf{\bibinfo{volume}{131}},
  \bibinfo{pages}{784} (\bibinfo{year}{2007}).

\bibitem[{\citenamefont{Huxley}(1957)}]{a:Huxley1957}
\bibinfo{author}{\bibfnamefont{A.~F.} \bibnamefont{Huxley}},
  \bibinfo{journal}{Prog. Biophys.} \textbf{\bibinfo{volume}{7}},
  \bibinfo{pages}{255} (\bibinfo{year}{1957}).

\bibitem[{\citenamefont{Huxley and Simmons}(1971)}]{a:HuxleySimmons1971}
\bibinfo{author}{\bibfnamefont{A.~F.} \bibnamefont{Huxley}} \bibnamefont{and}
  \bibinfo{author}{\bibfnamefont{R.~M.} \bibnamefont{Simmons}},
  \bibinfo{journal}{Nature} \textbf{\bibinfo{volume}{233}},
  \bibinfo{pages}{533} (\bibinfo{year}{1971}).

\bibitem[{\citenamefont{Geeves and Holmes}(2005)}]{a:GeevesHolmes2005}
\bibinfo{author}{\bibfnamefont{M.~A.} \bibnamefont{Geeves}} \bibnamefont{and}
  \bibinfo{author}{\bibfnamefont{K.~C.} \bibnamefont{Holmes}},
  \bibinfo{journal}{Adv. Protein Chem.} \textbf{\bibinfo{volume}{71}},
  \bibinfo{pages}{161} (\bibinfo{year}{2005}).

\bibitem[{\citenamefont{Leibler and Huse}(1993)}]{a:LeiblerHuse1993}
\bibinfo{author}{\bibfnamefont{S.}~\bibnamefont{Leibler}} \bibnamefont{and}
  \bibinfo{author}{\bibfnamefont{D.~A.} \bibnamefont{Huse}},
  \bibinfo{journal}{J. Cell Biol.} \textbf{\bibinfo{volume}{121}},
  \bibinfo{pages}{1357} (\bibinfo{year}{1993}).

\bibitem[{\citenamefont{Duke}(1999)}]{a:Duke1999}
\bibinfo{author}{\bibfnamefont{T.~A.~J.} \bibnamefont{Duke}},
  \bibinfo{journal}{Proc. Natl. Acad. Sci. U.S.A.}
  \textbf{\bibinfo{volume}{96}}, \bibinfo{pages}{2770} (\bibinfo{year}{1999}).

\bibitem[{\citenamefont{Veigel et~al.}(2003)\citenamefont{Veigel, Molloy,
  Schmitz, and Kendrick-Jones}}]{a:VeigelEtAl2003}
\bibinfo{author}{\bibfnamefont{C.}~\bibnamefont{Veigel}},
  \bibinfo{author}{\bibfnamefont{J.~E.} \bibnamefont{Molloy}},
  \bibinfo{author}{\bibfnamefont{S.}~\bibnamefont{Schmitz}}, \bibnamefont{and}
  \bibinfo{author}{\bibfnamefont{J.}~\bibnamefont{Kendrick-Jones}},
  \bibinfo{journal}{Nature Cell Biology} \textbf{\bibinfo{volume}{5}},
  \bibinfo{pages}{980} (\bibinfo{year}{2003}).

\bibitem[{\citenamefont{Guo and Guilford}(2006)}]{a:GuoGuilford2006}
\bibinfo{author}{\bibfnamefont{B.}~\bibnamefont{Guo}} \bibnamefont{and}
  \bibinfo{author}{\bibfnamefont{W.~H.} \bibnamefont{Guilford}},
  \bibinfo{journal}{Proc. Natl. Acad. Sci. U.S.A.}
  \textbf{\bibinfo{volume}{103}}, \bibinfo{pages}{9844} (\bibinfo{year}{2006}).

\bibitem[{\citenamefont{Vicente-Manzanares
  et~al.}(2009)\citenamefont{Vicente-Manzanares, Ma, Adelstein, and
  Horwitz}}]{a:VicenteEtAl2009}
\bibinfo{author}{\bibfnamefont{M.}~\bibnamefont{Vicente-Manzanares}},
  \bibinfo{author}{\bibfnamefont{X.}~\bibnamefont{Ma}},
  \bibinfo{author}{\bibfnamefont{R.~S.} \bibnamefont{Adelstein}},
  \bibnamefont{and} \bibinfo{author}{\bibfnamefont{A.~R.}
  \bibnamefont{Horwitz}}, \bibinfo{journal}{Nature Reviews Molecular Cell
  Biology} \textbf{\bibinfo{volume}{10}}, \bibinfo{pages}{778}
  (\bibinfo{year}{2009}).

\bibitem[{\citenamefont{Gupton and
  Waterman-Storer}(2006)}]{a:GuptonWatermanStorer2006}
\bibinfo{author}{\bibfnamefont{S.~L.} \bibnamefont{Gupton}} \bibnamefont{and}
  \bibinfo{author}{\bibfnamefont{C.~M.} \bibnamefont{Waterman-Storer}},
  \bibinfo{journal}{Cell} \textbf{\bibinfo{volume}{125}}, \bibinfo{pages}{1361}
  (\bibinfo{year}{2006}).

\bibitem[{\citenamefont{Wakatsuki et~al.}(2003)\citenamefont{Wakatsuki,
  Wysolmerski, and Elson}}]{a:WakatsukiEtAl2003}
\bibinfo{author}{\bibfnamefont{T.}~\bibnamefont{Wakatsuki}},
  \bibinfo{author}{\bibfnamefont{R.~B.} \bibnamefont{Wysolmerski}},
  \bibnamefont{and} \bibinfo{author}{\bibfnamefont{E.~L.} \bibnamefont{Elson}},
  \bibinfo{journal}{J. Cell Sci.} \textbf{\bibinfo{volume}{116}},
  \bibinfo{pages}{1617} (\bibinfo{year}{2003}).

\bibitem[{\citenamefont{Paluch et~al.}(2005)\citenamefont{Paluch, Piel, Prost,
  Bornens, and Sykes}}]{a:PaluchEtAl2005}
\bibinfo{author}{\bibfnamefont{E.}~\bibnamefont{Paluch}},
  \bibinfo{author}{\bibfnamefont{M.}~\bibnamefont{Piel}},
  \bibinfo{author}{\bibfnamefont{J.}~\bibnamefont{Prost}},
  \bibinfo{author}{\bibfnamefont{M.}~\bibnamefont{Bornens}}, \bibnamefont{and}
  \bibinfo{author}{\bibfnamefont{C.}~\bibnamefont{Sykes}},
  \bibinfo{journal}{Biophysical Journal} \textbf{\bibinfo{volume}{89}},
  \bibinfo{pages}{724} (\bibinfo{year}{2005}).

\bibitem[{\citenamefont{Pellegrin and Mellor}(2007)}]{a:PellegrinMellor2007}
\bibinfo{author}{\bibfnamefont{S.}~\bibnamefont{Pellegrin}} \bibnamefont{and}
  \bibinfo{author}{\bibfnamefont{H.}~\bibnamefont{Mellor}},
  \bibinfo{journal}{J. Cell Sci.} \textbf{\bibinfo{volume}{120}},
  \bibinfo{pages}{3491} (\bibinfo{year}{2007}).

\bibitem[{\citenamefont{Peterson et~al.}(2004)\citenamefont{Peterson, Rajfur,
  Maddox, Freel, Chen, Edlund, Otey, and Burridge}}]{a:PetersonEtAl2004}
\bibinfo{author}{\bibfnamefont{L.~J.} \bibnamefont{Peterson}},
  \bibinfo{author}{\bibfnamefont{Z.}~\bibnamefont{Rajfur}},
  \bibinfo{author}{\bibfnamefont{A.~S.} \bibnamefont{Maddox}},
  \bibinfo{author}{\bibfnamefont{C.~D.} \bibnamefont{Freel}},
  \bibinfo{author}{\bibfnamefont{Y.}~\bibnamefont{Chen}},
  \bibinfo{author}{\bibfnamefont{M.}~\bibnamefont{Edlund}},
  \bibinfo{author}{\bibfnamefont{C.}~\bibnamefont{Otey}}, \bibnamefont{and}
  \bibinfo{author}{\bibfnamefont{K.}~\bibnamefont{Burridge}},
  \bibinfo{journal}{Molecular Biology of the Cell}
  \textbf{\bibinfo{volume}{15}}, \bibinfo{pages}{3497} (\bibinfo{year}{2004}).

\bibitem[{\citenamefont{Verkhovsky and Borisy}(1993)}]{a:VerkhovskyBorisy1993}
\bibinfo{author}{\bibfnamefont{A.~B.} \bibnamefont{Verkhovsky}}
  \bibnamefont{and} \bibinfo{author}{\bibfnamefont{G.~G.}
  \bibnamefont{Borisy}}, \bibinfo{journal}{J. Cell Biol.}
  \textbf{\bibinfo{volume}{123}}, \bibinfo{pages}{637} (\bibinfo{year}{1993}).

\bibitem[{\citenamefont{Thoresen et~al.}(2011)\citenamefont{Thoresen, Lenz, and
  Gardel}}]{a:ThoresenEtAl2011}
\bibinfo{author}{\bibfnamefont{T.}~\bibnamefont{Thoresen}},
  \bibinfo{author}{\bibfnamefont{M.}~\bibnamefont{Lenz}}, \bibnamefont{and}
  \bibinfo{author}{\bibfnamefont{M.~L.} \bibnamefont{Gardel}},
  \bibinfo{journal}{Biophysical Journal} \textbf{\bibinfo{volume}{100}},
  \bibinfo{pages}{2698} (\bibinfo{year}{2011}).

\bibitem[{\citenamefont{Thoresen et~al.}(2013)\citenamefont{Thoresen, Lenz, and
  Gardel}}]{a:ThoresenEtAl2013}
\bibinfo{author}{\bibfnamefont{T.}~\bibnamefont{Thoresen}},
  \bibinfo{author}{\bibfnamefont{M.}~\bibnamefont{Lenz}}, \bibnamefont{and}
  \bibinfo{author}{\bibfnamefont{M.~L.} \bibnamefont{Gardel}},
  \bibinfo{journal}{Biophysical Journal} \textbf{\bibinfo{volume}{104}},
  \bibinfo{pages}{655} (\bibinfo{year}{2013}).

\bibitem[{\citenamefont{Soares~e Silva et~al.}(2011)\citenamefont{Soares~e
  Silva, Depken, Stuhrmann, Korsten, MacKintosh, and
  Koenderink}}]{a:SoaresEtAl2011}
\bibinfo{author}{\bibfnamefont{M.}~\bibnamefont{Soares~e Silva}},
  \bibinfo{author}{\bibfnamefont{M.}~\bibnamefont{Depken}},
  \bibinfo{author}{\bibfnamefont{B.}~\bibnamefont{Stuhrmann}},
  \bibinfo{author}{\bibfnamefont{M.}~\bibnamefont{Korsten}},
  \bibinfo{author}{\bibfnamefont{F.~C.} \bibnamefont{MacKintosh}},
  \bibnamefont{and} \bibinfo{author}{\bibfnamefont{G.~H.}
  \bibnamefont{Koenderink}}, \bibinfo{journal}{Proc. Natl. Acad. Sci. U.S.A.}
  \textbf{\bibinfo{volume}{108}}, \bibinfo{pages}{9408} (\bibinfo{year}{2011}).

\bibitem[{\citenamefont{Finer et~al.}(1994)\citenamefont{Finer, Simmons, and
  Spudich}}]{a:FinerEtAl1994}
\bibinfo{author}{\bibfnamefont{J.~T.} \bibnamefont{Finer}},
  \bibinfo{author}{\bibfnamefont{R.~M.} \bibnamefont{Simmons}},
  \bibnamefont{and} \bibinfo{author}{\bibfnamefont{J.~A.}
  \bibnamefont{Spudich}}, \bibinfo{journal}{Nature}
  \textbf{\bibinfo{volume}{368}}, \bibinfo{pages}{113} (\bibinfo{year}{1994}).

\bibitem[{\citenamefont{Debold et~al.}(2005)\citenamefont{Debold, Patlak, and
  Warshaw}}]{a:DeboldEtAl2005}
\bibinfo{author}{\bibfnamefont{E.~P.} \bibnamefont{Debold}},
  \bibinfo{author}{\bibfnamefont{J.~B.} \bibnamefont{Patlak}},
  \bibnamefont{and} \bibinfo{author}{\bibfnamefont{D.~M.}
  \bibnamefont{Warshaw}}, \bibinfo{journal}{Biophysical Journal}
  \textbf{\bibinfo{volume}{89}}, \bibinfo{pages}{L34} (\bibinfo{year}{2005}).

\bibitem[{\citenamefont{Mizuno et~al.}(2007)\citenamefont{Mizuno, Tardin,
  Schmidt, and MacKintosh}}]{a:MizunoEtAl2007}
\bibinfo{author}{\bibfnamefont{D.}~\bibnamefont{Mizuno}},
  \bibinfo{author}{\bibfnamefont{C.}~\bibnamefont{Tardin}},
  \bibinfo{author}{\bibfnamefont{C.~F.} \bibnamefont{Schmidt}},
  \bibnamefont{and} \bibinfo{author}{\bibfnamefont{F.~C.}
  \bibnamefont{MacKintosh}}, \bibinfo{journal}{Science}
  \textbf{\bibinfo{volume}{315}}, \bibinfo{pages}{370} (\bibinfo{year}{2007}).

\bibitem[{\citenamefont{Duke et~al.}(1995)\citenamefont{Duke, Holy, and
  Leibler}}]{a:DukeEtAl1995}
\bibinfo{author}{\bibfnamefont{T.}~\bibnamefont{Duke}},
  \bibinfo{author}{\bibfnamefont{T.~E.} \bibnamefont{Holy}}, \bibnamefont{and}
  \bibinfo{author}{\bibfnamefont{S.}~\bibnamefont{Leibler}},
  \bibinfo{journal}{Phys. Rev. Lett.} \textbf{\bibinfo{volume}{74}},
  \bibinfo{pages}{330} (\bibinfo{year}{1995}).

\bibitem[{\citenamefont{Pla{\c c}ais et~al.}(2009)\citenamefont{Pla{\c c}ais,
  Balland, Gu{\'e}rin, Joanny, and Martin}}]{a:PlacaisEtAl2009}
\bibinfo{author}{\bibfnamefont{P.-Y.} \bibnamefont{Pla{\c c}ais}},
  \bibinfo{author}{\bibfnamefont{M.}~\bibnamefont{Balland}},
  \bibinfo{author}{\bibfnamefont{T.}~\bibnamefont{Gu{\'e}rin}},
  \bibinfo{author}{\bibfnamefont{J.-F.} \bibnamefont{Joanny}},
  \bibnamefont{and} \bibinfo{author}{\bibfnamefont{P.}~\bibnamefont{Martin}},
  \bibinfo{journal}{Phys. Rev. Lett.} \textbf{\bibinfo{volume}{103}},
  \bibinfo{pages}{158102} (\bibinfo{year}{2009}).

\bibitem[{\citenamefont{Hexner and Kafri}(2009)}]{a:HexnerKafri2009}
\bibinfo{author}{\bibfnamefont{D.}~\bibnamefont{Hexner}} \bibnamefont{and}
  \bibinfo{author}{\bibfnamefont{Y.}~\bibnamefont{Kafri}},
  \bibinfo{journal}{Phys. Biol.} \textbf{\bibinfo{volume}{6}},
  \bibinfo{pages}{036016} (\bibinfo{year}{2009}).

\bibitem[{\citenamefont{Erdmann and Schwarz}(2012)}]{a:ErdmannSchwarz2012}
\bibinfo{author}{\bibfnamefont{T.}~\bibnamefont{Erdmann}} \bibnamefont{and}
  \bibinfo{author}{\bibfnamefont{U.~S.} \bibnamefont{Schwarz}},
  \bibinfo{journal}{Phys. Rev. Lett.} \textbf{\bibinfo{volume}{108}},
  \bibinfo{pages}{188101} (\bibinfo{year}{2012}).

\bibitem[{\citenamefont{J{\"u}licher and Prost}(1995)}]{a:JuelicherProst1995}
\bibinfo{author}{\bibfnamefont{F.}~\bibnamefont{J{\"u}licher}}
  \bibnamefont{and} \bibinfo{author}{\bibfnamefont{J.}~\bibnamefont{Prost}},
  \bibinfo{journal}{Phys. Rev. Lett.} \textbf{\bibinfo{volume}{75}},
  \bibinfo{pages}{2618} (\bibinfo{year}{1995}).

\bibitem[{\citenamefont{J{\"u}licher and Prost}(1997)}]{a:JuelicherProst1997}
\bibinfo{author}{\bibfnamefont{F.}~\bibnamefont{J{\"u}licher}}
  \bibnamefont{and} \bibinfo{author}{\bibfnamefont{J.}~\bibnamefont{Prost}},
  \bibinfo{journal}{Phys. Rev. Lett.} \textbf{\bibinfo{volume}{78}},
  \bibinfo{pages}{4510} (\bibinfo{year}{1997}).

\bibitem[{\citenamefont{Badoual et~al.}(2002)\citenamefont{Badoual,
  J{\"u}licher, and Prost}}]{a:BadoualEtAl2002}
\bibinfo{author}{\bibfnamefont{M.}~\bibnamefont{Badoual}},
  \bibinfo{author}{\bibfnamefont{F.}~\bibnamefont{J{\"u}licher}},
  \bibnamefont{and} \bibinfo{author}{\bibfnamefont{J.}~\bibnamefont{Prost}},
  \bibinfo{journal}{Proceedings of the National Academy of Sciences of the
  United States of America} \textbf{\bibinfo{volume}{99}},
  \bibinfo{pages}{6696} (\bibinfo{year}{2002}).

\bibitem[{\citenamefont{Vilfan et~al.}(1999)\citenamefont{Vilfan, Frey, and
  Schwabl}}]{a:VilfanEtAl1999}
\bibinfo{author}{\bibfnamefont{A.}~\bibnamefont{Vilfan}},
  \bibinfo{author}{\bibfnamefont{E.}~\bibnamefont{Frey}}, \bibnamefont{and}
  \bibinfo{author}{\bibfnamefont{F.}~\bibnamefont{Schwabl}},
  \bibinfo{journal}{EPL (Europhysics Letters)} \textbf{\bibinfo{volume}{45}},
  \bibinfo{pages}{283} (\bibinfo{year}{1999}).

\bibitem[{\citenamefont{Vilfan and Frey}(2005)}]{a:VilfanFrey2005}
\bibinfo{author}{\bibfnamefont{A.}~\bibnamefont{Vilfan}} \bibnamefont{and}
  \bibinfo{author}{\bibfnamefont{E.}~\bibnamefont{Frey}},
  \bibinfo{journal}{Journal of Physics: Condensed Matter}
  \textbf{\bibinfo{volume}{17}}, \bibinfo{pages}{S3901} (\bibinfo{year}{2005}).

\bibitem[{\citenamefont{Duke}(2000)}]{a:Duke2000}
\bibinfo{author}{\bibfnamefont{T.}~\bibnamefont{Duke}}, \bibinfo{journal}{Phil.
  Trans. Roy. Soc. Lond. B} \textbf{\bibinfo{volume}{355}},
  \bibinfo{pages}{529} (\bibinfo{year}{2000}).

\bibitem[{\citenamefont{Vilfan and Duke}(2003)}]{a:VilfanDuke2003b}
\bibinfo{author}{\bibfnamefont{A.}~\bibnamefont{Vilfan}} \bibnamefont{and}
  \bibinfo{author}{\bibfnamefont{T.}~\bibnamefont{Duke}},
  \bibinfo{journal}{Biophysical Journal} \textbf{\bibinfo{volume}{85}},
  \bibinfo{pages}{818} (\bibinfo{year}{2003}).

\bibitem[{\citenamefont{Walcott et~al.}(2012)\citenamefont{Walcott, Warshaw,
  and Debold}}]{a:WalcottEtAl2012}
\bibinfo{author}{\bibfnamefont{S.}~\bibnamefont{Walcott}},
  \bibinfo{author}{\bibfnamefont{D.~M.} \bibnamefont{Warshaw}},
  \bibnamefont{and} \bibinfo{author}{\bibfnamefont{E.~P.}
  \bibnamefont{Debold}}, \bibinfo{journal}{Biophysical Journal}
  \textbf{\bibinfo{volume}{103}}, \bibinfo{pages}{501} (\bibinfo{year}{2012}).

\bibitem[{\citenamefont{Chen and Gao}(2011)}]{a:ChenGao2011}
\bibinfo{author}{\bibfnamefont{B.}~\bibnamefont{Chen}} \bibnamefont{and}
  \bibinfo{author}{\bibfnamefont{H.}~\bibnamefont{Gao}},
  \bibinfo{journal}{Biophysical Journal} \textbf{\bibinfo{volume}{101}},
  \bibinfo{pages}{396} (\bibinfo{year}{2011}).

\bibitem[{\citenamefont{Sellers and Veigel}(2010)}]{a:SellersVeigel2010}
\bibinfo{author}{\bibfnamefont{J.~R.} \bibnamefont{Sellers}} \bibnamefont{and}
  \bibinfo{author}{\bibfnamefont{C.}~\bibnamefont{Veigel}},
  \bibinfo{journal}{Nature Structural and Molecular Biology}
  \textbf{\bibinfo{volume}{17}}, \bibinfo{pages}{590} (\bibinfo{year}{2010}).

\bibitem[{\citenamefont{Colombini et~al.}(2007)\citenamefont{Colombini, Bagni,
  Romano, and Cecchi}}]{a:ColombiniEtAl2007}
\bibinfo{author}{\bibfnamefont{B.}~\bibnamefont{Colombini}},
  \bibinfo{author}{\bibfnamefont{M.~A.} \bibnamefont{Bagni}},
  \bibinfo{author}{\bibfnamefont{G.}~\bibnamefont{Romano}}, \bibnamefont{and}
  \bibinfo{author}{\bibfnamefont{G.}~\bibnamefont{Cecchi}},
  \bibinfo{journal}{Proc. Natl. Acad. Sci. U.S.A.}
  \textbf{\bibinfo{volume}{104}}, \bibinfo{pages}{9284} (\bibinfo{year}{2007}).

\bibitem[{\citenamefont{Erdmann and
  Schwarz}(2004{\natexlab{a}})}]{a:ErdmannSchwarz2004a}
\bibinfo{author}{\bibfnamefont{T.}~\bibnamefont{Erdmann}} \bibnamefont{and}
  \bibinfo{author}{\bibfnamefont{U.~S.} \bibnamefont{Schwarz}},
  \bibinfo{journal}{Phys. Rev. Lett.} \textbf{\bibinfo{volume}{92}},
  \bibinfo{pages}{108102} (\bibinfo{year}{2004}{\natexlab{a}}).

\bibitem[{\citenamefont{Erdmann and
  Schwarz}(2004{\natexlab{b}})}]{a:ErdmannSchwarz2004c}
\bibinfo{author}{\bibfnamefont{T.}~\bibnamefont{Erdmann}} \bibnamefont{and}
  \bibinfo{author}{\bibfnamefont{U.~S.} \bibnamefont{Schwarz}},
  \bibinfo{journal}{J. Chem. Phys.} \textbf{\bibinfo{volume}{121}},
  \bibinfo{pages}{8997} (\bibinfo{year}{2004}{\natexlab{b}}).

\bibitem[{\citenamefont{Erdmann and Schwarz}(2006)}]{a:ErdmannSchwarz2006}
\bibinfo{author}{\bibfnamefont{T.}~\bibnamefont{Erdmann}} \bibnamefont{and}
  \bibinfo{author}{\bibfnamefont{U.~S.} \bibnamefont{Schwarz}},
  \bibinfo{journal}{Biophysical Journal} \textbf{\bibinfo{volume}{91}},
  \bibinfo{pages}{L60} (\bibinfo{year}{2006}).

\bibitem[{\citenamefont{van Kampen}(2003)}]{b:VanKampen2003}
\bibinfo{author}{\bibfnamefont{N.~G.} \bibnamefont{van Kampen}},
  \emph{\bibinfo{title}{{Stochastic Processes in Physics and Chemistry}}}
  (\bibinfo{publisher}{Elsevier Ltd}, \bibinfo{address}{Amsterdam},
  \bibinfo{year}{2003}).

\bibitem[{\citenamefont{Gillespie}(1976)}]{a:Gillespie1976}
\bibinfo{author}{\bibfnamefont{D.~T.} \bibnamefont{Gillespie}},
  \bibinfo{journal}{J. Chem. Phys.} \textbf{\bibinfo{volume}{22}},
  \bibinfo{pages}{403} (\bibinfo{year}{1976}).

\bibitem[{\citenamefont{Hill}(1939)}]{a:Hill1939b}
\bibinfo{author}{\bibfnamefont{A.~V.} \bibnamefont{Hill}},
  \bibinfo{journal}{Proc. R. Soc. Lond. B} \textbf{\bibinfo{volume}{127}},
  \bibinfo{pages}{434} (\bibinfo{year}{1939}).

\bibitem[{\citenamefont{Berger et~al.}(2012)\citenamefont{Berger, Keller,
  Klumpp, and Lipowsky}}]{a:BergerEtAl2012}
\bibinfo{author}{\bibfnamefont{F.}~\bibnamefont{Berger}},
  \bibinfo{author}{\bibfnamefont{C.}~\bibnamefont{Keller}},
  \bibinfo{author}{\bibfnamefont{S.}~\bibnamefont{Klumpp}}, \bibnamefont{and}
  \bibinfo{author}{\bibfnamefont{R.}~\bibnamefont{Lipowsky}},
  \bibinfo{journal}{Phys. Rev. Lett.} \textbf{\bibinfo{volume}{108}},
  \bibinfo{pages}{208101} (\bibinfo{year}{2012}).

\bibitem[{\citenamefont{Kunwar and Mogilner}(2010)}]{a:KunwarMogilner2010}
\bibinfo{author}{\bibfnamefont{A.}~\bibnamefont{Kunwar}} \bibnamefont{and}
  \bibinfo{author}{\bibfnamefont{A.}~\bibnamefont{Mogilner}},
  \bibinfo{journal}{Phys. Biol.} \textbf{\bibinfo{volume}{7}},
  \bibinfo{pages}{016012} (\bibinfo{year}{2010}).

\bibitem{SuppText} See Supplementary Material Document
No.~\rule{12mm}{0.2mm} for additional results with the same set of
parameters as used in the main text and further results demonstrating
the effect of changing the parameter values, in particular the single
motor duty ratio. For information on Supplementary Material, see
\url{http://www.aip.org/pubservs/epaps.html}.

\bibitem[{\citenamefont{Caruel et~al.}(2013)\citenamefont{Caruel, Allain, and
  Truskinovsky}}]{a:CaruelEtAl2013}
\bibinfo{author}{\bibfnamefont{M.}~\bibnamefont{Caruel}},
  \bibinfo{author}{\bibfnamefont{J.~M.} \bibnamefont{Allain}},
  \bibnamefont{and}
  \bibinfo{author}{\bibfnamefont{L.}~\bibnamefont{Truskinovsky}},
  \bibinfo{journal}{Physical Review Letters} \textbf{\bibinfo{volume}{110}},
  \bibinfo{pages}{248103} (\bibinfo{year}{2013}).

\bibitem[{\citenamefont{Laakso et~al.}(2008)\citenamefont{Laakso, Lewis,
  Shuman, and Ostap}}]{a:LaaksoEtAl2008}
\bibinfo{author}{\bibfnamefont{J.~M.} \bibnamefont{Laakso}},
  \bibinfo{author}{\bibfnamefont{J.~H.} \bibnamefont{Lewis}},
  \bibinfo{author}{\bibfnamefont{H.}~\bibnamefont{Shuman}}, \bibnamefont{and}
  \bibinfo{author}{\bibfnamefont{E.~M.} \bibnamefont{Ostap}},
  \bibinfo{journal}{Science} \textbf{\bibinfo{volume}{321}},
  \bibinfo{pages}{133} (\bibinfo{year}{2008}).


\end{thebibliography}


\newpage
\appendix




\section{Supplemental results for standard parameters}

\subsection{Constant load}


\FIG{FigS01}{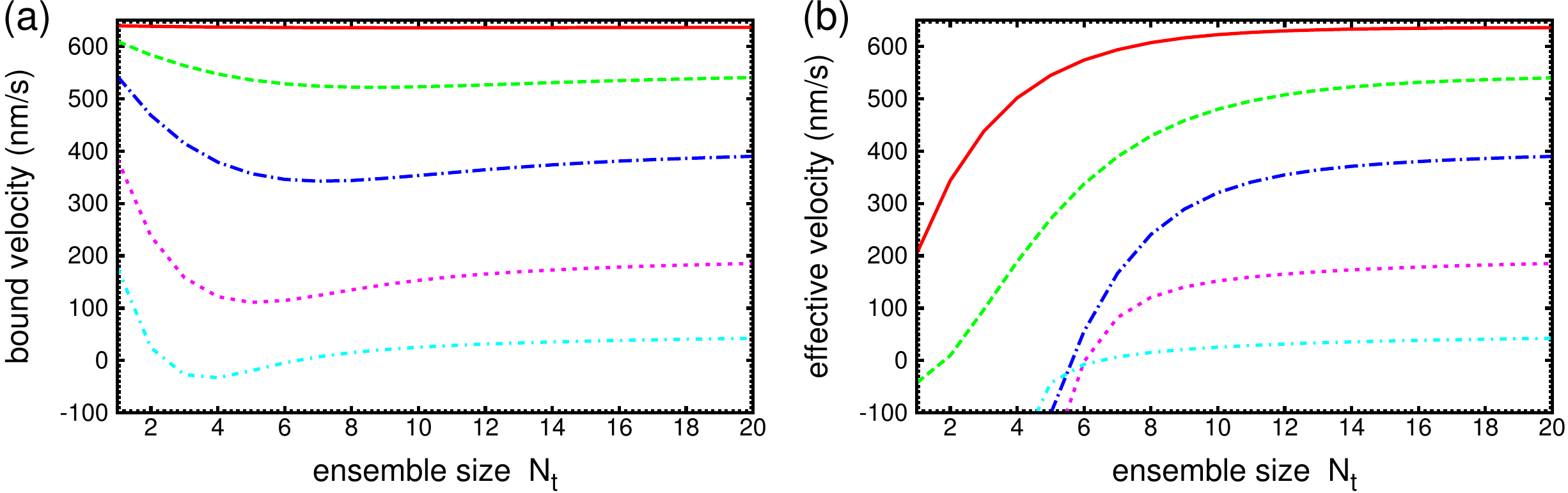}{Analytical results for the parallel cluster model
with constant external load. (a) Bound velocity $\vb$ and (b) effective
velocity $\veff$ as function of ensemble size $\nt$ for values $\fext/
\nt = 0.0126\pN$, $1.262\pN$, $3.787\pN$ and $8.835\pN$ (top to bottom
curves) of the external load per motor. For $\veff$ the viscous mobility
of detached ensembles is $\eta = 10^3 \mob$. Constant parameters are
listed in \tab{Tab01} of the main text.}

\fig{FigS01}(a) plots the bound velocity $\vb$ as function of ensemble
size $\nt$ for different values of the external load per motor $\fext/
\nt$. For finite external load, $\vb$ varies markedly only at small
$\nt$ but approaches a constant value for large $\nt$. The variation is
non-monotonic at large external loads and the bound velocity can be
negative for intermediate values of $\nt$. \fig{FigS01}(b) plots the
effective velocity $\veff$ as function of $\nt$ for different $\fext/
\nt$. For all values of the force, $\veff$ increases quickly at small
$\nt$ and approaches constant for $\nt \geqslant 15$.

\subsection{Validation of the parallel cluster model: Kramers' type off-rate} \label{app:AppB}

In the simulations for the validation of the parallel cluster model, we
have used an off-rate from the post-power-stroke state which decreases
for negative strain but remained constant for positive strain of the
motor (see \eq{OffStrain}). Here, we demonstrate the effect of a
Kramers' type off-rate
\begin{equation}\label{eq:Kramers}
k_{20}(\xin) = k_{20}^0 \exp(-\kf\xin)\,,
\end{equation}
for all values of the strain $\xin$ on the dynamics of the ensemble.
Because within the PCM, the strain of the post-power-stroke motors is
always positive, the behavior of $k_{20}(\xin)$ for $\xin > 0$ is not
relevant for the PCM.

\FIG{FigS02}{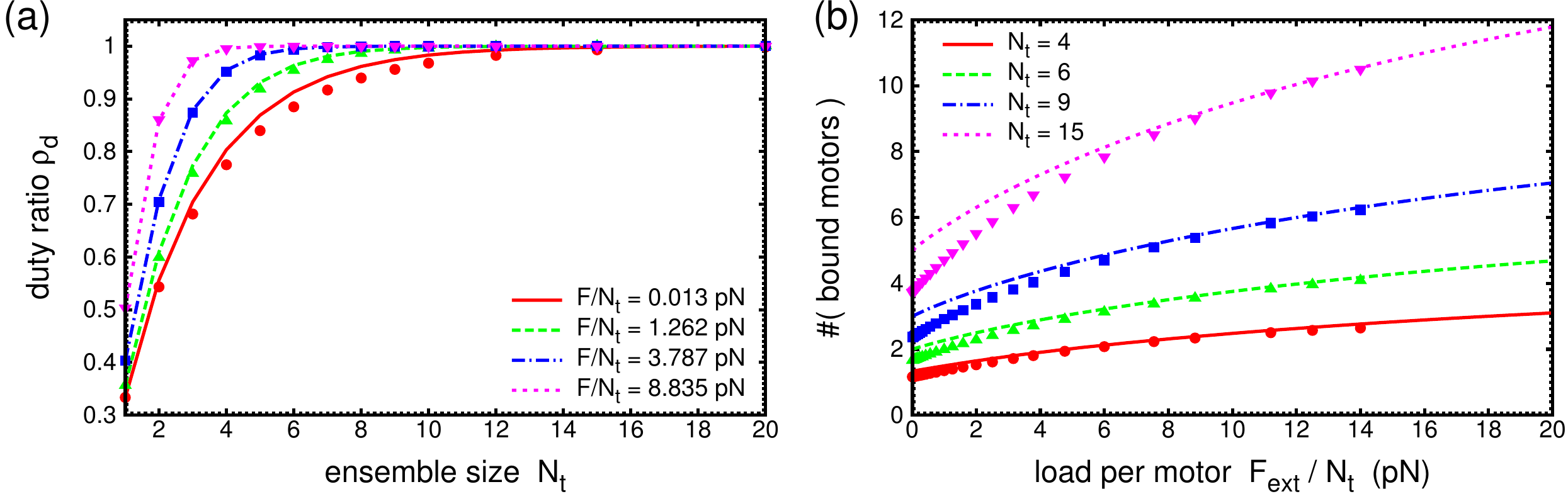}{Validation of the parallel cluster model:
comparison of analytical results using the PCM (lines) with computer
simulations with individual motor strains (symbols). (a) Duty ratio
$\rd$ as function of ensemble size $\nt$ for the values $\fext/\nt =
0.0126\pN$, $1.262\pN$, $3.787\pN$ and $8.835\pN$ of the external load
per motor. (b) Average number of bound motors $\nb$ as function of
external load per motor $\fext/\nt $ for ensemble sizes $\nt=4$, $6$,
$9$ and $15$. Constant parameters are listed in \tab{Tab01}.}

\fig{FigS02}(a) compares analytical results for the average number of
bound motors as function of ensemble size for different values of the
constant external load $\fext/\nt$ to results of simulations without PCM
or LTE with the off-rate from \eq{Kramers} from the post-power-stroke
state. While the average number of bound motors with the asymmetric
off-rate from \eq{OffStrain} was increased, the Kramers' type off-rate
from \eq{Kramers} leads to a clearly reduced number of bound motors at
small values of the external load. This is confirmed by \fig{FigS02}(b)
which plots $\nb$ as function of $\fext/\nt$ for different values of
$\nt$. At small $\fext/\nt$, $\nb$ is reduced in comparison to the
results from the PCM. This decrease is caused by the internal strains of
motors at different positions which are working against each other. For
the rate in \eq{OffStrain}, this induced an increase of $\nb$ compared
to the PCM. For the Kramers' type rate in \eq{Kramers}, the increase of
the off-rate for negative strain dominates so that $\nb$ is reduced. For
large external load, most motors are subject to negative strain so that
the effect of the internal strains is mitigated.

\FIG{FigS03}{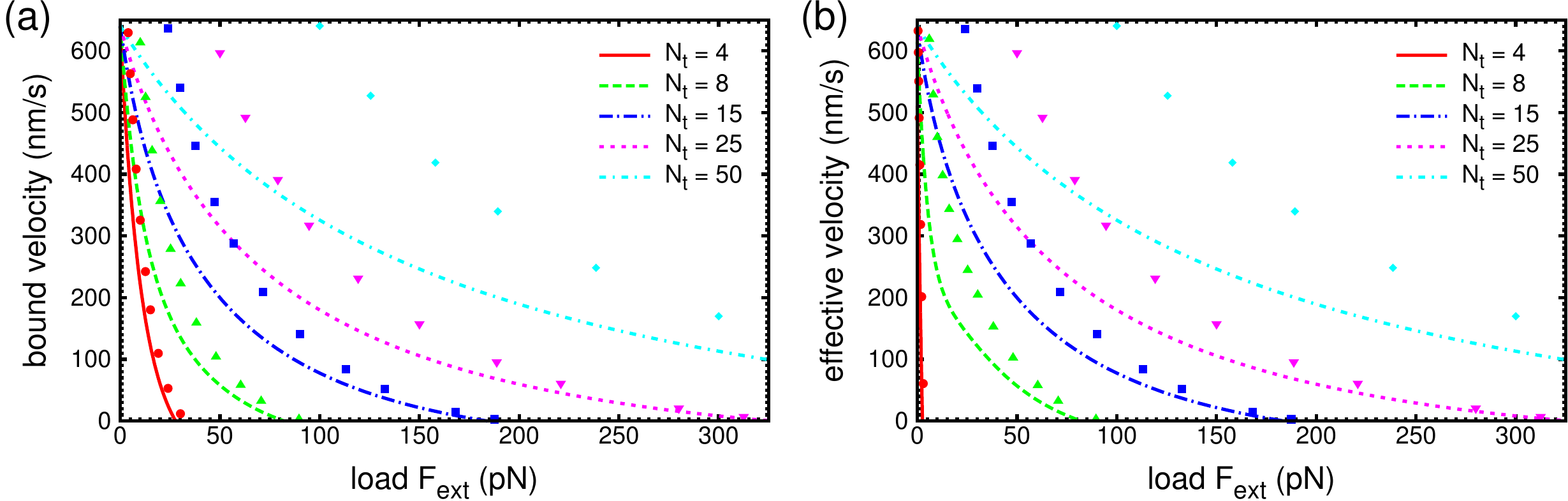}{Validation of the parallel cluster model:
comparison of analytical results using the PCM (lines) with computer
simulations with individual motor strains (symbols). Analytical results
for the parallel cluster model with constant external load. (a) Average
bound velocity $\vb$ and (b) average effective velocity $\veff$ as
function of the external load per motor $\fext/\nt$ for ensemble sizes
$\nt = 4$, $8$, $15$, $25$ and $50$. The mobility of the free ensemble
entering $\veff$ is $\eta = 10^3 \mob$. Constant parameters are listed
in \tab{Tab01}.}

\fig{FigS03} plots the bound velocity in (a) and the effective velocity
in (b) as function of the constant external load $\fext$. Due to the
smaller number of bound motors at small values of the external load, the
bound velocity as well as the effective velocity are strongly increased
in comparison with the results from the PCM. Because the effect of the
internal strains becomes negligible for large external load, the stall
force is unchanged by the different choice of the off-rate.

\subsection{Linear load}


\FIG{FigS04}{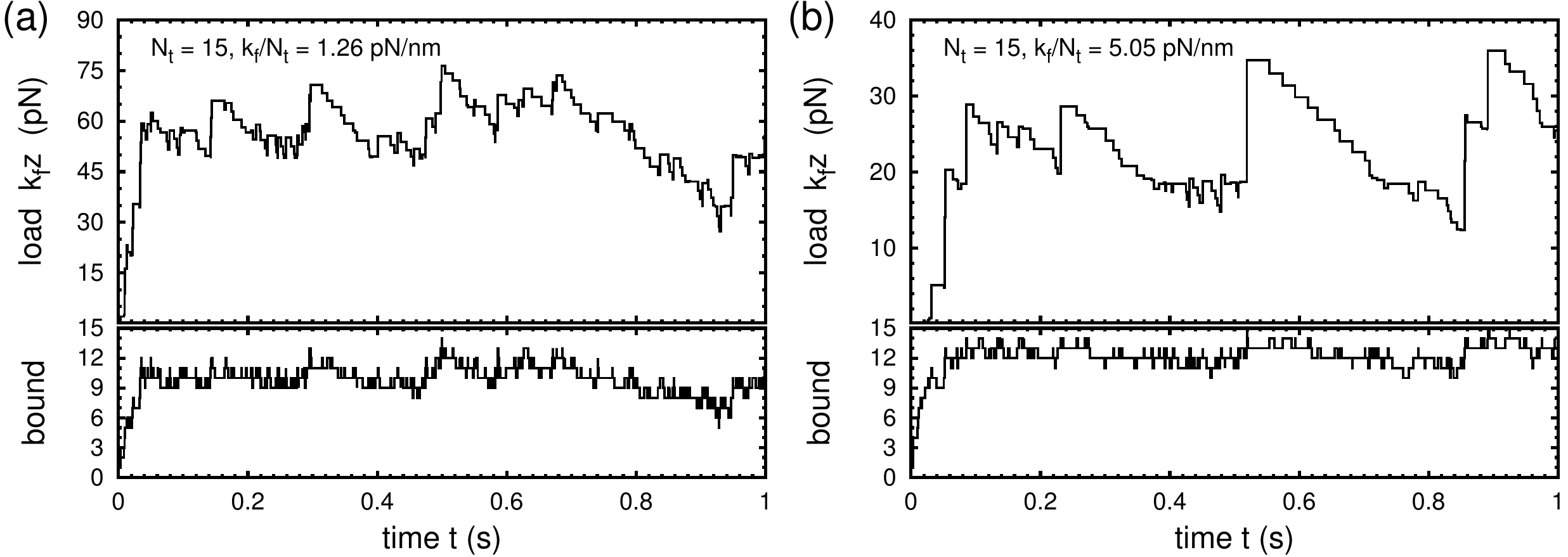}{Stochastic trajectories for linear external load.
External load $\kf z$ on the ensemble (upper panel) and number $i$ of
bound motors (lower panel) as function of time $t$ for ensemble size
$\nt=15$ and external elastic constant per motor (a) $\kf/\nt = 1.26
\pN\nm^{-1}$ and (b) $\kf/\nt = 5.05 \pN\nm^{-1}$. In (a) and (b)
detached ensembles are stationary with mobility $\eta=0$. Constant
parameters are listed in \tab{Tab01}.}

\fig{FigS04}(a) shows a trajectory for an ensemble with $\nt=15$ motors
working against the external elastic constant $\kf/\nt = 1.26
\pN\nm^{-1}$ (that is $\kf = 18.9 \pN\nm^{-1}$). For such a large
ensemble size, detachment is irrelevant even for $z=0$. Due to the large
external elastic constant, the transient towards the isometric state is
very short. The typical load $\kf z \simeq 60\pN$ in the isometric state
corresponds to the ensemble position $z \simeq 3.2\nm$ and the critical
elastic constant $\kfc/i \simeq 2 \pN\nm^{-1}$. For $\kf = 18.9
\pN\nm^{-1}$, the LTE distribution shifts towards the weakly-bound state
for $i \leqslant 9$ so that only states with $i > 9$ can contribute to
ensemble movement. Because of the slow unbinding from the weakly-bound
state, values of $i \leqslant 9$ are only observed when $z$ fluctuates
to values below $z \simeq 3.2\nm$. \fig{FigS04}(b) shows a trajectory
for $\nt=15$ with $\kf/\nt = 5.05 \pN\nm^{-1}$ (that is $\kf = 75.8
\pN\nm^{-1}$). Even for $i=\nt$, this $\kf$ is close to the critical
elastic constant for $z=0$ and the load in the isometric state is
significantly reduced in comparison to (a). The typical value of $\kf z
\simeq 19\pN$ corresponds to an ensemble position $z \simeq 0.25\nm$ for
which the critical elastic constant for the LTE transition is $\kfc/i
\simeq 6 \pN\nm^{-1}$. Therefore, the LTE distribution shifts towards
the weakly-bound state for $i \leqslant 12$. Because unbinding is slowed
down for a larger number of states, the typical number of bound motors
increases relative to (a). The isometric load, on the other hand,
reduces because only a few states with $i > 12$ can contribute to the
forward movement. For the large ensemble sizes shown in \fig{FigS04},
the alternating forward and backward motion in the isometric state
displays a characteristic pattern of rapid increase of $\kf z$
concomitant with an increase of $i$, followed by a gradual decrease of
$\kf z$ accompanied by a decrease of $i$. This pattern, which becomes
more pronounced for larger values of the external elastic constant, is
reminiscent of oscillation pattern for ensembles of motors working
against an elastic element \cite{a:JuelicherProst1997}. A Fourier
analysis, however, has not shown any characteristic time scale for the
fluctuations.

\FIG{FigS05}{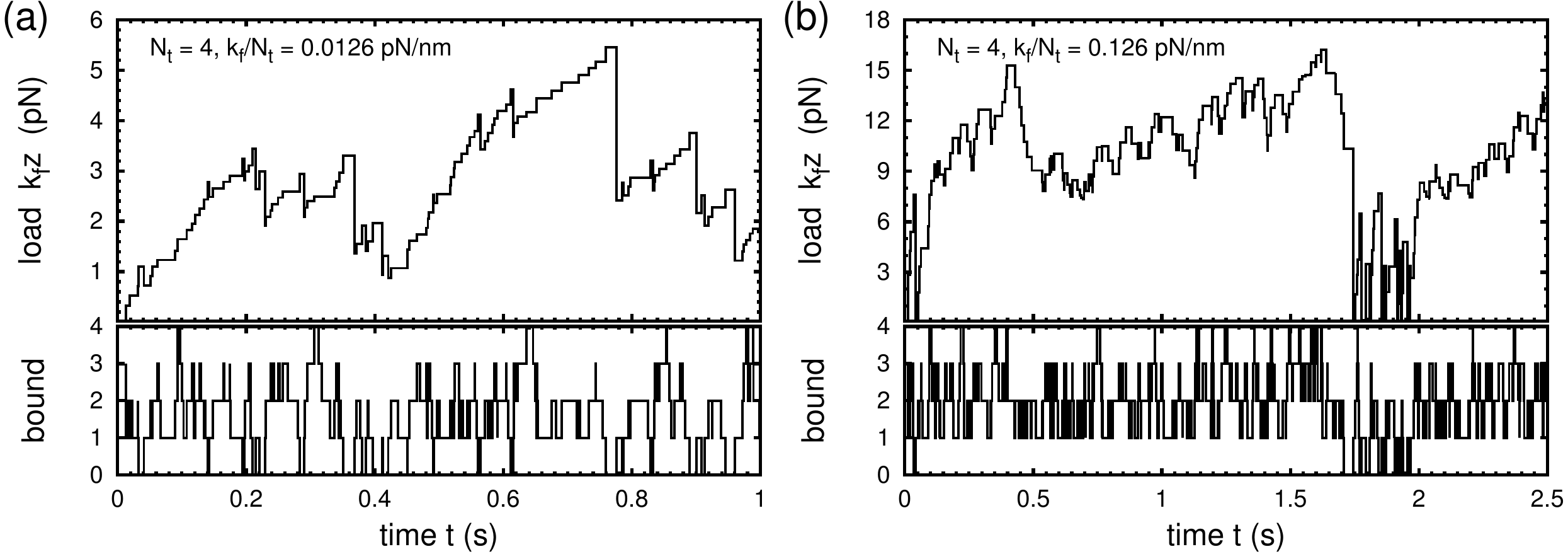}{Stochastic trajectories for linear external load.
Elastic load $\kf z$ on the ensemble (upper panel) and number $i$ of
bound motors (lower panel) as function of time $t$ for ensemble size
$\nt=4$ and external elastic constant per motor (a) $\kf /\nt = 0.126
\pN\nm^{-1}$ and (b) $\kf/\nt = 1.26 \pN\nm^{-1}$. In (a) and (b)
detached ensembles slide backwards with mobility $\eta = 10^3 \mob$.
Constant parameters are listed in \tab{Tab01}.}

\fig{FigS05} plots two trajectories for ensembles with finite mobility
$\eta = 10^3 \mob$. The other parameters are as in \fig{Fig16} of the
main text. For a small filament stiffness, the detached ensemble is
usually rescued before $z=0$ is reached. The typical external load is
thus larger than for $\eta \to \infty$ but still smaller than the
isometric load. For a larger external elastic constant, the trajectory
is identical to the one for $\eta \to \infty$ because the backsteps are
large and reset the ensemble to $z=0$.


\FIG{FigS06}{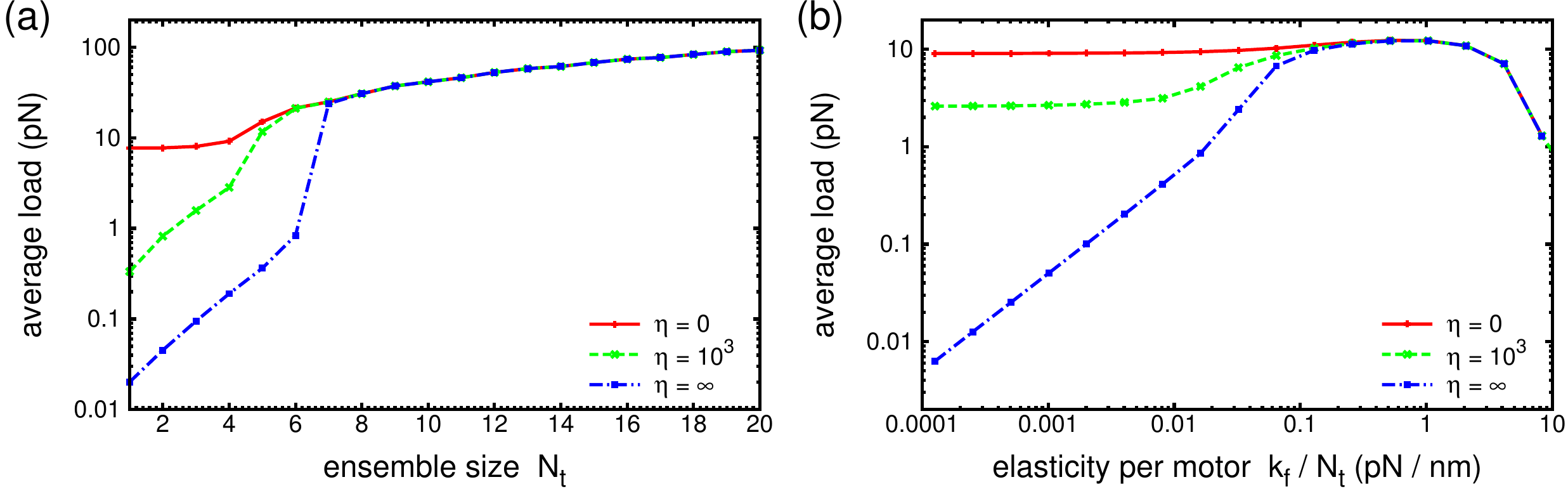}{Numerical results for the parallel cluster model
with linear external load: average external load $\favg$ for $\eta=0$,
$\eta = 10^3 \mob$ and $\eta \to \infty$. (a) $\favg$ as function of
ensemble size $\nt$ for $\kf/\nt = 0.00379 \pN\nm^{-1}$ of the external
elastic constant. (b) $\favg$ as function of external elastic constant
for $\nt=4$. Constant parameters are listed in \tab{Tab01}.}

\fig{FigS06} compares the dependence of the average external load on
ensemble size $\nt$ and external elastic constant $\kf/\nt$ for finite
mobility $\eta = 10^3 \mob$ to the limits $\eta=0$ and $\eta \to
\infty$. As expected, the average external load for finite mobility is
intermediate between the extreme limits. As function of $\nt$ for
constant $\kf/\nt$, the dependence of $\favg$ at finite $\eta$ resembles
that of $\favg$ for $\eta \to \infty$: it increases linearly for small
$\nt$ and jumps to the isometric value at smaller $\kf/\nt$ that $\favg$
for $\eta \to \infty$. As function of $\kf/\nt$ at given $\nt$, the
average external load for finite $\eta$ is constant as for $\eta=0$ but
smaller than the isometric load. The transition to the isometric value
occurs at the same $\nt$ as for $\eta \to \infty$.

\section{Variation of model parameters}\label{app:AppA}

In the main body of the manuscript, one set of parameters was used to
describe the dynamics of \myo ensembles and only the dependence on
$\fext$, $\kf$ and $\eta$ was investigated. There are, however, many
different types of \myo motors for which the exact parameter values vary
and indeed different sets of parameters are used in models (see
\tab{Tab01}). The discussion of the force-velocity relation for constant
external load following \fig{Fig10} has already demonstrated how
experimentally accessible quantities such as load free velocity
$\vb(F=0)$ and stall force $\fs$ are determined by our model parameters.
A particularly characteristic quantity for different types of \myo is
the duty ratio $\rds$ of individual motors.  For example, the duty ratio
of smooth muscle \myo has been measured as $\rds \simeq 0.04$ (see
Ref.~\cite{a:ThoresenEtAl2011}) while for non-muscle \myo a duty ratio
of $\rds \simeq 0.23$ was found (see Ref.~\cite{a:SoaresEtAl2011}).
Moreover, the single motor duty ratio depends on \atp concentration
through the off-rate from the post-power-stroke state, which is a first
order function of \atp concentration \cite{a:WalcottEtAl2012}. While the
\atp concentration is rather stable \emph{in vivo}, it can be varied
\emph{in vitro} over several orders of magnitude, thus varying $\rds$.
For our set of model parameters, the single motor duty ratio for
vanishing load was $\rds \simeq 0.33$. In this appendix, we discuss
analytical results for constant external load for $\rds \simeq 0.1 <
0.33$ and $\rds \simeq 0.67 > 0.33$. The single motor duty ratio is
change through the unloaded off-rate $k_{20}^0$ from the
post-power-stroke state, because it is accessible experimentally and it
is the most convenient determinant of $\rds$ in our model. Changing the
off-rate $k_{10}$ from the weakly-bound state is not very efficient,
because unbinding proceeds mainly from the post-power-stroke state.
Changing the on-rate $k_{01}$, on the other hand, would also change the
unloaded velocity $\vb(F=0)$ directly and not only through the ensemble
duty ratio.

\subsection{Small duty ratio}\label{app:small}

In this section, we demonstrate the effect of a small duty ratio on the
analytical results for constant external load. The unloaded off-rate
from the post-power-stroke state is set to $k_{20}^0 \simeq 360\Hz$ so
that the single motor duty ratio is $\rds \simeq 0.1$. Other parameters
are as in \tab{Tab01}.

\FIG{FigS07}{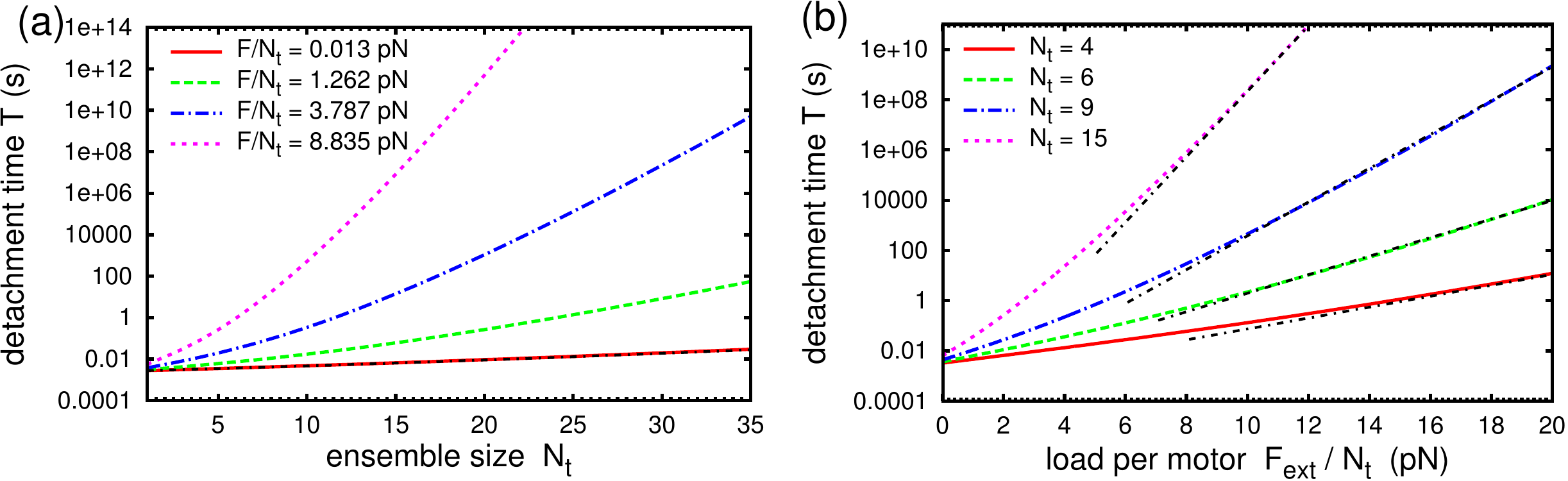}{Analytical results for the parallel cluster model
with constant external load at reduced duty ratio of individual motors:
average detachment time $\tdet$. (a) $\tdet$ as function of ensemble
size $\nt$ for the values $\fext/\nt = 0.0126\pN$, $1.262\pN$,
$3.787\pN$ and $8.835\pN$ of the external load per motor. The black,
dash-dotted curve is the approximation of \eq{BoundTimeApprx} for
$\fext/\nt = 0$.  (b) $\tdet$ as function of the external load per motor
$\fext/\nt$ for ensemble sizes $\nt=4$, $6$, $9$ and $15$. Black,
dash-dotted curves are exponential approximations. The unloaded off-rate
from the post-power-stroke state is $k_{20}^0 = 360\s^{-1}$ so that the
duty ratio of a single, unloaded motor is $\rds \simeq 0.1$. Other
parameters are as listed in \tab{Tab01}.}

\fig{FigS07}(a) plots the detachment time as function of ensemble size
for different values of the external load per motor. The qualitative
dependence of $\tdet$ on $\nt$ is the same as for $\rds \simeq 0.33$
(see \fig{Fig05}), although $\tdet$ is significantly smaller at $\nt=1$
and the increase of $\tdet$ with $\nt$ is weaker. This is particularly
striking for $\fext/\nt \simeq 0.013$ were $\tdet$ increases by less
than an order of magnitude from $\nt=1$ to $\nt=35$. Nevertheless,
$\tdet$ at $\fext/\nt \simeq 0.013\pN$ is very well described by the
approximation for vanishing load in \eq{BoundTimeApprx}. \fig{FigS07}(b)
plots the detachment time $\tdet$ as function of $\fext/\nt$ for
different $\nt$.  Again, the increase of $\tdet$ with $\fext/\nt$ is
weaker than for $\rds \simeq 0.33$ but does proceed exponentially for
large enough $\fext/\nt$.

\FIG{FigS08}{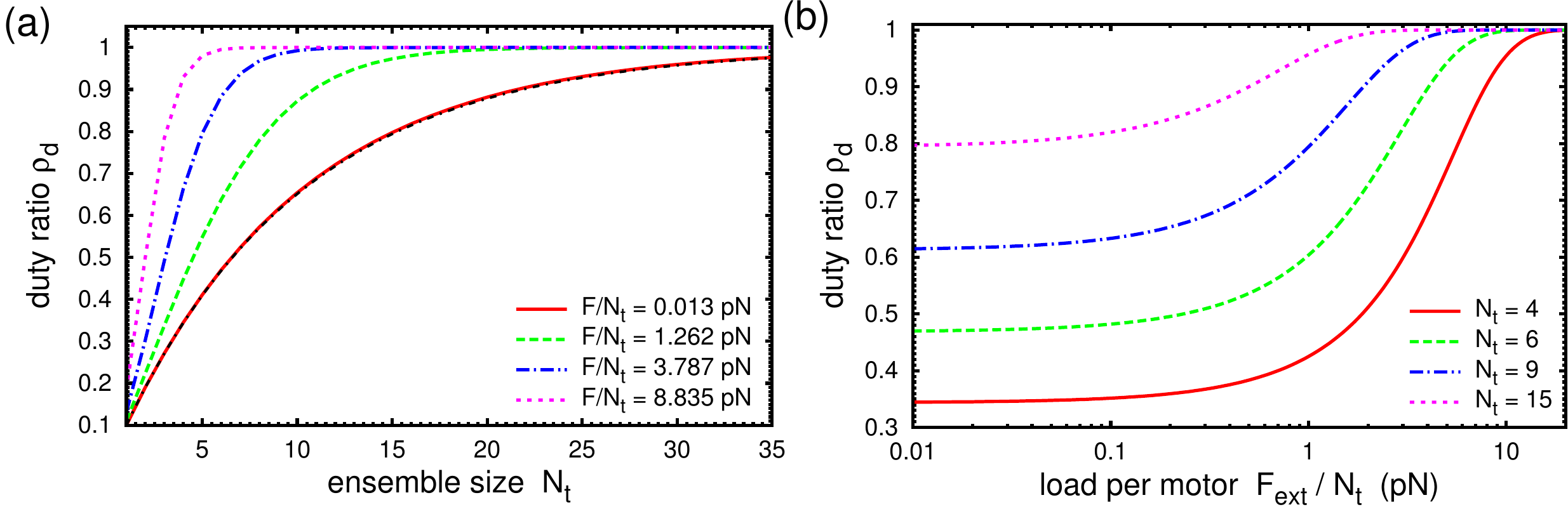}{Analytical results for the parallel cluster model
with constant external load at reduced duty ratio of individual motors:
ensemble duty ratio $\rd$. (a) $\rd$ as function of ensemble size $\nt$
for the values $\fext/\nt = 0.0126\pN$, $1.262\pN$, $3.787\pN$ and
$8.835\pN$ of the external load per motor. The black, dash-dotted curve
is the approximation of \eq{DutyRatioApprx}. (b) $\rd$ as function of
the external load per motor $\fext/\nt $ for ensemble sizes $\nt=4$,
$6$, $9$ and $15$. The unloaded off-rate from the post-power-stroke
state is $k_{20}^0 = 360\Hz$ so that the duty ratio of a single,
unloaded motor is $\rds \simeq 0.1$. Other parameters are as listed in
\tab{Tab01}.}

\fig{FigS08}(a) plots the ensemble duty ratio as function of $\nt$ for
different $\fext/\nt$. Due to the small detachment time, $\rd$ at
$\nt=1$ is reduced in comparison to the case of $\rds \simeq 0.33$ (see
\fig{Fig06}). For $\fext/\nt = 0.013\pN$, $\rd$ follows the
approximation for vanishing load from \eq{DutyRatioApprx}: it relaxes
exponentially from the single motor duty ratio $\rds \simeq 0.1$ and
reaches $\rd \simeq 1$ for ensemble sizes $\nt > 35$. For $\rds \simeq
0.33$, permanent attachment was achieved already for $\nt \geqslant 15$.
With increasing external load, the duty ratio increases more rapidly
with $\nt$. For very large $\fext/\nt$, $\rd \simeq 1$ is reached for
values of $\nt$ that are comparable to the case of $\rds \simeq 0.33$.
\fig{FigS08}(b) plots the ensemble duty ratio as function of
$\fext/\nt$. Although at small external load, the duty ratio is
significantly smaller than for $\rds \simeq 0.33$ in \fig{Fig06}, the
threshold for permanent attachment is reached at similar values of
$\fext/\nt \simeq 15\pN$.

\FIG{FigS09}{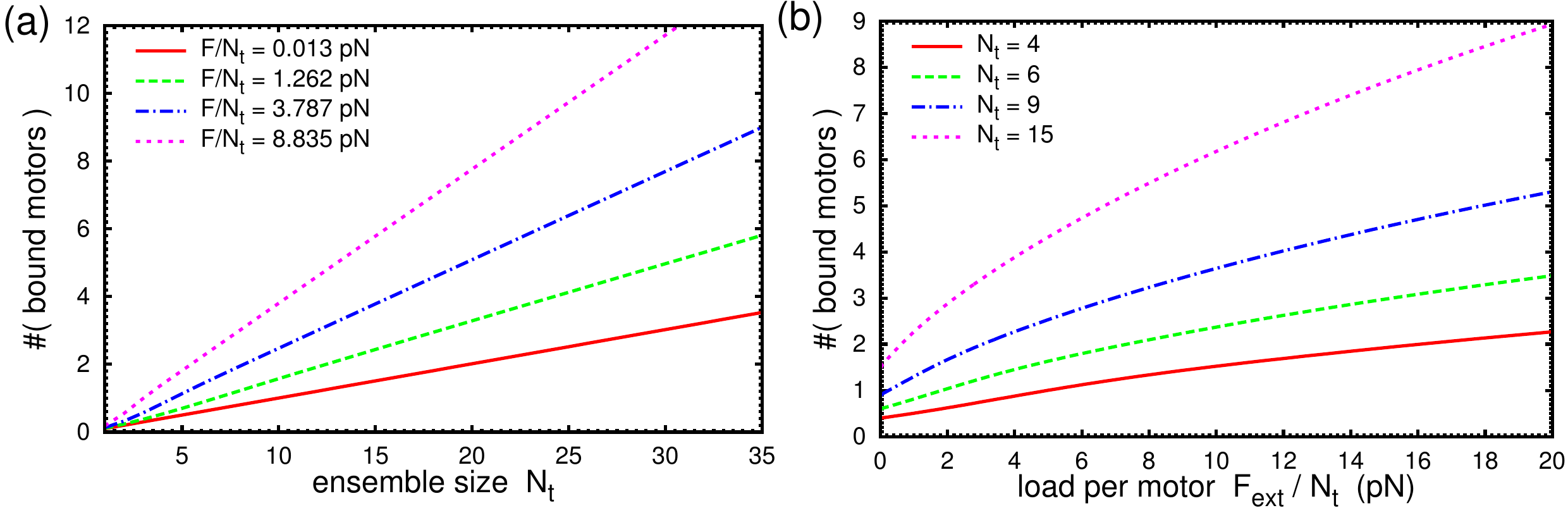}{Analytical results for the parallel cluster model
with constant external load at reduced duty ratio of individual motors:
average number of bound motors $\nb$. (a) $\nb$ as function of ensemble
size $\nt$ for the values $\fext/\nt = 0.0126\pN$, $1.262\pN$,
$3.787\pN$ and $8.835\pN$ of the external load per motor. (b) $\nb$ as
function of external load per motor $\fext/\nt$ for ensemble sizes
$\nt=4$, $6$, $9$ and $15$. The unloaded off-rate from the
post-power-stroke state is $k_{20}^0 = 360\Hz$ so that the duty ratio of
a single, unloaded motor is $\rds \simeq 0.1$. Other parameters are as
listed in \tab{Tab01}.}

\fig{FigS09} plots the average number of bound motors as function of
ensemble size for different values of the external load per motor in (a)
and as function of $\fext/\nt$ for different $\nt$ in (b). As function
of $\nt$, the average number of bound motors increases linearly. The
slope increases with increasing $\fext/\nt$. This increase is more
pronounced than for $\rds \simeq 0.33$ (see \fig{Fig07}). This is
confirmed by the plot of $\nb$ on $\fext/\nt$. Over the range of
external load shown in \fig{FigS09}(b), $\nb$ increases about $6$ fold
while in \fig{Fig07} it increased about $2.5$ fold. This apparently
stronger mechanosensitive response of \myo for smaller $\rds$ is due to
the saturation of $\nb$ towards $\nt$, which has a stronger effect for
large duty ratios.

\FIG{FigS10}{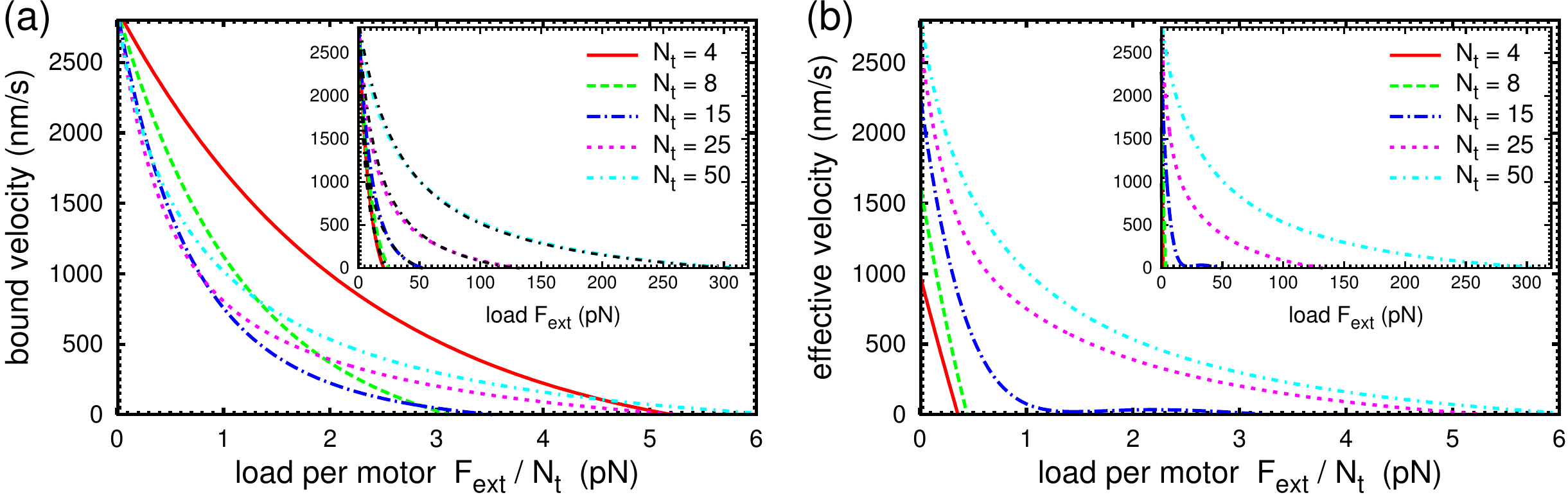}{Analytical results for the parallel cluster model
with constant external load at reduced duty ratio of individual motors.
(a) Average bound velocity $\vb$ and (b) average effective velocity
$\veff$ as function of the external load per motor $\fext/\nt$ for
ensemble sizes $\nt=4$, $8$, $15$, $25$ and $50$. The mobility of the
free ensemble entering $\veff$ is $\eta = 10^3 \mob$. The insets show
$\vb$ and $\veff$ as function of external load $\fext$. Black,
dash-dotted curves in the inset of (a) show the Hill-relation from
\eq{HillRelation} with $\alpha = 0.46\pN$, $0.25\pN$, $0.14\pN$,
$0.12\pN$ and $0.12\pN$ for $\nt=4$, $8$, $15$, $25$ and $50$. The
unloaded off-rate from the post-power-stroke state is $k_{20}^0 =
360\Hz$ so that the duty ratio of a single, unloaded motor is $\rds
\simeq 0.1$. Other parameters are as listed in \tab{Tab01}.}

\fig{FigS10}(a) plots the average bound velocity as function of the
external load per motor for different ensemble sizes. Due to the smaller
number of bound motors at vanishing load, $\nb \simeq 0.1\nt$, the load
free velocity $\vb(\fext=0) \simeq \left[(\nt-\nb) / \nb\right] k_{01} =
9 d k_{01} = 2880 \nm\s^{-1}$ is significantly larger than for $\rds
\simeq 0.33$ (see \fig{Fig10}). Because $\nb$ is also smaller at the
stall force, $\fs/\nt$ is reduced to $\fs/\nt \simeq 6\pN$ for large
$\nt$.  The inset shows that $\vb$ can again be fitted to the Hill
relation from \eq{HillRelation}. The fit parameter $\alpha$ for large
$\nt$ is smaller than in \fig{Fig10}, indicating a stronger curvature of
$\vb(\fext)$. Due to the more frequent detachment of ensembles, the
influence of the off-step $\dzoff_{1j}$ from \eq{dzoffij} on the bound
velocity is already observed for $\nt \leqslant 25$, which reveals
itself by the larger values of $\alpha$. \fig{FigS10}(b) plots the
effective velocity $\veff$ of an ensemble as function of the external
load per motor, $\fext/\nt$. Due to the more frequent detachment of
ensembles with smaller $\rds$, the deviations of $\veff$ from $\vb$ due
to the backward slips are larger than for $\rds \simeq 0.33$ in
\fig{Fig10}. The differences between effective and bound velocity become
negligible only for $\nt > 25$.

\subsection{Large duty ratio}

In this section, we discuss the effect of a large duty ratio on the
analytical results for constant external load. The unloaded off-rate
from the post-power-stroke state is set to $k_{20}^0 \simeq 20\Hz$ so
that the single motor duty ratio is $\rds \simeq 0.67$. All other
parameters are as in \tab{Tab01}.

\FIG{FigS11}{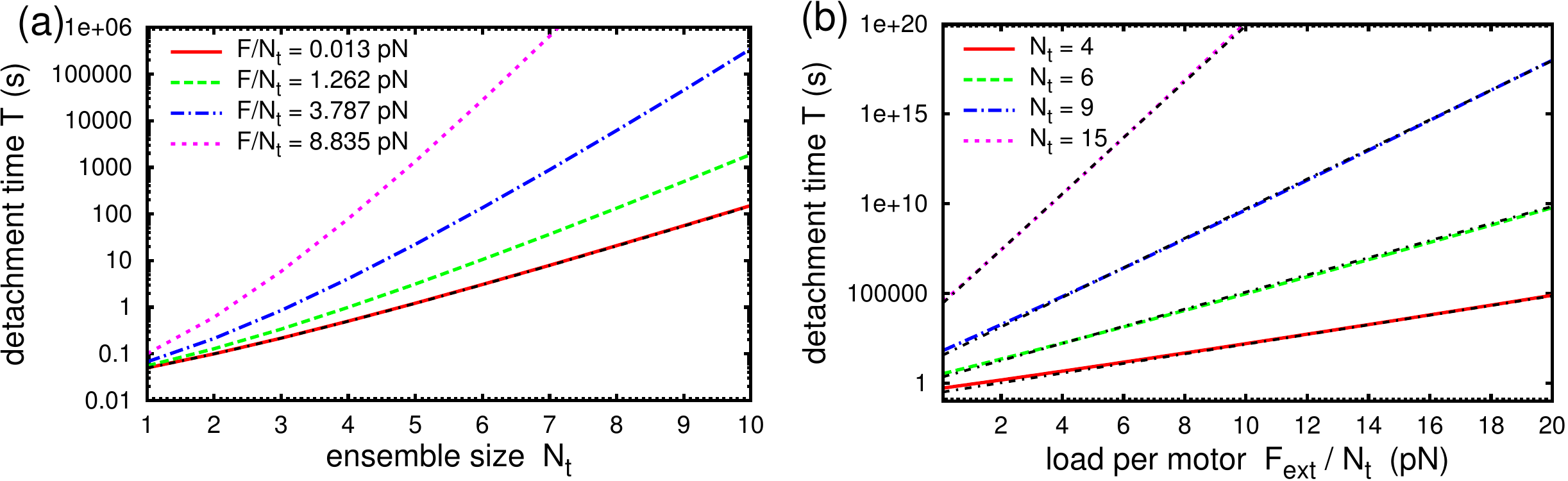}{Analytical results for the parallel cluster model
with constant external load at increased duty ratio of individual
motors: average detachment time $\tdet$. (a) $\tdet$ as function of
ensemble size $\nt$ for the values $\fext/\nt = 0.0126\pN$, $1.262\pN$,
$3.787\pN$ and $8.835\pN$ of the external load per motor. The black,
dash-dotted curve is the approximation of \eq{BoundTimeApprx} for
$\fext/\nt = 0$.  (b) $\tdet$ as function of the external load per motor
$\fext/\nt$ for ensemble sizes $\nt=4$, $6$, $9$ and $15$. Black,
dash-dotted curves are exponential approximations. The unloaded off-rate
from the post-power-stroke state is $k_{20}^0 = 20\Hz$ so that the duty
ratio of a single, unloaded motor is $\rds \simeq 0.67$. Other
parameters are as listed in \tab{Tab01}.}

\fig{FigS11}(a) plots the detachment time as function of ensemble size
for different values of the external load per motor. Despite the larger
values of $\rd$ at $\nt=1$ and the steeper increase with increasing
$\nt$, $\tdet$ has the same qualitative dependence on $\nt$ as for $\rds
< 0.67$. For $\fext/\nt \simeq 0.013$, $\tdet$ is very well described by
the approximation for vanishing load from \eq{BoundTimeApprx}.
\fig{FigS11}(b) plots the detachment time $\tdet$ as function of
$\fext/\nt$ for different $\nt$. The increase of $\tdet$ with
$\fext/\nt$ is significantly more rapid than for smaller $\rds$ and is
described by an exponential over the whole range of $\fext/\nt$.

\FIG{FigS12}{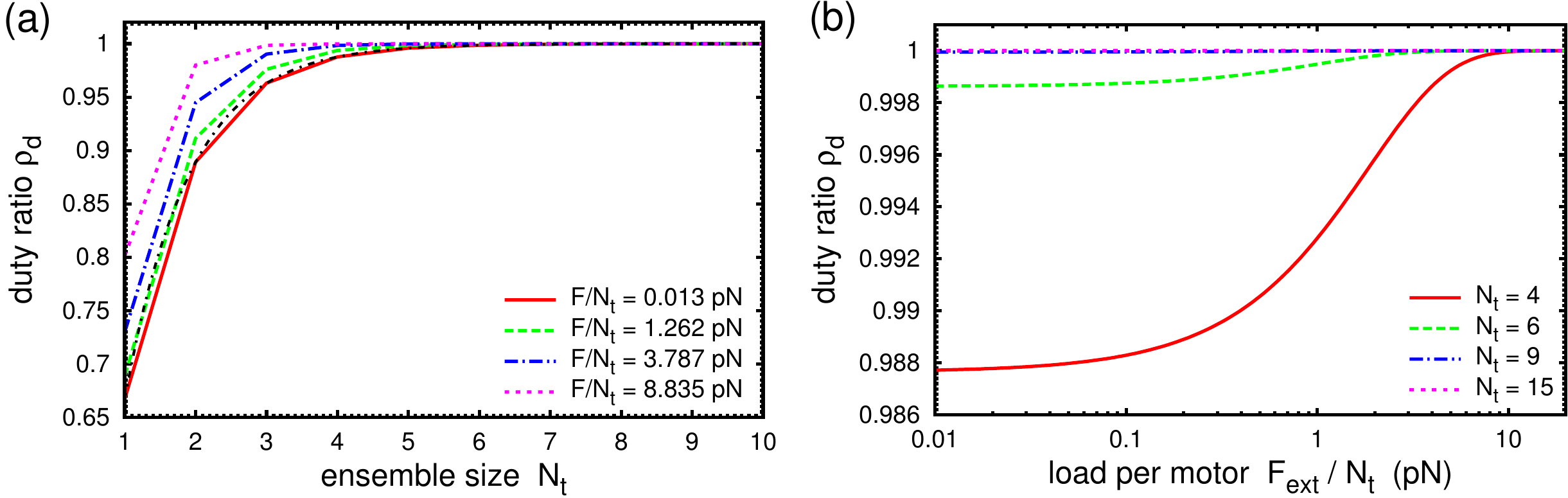}{Analytical results for the parallel cluster model
with constant external load at increased duty ratio of individual
motors: ensemble duty ratio $\rd$. (a) $\rd$ as function of ensemble
size $\nt$ for the values $\fext/\nt = 0.0126\pN$, $1.262\pN$,
$3.787\pN$ and $8.835\pN$ of the external load per motor. The black,
dash-dotted curve is the approximation of \eq{DutyRatioApprx}.  (b)
$\rd$ as function of the external load per motor $\fext/\nt $ for
ensemble sizes $\nt=4$, $6$, $9$ and $15$. The unloaded off-rate from
the post-power-stroke state is $k_{20}^0 = 20\Hz$ so that the duty ratio
of a single, unloaded motor is $\rds \simeq 0.67$. Other parameters are
as listed in \tab{Tab01}.}

\fig{FigS12}(a) plots the ensemble duty ratio as function of $\nt$ for
different values of $\fext/\nt$. For all values of $\fext/\nt$, the same
qualitative behavior of $\tdet$ is observed as for $\rds\simeq 0.33$
(see \fig{Fig06}. For small load, $\rd$ follows the approximation for
vanishing load in \eq{DutyRatioApprx} and reaches $\rd \simeq 1$ already
for $\nt \simeq 5$. \fig{FigS08}(b) plots the ensemble duty ratio as
function of $\fext/\nt$. For all values of $\nt$, the ensemble duty
ratio is practically unity over the whole range of $\fext/\nt$.

\FIG{FigS13}{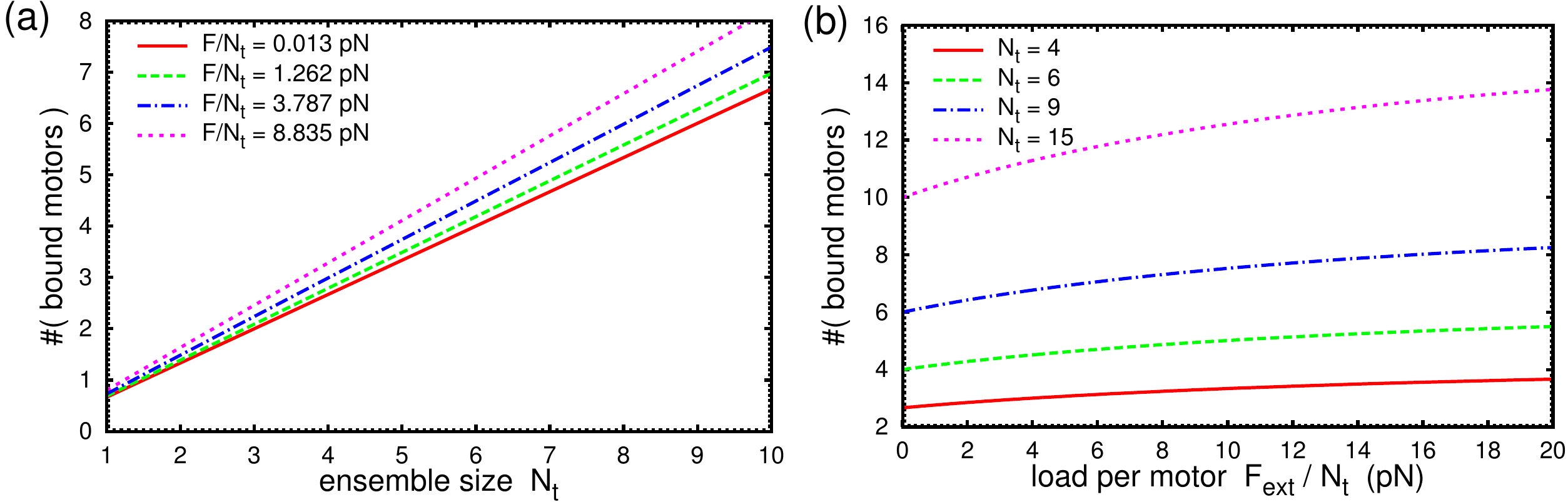}{Analytical results for the parallel cluster model
with constant external load at increased duty ratio of individual
motors: average number of bound motors $\nb$. (a) $\nb$ as function of
ensemble size $\nt$ for the values $\fext/\nt = 0.0126\pN$, $1.262\pN$,
$3.787\pN$ and $8.835\pN$ of the external load per motor. (b) $\nb$ as
function of external load per motor $\fext/\nt $ for ensemble sizes $\nt
= 4$, $6$, $9$ and $15$. The unloaded off-rate from the
post-power-stroke state is $k_{20}^0 = 20\Hz$ and the duty ratio of a
single, unloaded motor is $\rds \simeq 0.67$. Other parameters are as
listed in \tab{Tab01}.}

\fig{FigS13} plots the average number of bound motors $\nb$ as function
of ensemble size $\nt$ for different values of the external load per
motor $\fext/\nt$ in (a) and as function of $\fext/\nt$ for different
$\nt$ in (b). As function of $\nt$, the average number of bound motors
increases linearly but the slope shows a weak dependence on $\fext/\nt$.
This is confirmed by the plot of $\nb$ on $\fext/\nt$. Due to the larger
number of bound motors at $\fext/\nt=0$, the $\nb$ can only increase
weakly with $\fext/\nt$ so that the ensemble with large $\rds$ appears
to be weakly mechanosensitive.

\FIG{FigS14}{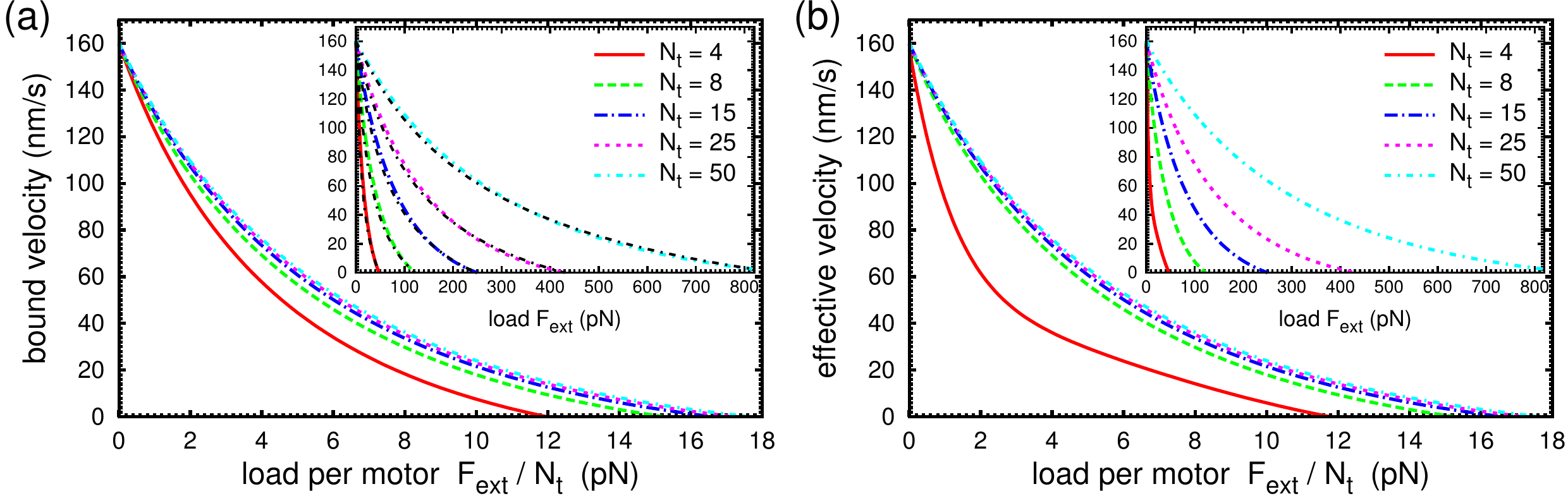}{Analytical results for the parallel cluster model
with constant external load at increased duty ratio of individual
motors. (a) Average bound velocity $\vb$ and (b) average effective
velocity $\veff$ as function of the external load per motor $\fext/\nt$
for ensemble sizes $\nt=4$, $8$, $15$, $25$ and $50$. The mobility of
the free ensemble entering $\veff$ is $\eta = 10^3 \mob$. The insets
show $\vb$ and $\veff$ as function of external load $\fext$. Black,
dash-dotted curves in the inset of (a) show the Hill-relation from
\eq{HillRelation} with $\alpha = 0.28\pN$, $0.28\pN$, $0.30\pN$,
$0.32\pN$ and $0.34\pN$ for $\nt=4$, $8$, $15$, $25$ and $50$. The
unloaded off-rate from the post-power-stroke state is $k_{20}^0 = 20\Hz$
so that the duty ratio of a single, unloaded motor is $\rds \simeq
0.67$. Other parameters are as listed in \tab{Tab01}.}

\fig{FigS14}(a) plots the average bound velocity as function of external
load per motor for different ensemble sizes. Due to the larger number of
bound motors at $\fext/\nt = 0$, the load free velocity $\vb(\fext=0)
\simeq \left[(\nt - \nb) / \nb\right] k_{01} = 0.5 d k_{01} = 160
\nm\s^{-1}$ is significantly smaller than for $\rds \simeq 0.33$ (see
\fig{Fig10}). Because of the larger number of bound motors at the stall
force, $\fs/\nt$ is increased to $\fs/\nt \simeq 16\pN$ for large $\nt$.
The force-velocity relation is constant for ensemble sizes above $\nt
\geqslant 8$. The inset shows that $\vb$ can again be fitted to the Hill
relation from \eq{HillRelation}. The fit parameter $\alpha$ is generally
larger than in \fig{Fig10}, indicating a smaller curvature of
$\vb(\fext)$.  Because detachment is rare even for $\nt=4$, there are no
significant deviations from the shape of the force-velocity curve for
large $\nt$. Also the effective velocity deviates from the bound
velocity only for $\nt=4$.

The comparison of the results for constant external load for different
values of the single motor duty ratio $\rds$ has shown that the
qualitative behavior of the motor ensembles is robust to changes of
$\rds$, although the quantitative results do change. In particular the
stochastic effects from ensemble detachment become more pronounced for
smaller $\rds$. The results for the force-velocity relation confirm the
dependence of the load free velocity and the stall force on the model
parameters.

\end{document}